\documentclass[useAMS,usenatbib]{mn2e}
\usepackage{graphicx}
\usepackage{natbib}
\title[On the deceleration of FR\,I jets: mass loading by stellar-winds]{On the deceleration of FR\,I jets: mass loading by stellar-winds}
\author[Perucho et al.]{M. Perucho$^{1}$\thanks{E-mail:
manel.perucho@uv.es}, J. M. Mart\'{\i}$^{1}$, R. A. Laing$^{2}$, P. E. Hardee$^{3}$\\
$^{1}$Departament d'Astronomia i Astrof\'{\i}sica. Universitat de Val\`encia. C/ Dr. Moliner 50, 46100 Burjassot (Val\`encia), Spain\\
$^{2}$European Southern Observatory, Karl-Schwarzschild-Strasse 2, D-85748 Garching-bei-M\"unchen, Germany\\
$^{3}$Department of Physics \& Astronomy, The University of Alabama, Tuscaloosa, AL 35487, USA}

\begin{document}
\date{Released 2012 Xxxxx XX}

\pagerange{\pageref{firstpage}--\pageref{lastpage}} \pubyear{2014}

\label{firstpage}

\maketitle

\begin{abstract}
Jets in low-luminosity radio galaxies are known to decelerate from relativistic speeds on parsec scales to mildly or sub-relativistic speeds on kiloparsec scales. Several mechanisms have been proposed to explain this effect, including strong reconfinement shocks and the growth of instabilities (both leading to boundary-layer entrainment) and mass loading from stellar winds or molecular clouds. We have performed a series of axisymmetric simulations of the early evolution of jets in a realistic ambient medium to probe the effects of mass loading from stellar winds using the code {\emph Ratpenat}. We study the evolution of Fanaroff-Riley Class I (FR\,I)  jets, with kinetic powers $L_{\rm j}\sim 10^{41}-10^{44}\,{\rm erg\, s^{-1}}$, within the first 2~kpc of their evolution, where deceleration by stellar mass loading should be most effective. Mass entrainment rates consistent with present models of stellar mass loss in elliptical galaxies produce deceleration and effective decollimation of weak FR\,I jets within the first kiloparsec. However, powerful FR\,I jets are not decelerated significantly. In those cases where the mass loading is important, the jets show larger opening angles and decollimate at smaller distances, but the overall structure and dynamics of the bow-shock are similar to those of unloaded jets with the same power and thrust. According to our results, the flaring observed on kpc scales is initiated by mass loading in the weaker FR\,I jets and by reconfinement shocks or the growth of instabilities in the more powerful jets. The final mechanism of decollimation and deceleration is always the development of disruptive pinching modes.

\end{abstract}


\section{Introduction}
%            %%%%%%%%%%%
\label{intro}

Jets from radio-loud AGN show two characteristic morphologies,
associated with the FR\,I and FR\,II classes defined by
\citet{fr74}. Jets in FR\,I sources \citep[e.g., 3C~31,][]{lb02a} expand rapidly on 
kiloparsec scales, whereas
those in the more powerful FR\,II sources \citep[e.g., Cyg~A,][]{cb96} are
highly collimated until they terminate in compact hot-spots. FR\,I and FR\,II jets appear morphologically similar on parsec scales and
both show evidence for relativistic speeds, although the former appear
to be somewhat slower and to have significant velocity gradients
\citep{gi01,cg08,me11}.  The current paradigm for FR\,I jets is that, unlike FR\,II
jets, they are decelerated by entrainment of gas \citep{bi84,la93,la96}. Whether this susceptibility to entrainment is solely a
function of the power of the jet and its external environment or is
due to a more fundamental difference is currently a matter of debate.
 
Two main processes have been invoked to explain entrainment in FR\,I 
jets: (i) mixing in a turbulent shear layer between
the jet and the ambient \citep{dy86,dy93,bi94,wa09}, and (ii) injection from
stellar mass loss \citep{ph83,ko94,BL96,lb02b,HB06}. Concerning the process
of entrainment through a turbulent shear layer, \citet[hereafter PM07]{PM07} showed, via a 2D axisymmetric simulation, that a
recollimation shock in a light jet formed in reaction to steep interstellar density and pressure
gradients may trigger nonlinear
perturbations that lead to jet disruption and mixing with the external
medium. \citet*{me08} and \citet{ro08} discussed the deceleration of FR\,I
jets at discontinuities in the ambient medium and by the growth of
helical instabilities in a jet propagating through a homogeneous
ambient medium, respectively. 

%Also \cite{me08} have studied the influence of a
%discontinuous jump in the external density gas on the jet disruption in
%hybrid FR\,I/FR\,II sources (HYMORS), where the jet and counter-jet present
%different morphology and this is ascribed to such environmental
%irregularities. However, this work is focused on those specific cases in
%which the jet could cross a satellite galaxy of its host. Thus, in
%general, the triggering of nonlinear instabilities by reconfinement
%shocks induced by pressure mismatch with the ambient seems to be the
%most probable reason for entrainment through mixing and deceleration of
%FR\,I jets.

The influence of mass loading by stellar winds in FR\,I jets was studied by
\cite{ko94}, who showed that this problem can be treated as a
hydrodynamical one, as the gyroradii of the particles are much smaller
than the size of the interaction region between the jet and the stellar
wind. Thus,  this problem can be reduced to that of a distributed source
of mass that is injected into and thereafter advected with the jet
flow. \citet*[hereafter BLK]{BL96} studied the
effect of mass loading by a typical stellar population 
on jets with different properties by solving the equations of evolution
in steady state. In particular, they focused on light and hot electron/proton jets. They
concluded that these jets can be efficiently decelerated by this
mechanism, with some differences depending on the thermodynamical
properties of the jets: hotter jets cool down due to entrainment of the cold wind
particles, whereas relatively colder jets gain temperature in the
entrainment region due to dissipation. \cite{HB06} studied the different
ranges of stellar mass-loss rates and jet powers that could imply
efficient jet deceleration within the host galaxy, and concluded that
the stellar wind from a single Wolf-Rayet star  could be enough to
decelerate a weak FR\,I jet.

\citet{lb02b} constructed a one-dimensional model of the jet in 3C~31
using the basic conservation laws and the velocity field inferred by
\citet{lb02a}, which includes a substantial deceleration of the jet
within the flaring region, located at 1--3 kpc from the central
engine. Close to the outer boundary of this region, at $3.2$\,kpc,
\cite{lb02b} found maxima in the mass entrainment rate (per unit length
of the jet) both for their reference model and for the stellar mass
input. However, the latter is much smaller. Although the expected mass
injection by stellar winds seems to be enough to counterbalance the
effects of adiabatic expansion and to keep the velocity fairly constant
at the beginning of the flaring region, the continuous deceleration in
the jet indicated by \citet{lb02a}'s results requires a monotonic
increase of the entrainment rate at large distances. This cannot be
the result of mass loss from stars, whose density falls rapidly with
increasing radius. \citet{lb02b} concluded that entrainment from the
galactic atmosphere across the boundary layer of the jet is the
dominant mass input process far from the nucleus in this powerful
($10^{44}$\,erg\,s$^{-1}$) FR\,I jet, but that stellar mass loss might 
also contribute near the flaring point.

\citet{lb14} have recently presented observations and kinematic models
for ten FR\,I radio galaxies. They concluded that the deceleration of
FR\,I jets after the flaring point is a gradual process in which the
transverse velocity profile evolves from constant to slower at the
edges than on-axis. Mass loading by stars is distributed throughout the jet volume and is therefore not expected to create
large transverse velocity gradients. It is therefore unlikely to be
the dominant cause of deceleration on these scales: boundary-layer
entrainment is more plausible. A contribution 
from stellar mass loading, particularly on smaller scales, is not excluded, however. 
This conclusion was based on models for the
six sources with adequately determined transverse velocity
profiles. It is consistent with, but not required by the models for
the remaining four. All of the cases with well-determined profiles are powerful FR\,I sources. It remains
possible that weaker sources such as M84 and NGC193, for which the models do not show conclusive evidence for velocity gradients, are decelerated primarily by stellar mass loading.

In PM07 the deceleration and decollimation of the jet was produced by
the development of a recollimation shock formed at a few kiloparsecs
from the central engine due to the density decrease in the ambient IGM. 
The expansion caused by the jet overpressure as it propagated 
outwards and the subsequent recollimation shock effectively
decelerated the jet after the shock, triggering the development of
Kelvin-Helmholtz (KH) instabilities and enhancing the mass entrainment
beyond this point. The main purposes of this paper are to extend the analysis of PM07 
to include the
combined effects of the mass loading of the jet by stellar winds and
the recollimation shock, to compare with the earlier
results and to cover a larger range of jet powers. We also investigate the effects of changing the boundary
conditions of the simulations.

We present simulations of jets with typical FR\,I powers, including
source terms in the equations of relativistic hydrodynamics that
account for the mass entrainment by winds from the stellar
populations expected for typical host galaxies. Our aim is to
understand the conditions under which FR\,I jets can be decelerated by
this type of entrainment and those for which either another process or a
different stellar population are required.

The paper is structured as follows. The setup of the simulations,
together with the parameters used are presented in Section~\ref{s:sim}, and the
results are given in Section~\ref{s:res}. Section~\ref{disc} discusses the implications.
A summary and the conclusions of this work are given in
Section~\ref{conc}.

\section{Simulations}
%            %%%%%%%%%%%
\label{s:sim}

\subsection{Ambient medium and stellar mass-loss profile}
%                  ---------------------------------------------
\label{ss:ambient}

\begin{figure*}
 \includegraphics[width=0.45\textwidth]{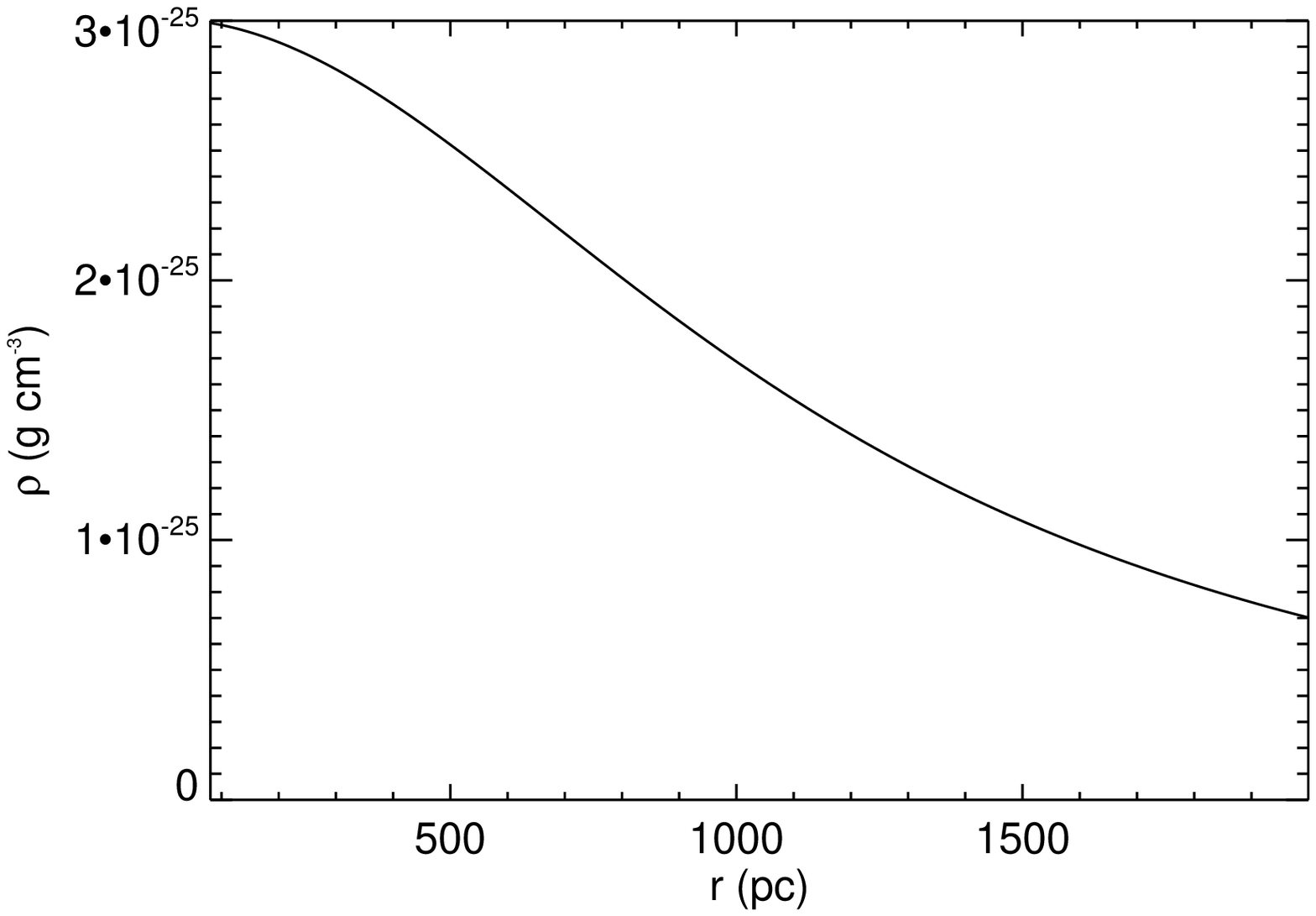}
 \includegraphics[width=0.45\textwidth]{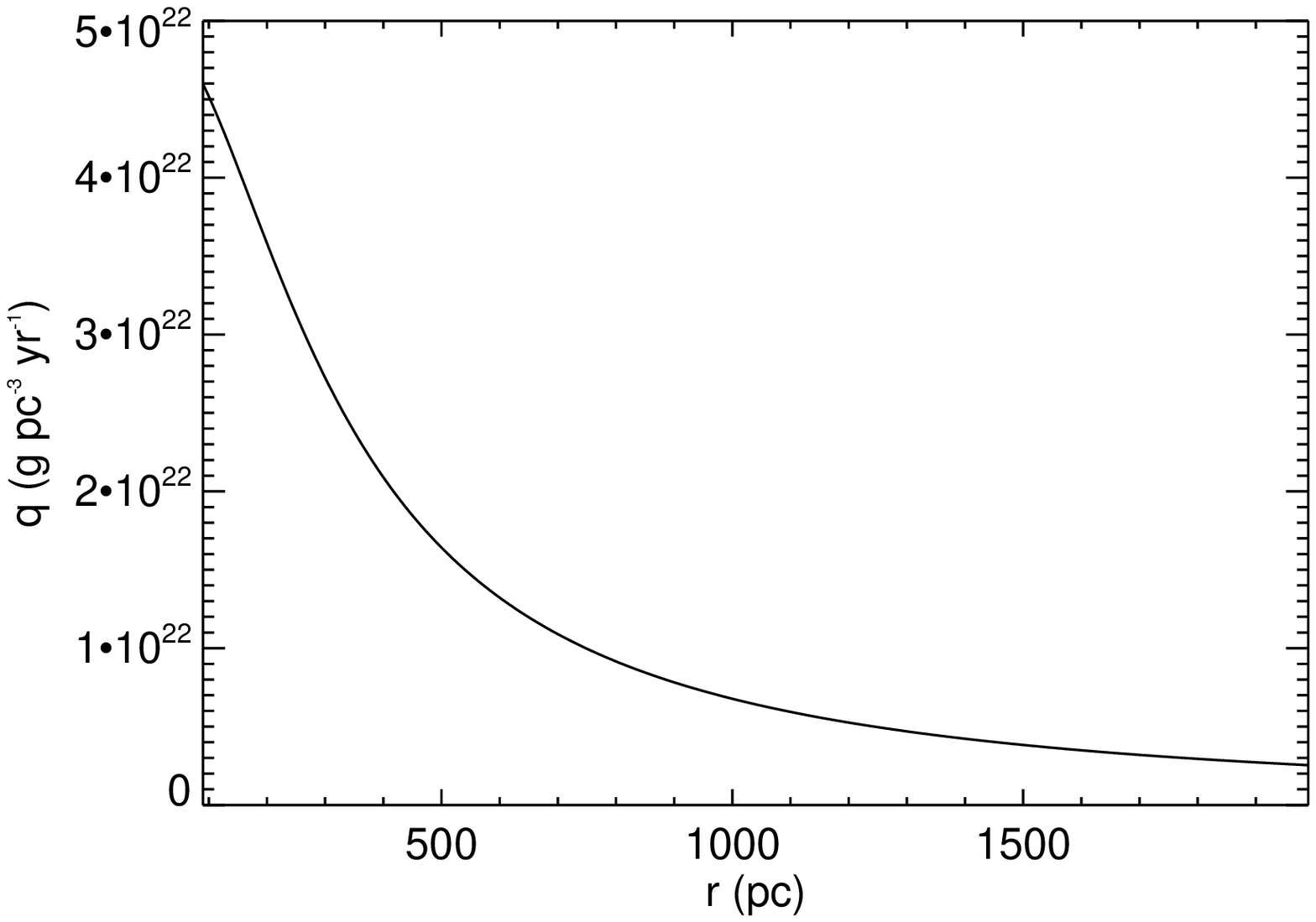}
 \caption{Radial profiles of ambient density (left panel) and stellar mass-loss rate per unit volume (right panel) in the simulated region for models Pr, Po, A, B and C. In the case of model D, the normalization factor for the mass-loss rate is smaller by one order of magnitude.}
 \label{fig:setup}
 \end{figure*}

  In the simulations, the ambient medium is composed of a decreasing
density atmosphere of ionized hydrogen in hydrostatic equilibrium. The
profile for the number density of such a medium is \citep[][PM07]{ha02}:
\begin{equation}\label{next}
  n_{\rm ext} = n_0 \left[1 +
\left(\frac{r}{r_{\rm c}}\right)^2\right]^{-3\beta_{\rm atm}/2},
\end{equation}
with $n_0 = 0.18$ cm$^{-3}$, $r_{\rm c} = 1.2$ kpc and  $\beta_{\rm atm} =
0.73$, where the contribution from the galaxy group atmosphere as  used by PM07 is not included for simplicity.\footnote{FR\,I radio sources are found in galaxies without group components: an example is NGC~315 (\citealt{cr08}).}  This profile, in the simulated region (between 80~pc and 2~kpc), is plotted in Fig.~\ref{fig:setup}. The temperature profile \citep[][PM07]{ha02} is close to constant with $T_{\rm ext} = 4.9 \times 10^6$\,K over the simulated region.
 The external
pressure is derived from the number density and temperature 
assuming pure ionized hydrogen \citep[][PM07]{ha02}:
\begin{equation}\label{pext}
  p_{\rm ext} = \frac{k_{\rm B} T_{\rm ext}}{\mu X} n_{\rm ext},
\end{equation}
where $\mu= 0.5$ is the mass per particle in amu, $X=1$ is the abundance
of hydrogen per mass unit, and $k_{\rm B}$ is Boltzmann's constant.

  In addition to this external medium, we have included a law that accounts
for mass entrainment into the jets from stellar winds, derived from
a Nuker distribution of surface brightness in an elliptical galaxy \citep{la07}:
\begin{equation}\label{nuker}
   Q = Q_0 \left(\frac{r_{\rm b}}{r}\right)^\gamma
\left[1+\left(\frac{r}{r_{\rm b}}\right)^\alpha\right]^{(\gamma-\beta)/\alpha},
\end{equation}
where $Q_0$ is the mass loss rate per unit area at $r = 0$, ${r_{\rm b}}$ is the
characteristic radius of the profile and $\alpha$, $\beta$ and $\gamma$
are constants. The profile used in the simulations corresponds to the 
deprojection \citep{bm98} of the Nuker profile with $r_{\rm b}=265\,{\rm pc}$, $\alpha=2.0$, $\beta=0.46$ and $\gamma=0.0$, which fall well within the observed range \citep{la07}. This profile has been deprojected using a stellar distribution extending out to 20 kpc. The resulting mass-loss rate per unit volume in the simulated region is plotted against distance in Fig.~\ref{fig:setup}. This profile results in an increasing total mass load per unit distance up to the end of the simulated grid. For other values of the parameters,
the mass-loading rate can turn over within the 2 kpc simulation
distance.

We parameterize the  mass input by the central mass-loss rate per unit volume for the deprojected profile, $q_0$.
We take $q_0 = 4.95\times 10^{22} {\rm g\,yr^{-1}\,pc^{-3}}$  (a factor of ten
smaller in the case of model D, see Table~\ref{tab1}). For comparison, the
central rate for the reference model of BLK is $q_0 = 2.36\times 10^{22} {\rm g\,yr^{-1}\,pc^{-3}}$.
Our assumed profile gives a mass input rate per unit time and
volume of $5.9\times 10^{21} {\rm g\,yr^{-1}\,pc^{-3}}$ at 1.1\,kpc 
(Fig.~\ref{fig:setup}). At the same location (the starting point of their model), \citet{lb02b} find a
very similar value of $6.4\times 10^{21} {\rm g\,yr^{-1}\,pc^{-3}}$
(their Figs~1 and 11); that used by BLK in their reference model (their equation 12, Tables 
1 and 2) is very slightly higher ($7.2\times 10^{21} {\rm
g\,yr^{-1}\,pc^{-3}}$).  These differences are well within the
uncertainties of the true mass input rate.

%%%%%%%%%%%%%%%%%%%%%%%%%%%%%%%%%%%%%%%%%%%%%%%%%%%%%%%%%%%%%%%%%%%%%%%%%%%%%%%%%%%%%%%%%%%%%%%%%%%%%%%%%%%%%%%%%%%%%%%%%%%%%%%
%
%
%%%%%%%%%%%%%%%%%%%%%%%%%%%%%%%%%%%%%%%%%%%%%%%%%%%%%%%%%%%%%%%%%%%%%%%%%%%%%%%%%%%%%%%%%%%%%%%%%%%%%%%%%%%%%%%%%%%%%%%%%%%%%%%

\subsection{Jet parameters}
%                  ---------------
\label{ss:jets} 

%%%%%%%%%%%%%%%%%%%%%%%%%%%%%%%%%%%%%%%%%%%%%%%%%%%%%%%%%%%%%%%%%%%%%%%%%%%%%%%%
%
\begin{table*}
  \begin{center}
  \caption{Parameters of the simulated jets. (1) model name; (2) injection velocity (3) injection density; (4) jet temperature at injection; (5) jet to
ambient pressure contrast at injection; (6) jet power; (7) central 
mass-entrainment rate; (8) jet length at the end of the simulation; (9) duration of the simulation.}
  \label{tab1}
 {\small
  \begin{tabular}{|l|c|c|c|c|c|c|c|c|}\hline
Model & Velocity & Density  & Temperature &
$P_{\rm j}/P_{\rm amb}$ & $L_{\rm j}$ & $q_0$ & Jet length &
$t_{\rm sim}$ \\
                   & [$c$]              & [g\,cm$^{-3}$]     &
[$K$]                        &                       &
[erg\,s$^{-1}$]                 & [${\rm g\,yr^{-1}\,pc^{-3}}$] & [kpc] &
[Myr]  \\  
\hline
Po &  0.99   & $9.65\times 10^{-30}$ & $3\times 10^{9}$ & $17.6$ &
$10^{44}$ & $4.95\times 10^{22}$ & 2.2 & 0.25  \\
Pr &  0.99   & $9.65\times 10^{-30}$ & $3\times 10^{9}$ & $17.6$  &
$10^{44}$ & $4.95\times 10^{22}$ & 2.2 & 0.25  \\
A0 &  0.95 & $3\times 10^{-33}$ & $3\times 10^{11}$  & $0.54$ & $5\times
10^{41}$ & $0$                            & 1.5 & 1.6  \\
A   &  0.95 & $3\times 10^{-33}$ & $3\times 10^{11}$  & $0.54$ &
$5\times 10^{41}$ & $4.95\times 10^{22}$ & 2.1 & 2.4\\
B    &  0.95 & $3\times 10^{-34}$ & $3\times 10^{12}$  & $0.54$ &
$5\times 10^{41}$ & $4.95\times 10^{22}$ & 2.0 & 2.1\\
C   &  0.95 & $3\times 10^{-35}$ & $3\times 10^{13}$  & $0.54$ &
$5\times 10^{41}$ & $4.95\times 10^{22}$ & 1.8 & 1.9\\
D   &  0.95 & $3\times 10^{-35}$ & $3\times 10^{13}$  & $0.54$ &
$5\times 10^{41}$ & $4.95\times 10^{21}$ & 1.8 & 1.8\\
\hline
 \end{tabular}
 }
 \end{center}

\end{table*}
%
%%%%%%%%%%%%%%%%%%%%%%%%%%%%%%%%%%%%%%%%%%%%%%%%%%%%%%%%%%%%%%%%%%%%%%%%%%%%%%%

Table~\ref{tab1} collects the parameters of the jets in our
simulations. All of the simulated jets are purely leptonic and are injected
at a distance from the galaxy nucleus of 80~pc with a radius $R_{\rm j} =
10$~pc. This assumes an opening angle of $7^\circ$, consistent with recent estimates of the opening angles of parsec-scale FR\,I jets \citep{mu11,an12}.   The ambient density at injection is $3 \times 10^{-25}$~g\,cm$^{-3}$, and the ambient pressure is $2.5 \times 10^{-10}$~erg\,cm$^{-3}$. 

We performed two simulations of jets with the same power as that in
PM07, but including mass loading. Real jets are bipolar, but in
order to save on computation time we have simulated only one
side. This raises the question of the effect of the boundary
conditions at injection. We have therefore made two simulations, identical except for the boundary conditions: {\it Powerful reflecting}
(Pr), for a reflecting condition at the injection boundary of the
numerical box, and {\it Powerful open} (Po), for an open boundary
condition there. These simulations were designed to match as
closely as possible the jet properties derived for 3C~31
\citep{lb02a,lb02b,ha02}, and the jets therefore have  kinetic
luminosities $L_{\rm j} = 10^{44}$ erg\,s$^{-1}$, at the upper end of the power distribution for 
FR\,I sources.

Although models Po and Pr have the same jet power as that studied in
PM07, they are simulated here for a shorter time ($\simeq$0.25~Myr
instead of 7.3~Myr) and over a correspondingly shorter distance
(2.2~kpc instead of 14.5~kpc).  Mass-loading terms are included,
however.  In PM07, the injection of the jet into the numerical grid
was done at 500~pc from the galactic nucleus, because the aim of the
simulation was to study the long-term evolution. In this work,
however, we are more interested in the influence of the mass loading
of jets by stellar winds, so it is important to bring the injection
point as close as possible to the centre of the galaxy, where the
stellar density is highest.  The initial cross-section must therefore
be smaller (10~pc versus 60~pc in PM07).  In order to keep the same
injection power as in PM07, this difference in cross-section has to be
compensated, resulting in an initially faster, denser and more
overpressured jet here than in PM07.  Models Po and Pr have jet
Lorentz factor $\simeq7$ and jet to ambient density ratio
$3.2\times10^{-5}$ at injection.  They are significantly faster,
denser and colder than the other simulated jets we discuss below. They
are also more overpressured with respect to the ambient.

The simulations of these models (to be discussed below) showed that
the entrained mass is not high enough to decelerate the jets
efficiently.  We therefore ran a set of models (A, A0, B, C and
D) with powers 200 times smaller. Reflecting boundary conditions were
used in all cases. These model jets are relatively fast (Lorentz
factor $3.2$), light (jet to ambient density ratio
$10^{-8}-10^{-10}$), hot (specific internal energy $1.5 \times 10^2 -
1.5 \times 10^4 \, c^2$) and slightly underpressured with respect to
the ambient (as a result of fixing the jet density and
temperature). The one-dimensional estimate for the jet advance speed
is $4.3 \times 10^{-3} \, c$. Model A0, exactly the same as model A
but without any mass entrainment, serves as a reference to identify
the effects of the mass entrainment on the jet collimation and
propagation. The parameters of these  models were selected to
fall within the range used by BLK.

\subsection{Computational setup}
%---------------------------------------------------
\label{ss:setup}

  The two-dimensional grid reproduces the ambient medium of an elliptical
galaxy in axisymmetric cylindrical coordinates. In the axial direction,
the grid starts at 80 pc from the galactic centre and typically ends at
2~kpc, depending on the simulation. In the transversal
direction, a homogeneous grid (with constant cell-size) covers up to 1~kpc, and is
expanded by an extra grid with geometrically increasing cell size from 1 to
2~kpc. The jet is injected into the numerical grid (at 80~pc) with radius
$R_{\rm j}=10\,{\rm pc}$. The numerical resolution is 16 cells across the jet 
radius at injection. This translates into a homogeneous grid size of
$1600 \times 3200$ cells (transversal and axial, respectively).

 For this work, we have modified the equations of conservation of mass,
momentum and energy in order to account for mass loading from stellar winds
and external gravity. The conservation equations for a relativistic flow
in two-dimensional cylindrical coordinates ($R,\, z$), assuming
axisymmetry and using units in which $c=1$, are:
\begin{equation}
  \frac{\partial \mathbf{U}}{\partial t} + \frac{1}{R}\frac{\partial R
\mathbf{F}^R}{\partial R} + \frac{\partial \mathbf{F}^z}{\partial z} =
\mathbf{S} ,
\end{equation}
with the vector of unknowns
\begin{equation}
  \mathbf{U}=(D,D_{\rm l},S^R,S^z,\tau)^T ,
\end{equation}
fluxes
\begin{equation}
  \mathbf{F}^R=(D v^R , D_{\rm l} v^R , S^R v^R + p , S^z v^R , S^R - D v^R)^T ,
\end{equation}
\begin{equation}
  \mathbf{F}^z=(D v^z , D_{\rm l} v^z , S^R v^z , S^z v^z + p, S^z -D v^z)^T ,
\end{equation}
and source terms
\begin{eqnarray}
  \mathbf{S} & = & (q W_{\rm w}, q_{\rm l} W_{\rm w}, q h_ w W_{\rm w}^2 v_{\rm w}^R + p/R + g^R, \\
  \nonumber
                   &     & q h_{\rm w} W_{\rm w}^2 v_{\rm w}^z + g^z, q W_{\rm w} (h_{\rm w} W_{\rm w} - 1)
+ v^R g^R + v^z g^z)^T .
\end{eqnarray}
 
 The five unknowns $D,D_{\rm l},S^R,S^z$ and $\tau$, refer to the densities of
five conserved quantities, namely the total and leptonic rest masses,
the radial and axial components of the momentum, and the energy
(excluding the rest mass energy). All five unknowns are defined in the laboratory
frame, and are related to the quantities in the local rest frame of the
fluid (primitive variables) according to:
\begin{equation}
  D = \rho W,
\end{equation}
\begin{equation}
  D_{\rm l} = \rho_{\rm l} W,
\end{equation}
\begin{equation}
  S^{R,z} = \rho h W^2 v^{R,z},
\end{equation}
\begin{equation}
  \tau=\rho h W^2\,-\,p\,-\,D,
\end{equation}
where $\rho$ and $\rho_{\rm l}$ are the total and the leptonic rest-mass
densities, respectively, $v^{R, z}$ are the components of the velocity
of the fluid. $W$ is the Lorentz factor [$W = (1-v^i v_i)^{-1/2}$, where
summation over repeated indices is implied], and $h$ is the specific
enthalpy defined as
\begin{equation}
  h = 1 + \varepsilon + p/\rho,
\end{equation}
where $\varepsilon$ is the specific internal energy and $p$ is the
pressure. Quantities $g^R$ and $g^z$ in the definition of the
source-term vector ${\bf S}$, are the components of an external gravity
force that keeps the atmosphere in equilibrium. %All these parameters represent a moderate size galaxy cluster
%with mass $10^{14}\,M_{\odot}$ and $\sim 1\, \rm{Mpc}$ virial radius.  

  The difference between these equations and those used in PM07 is just the
inclusion of the source term related to mass loading by stellar winds.
In the source-term vector, the subscript ${\rm w}$ refers to
 the stellar wind and $q$ represents the mass-loading rate
per unit volume defined by deprojecting the expression in
equation~(\ref{nuker}), as plotted in Fig.~\ref{fig:setup}. The wind is taken to be 
a cold electron-proton gas 
entrained with negligible velocity compared to the jet. Thus we neglect the terms
depending on the internal energy and temperature and take $h_{\rm w}=1$, $v_{\rm w}^R = v_{\rm w}^z = 0$, $W_{\rm w} = 1$, so
\begin{equation}
  \mathbf{S}=(q , q_{\rm l}, p/R + g^R , g^z, v^R g^R + v^z g^z)^T ,
\end{equation}
with $q_{\rm l} = q \, m_e/m_p$. The system is closed by means of the Synge equation of state
\citep{Sy57}, as described in Appendix~A of PM07.  This accounts for a mixture
of relativistic Boltzmann gases (in our case, electrons, positrons and
protons). The code also integrates an equation for the jet mass fraction, $f$. This quantity, set to 1 for the injected jet material and 0 otherwise, is used as a tracer of the jet material through the grid. 

The simulations presented in this paper use the finite-volume code {\emph Ratpenat}. This is a hybrid -- MPI + OpenMP -- parallel code that solves the equations of relativistic hydrodynamics in conservative form using high-resolution-shock-capturing methods \citep[see][and references therein]{PM10}: (i) primitive variables within numerical cells are reconstructed using PPM routines; (ii) numerical fluxes across cell interfaces are computed using the Marquina flux formula and (iii) advance in time is performed with third order TVD-preserving Runge-Kutta methods.

The evolution of the jets in simulations A, B, C and D has been followed
up to $1-2$\,Myr (see Table~\ref{tab1}), by which time the jets have
propagated between $1$ and $2$ kpc. This is far enough to capture the effect of
mass loading on jet flaring and disruption. Each simulation needed about 4
million time steps.

Simulations A, B, C and D were performed in Magerit, at the
Supercomputing and Visualization Center of Madrid, within the \emph{Red
Espa\~nola de Supercomputaci\'on} (Spanish Supercomputing Network),
with up to 48 processors. Processors were added as the jet evolved (starting
with 8 processors). These simulations required between $1.0\times10^5$
and $1.5\times10^5$ computational hours, depending on the model,
resulting in a total of around $5\times10^5$ hours.

\section{Results}
%            %%%%%%%
\label{s:res}

\subsection{The influence of boundary conditions on jet evolution}
\label{ss:boundary}

\begin{figure}
 \includegraphics[width=0.45\textwidth]{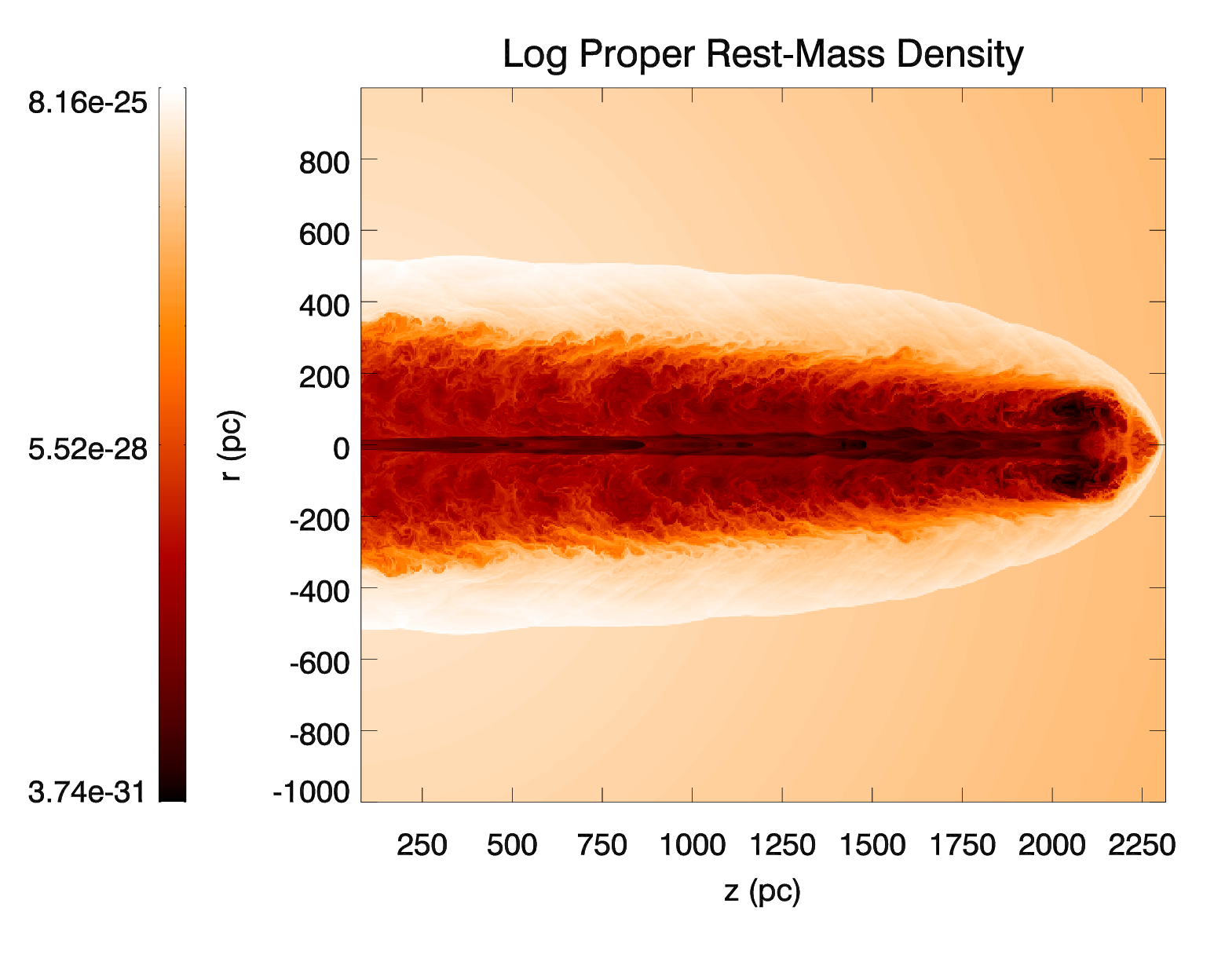}

 \includegraphics[width=0.45\textwidth]{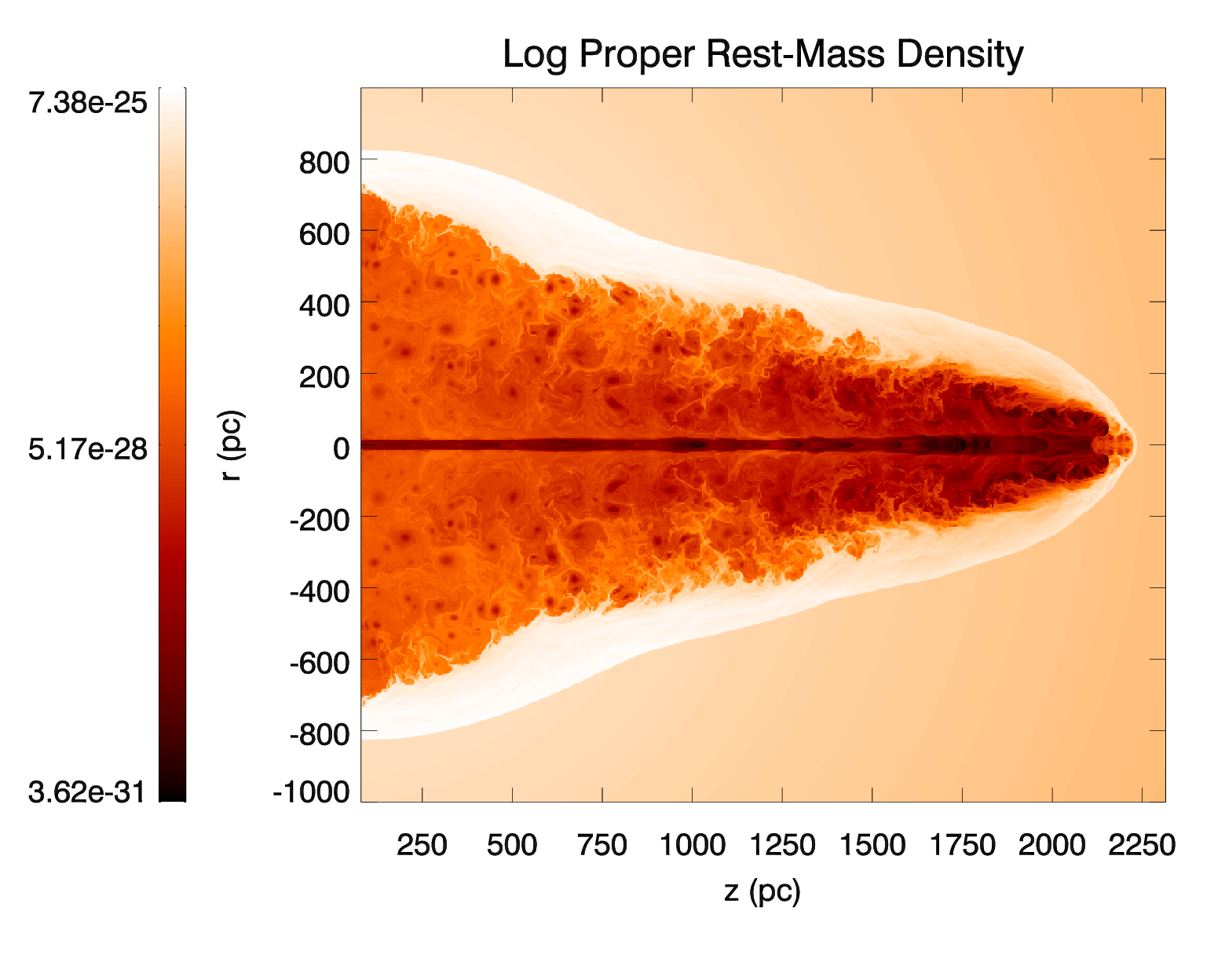}
 \caption{Maps of the logarithm of the rest-mass density at the end 
   time ($t\simeq 2.5\times 10^5$~yr) for simulations Po (upper panel) and
   Pr (lower panel).
 \label{fig:po-pr}}
 \end{figure}

 We first describe the effects resulting from the change of boundary conditions from
open (as used in PM07) to reflecting. Reflecting boundary
conditions mimic the presence of a counter-jet with the same properties
as the simulated jet and are therefore likely to be more realistic. Fig.~\ref{fig:po-pr} shows the last frames of both simulations at
$t\simeq2.5\times10^5$~yr. The images reveal the main differences
due to the boundary condition: the
shape of the bow-shock (the forward shock driven into the ambient by the
injection of the jet), the width of the jet and the prominence of
jet pinching. In Pr, the reflecting boundary condition causes 
both the bow-shock and the inner cocoon (the region with mixed
shocked ambient and jet material surrounding the jet) to be wider. 
The jet in Po is broader and shows stronger
recollimation (conical) shocks, which in turn make the head of the jet
broader. The jet in Pr remains more
collimated.

The choice of boundary condition only affects some aspects of the
simulations.  For example, the advance speeds of the jet terminal
shock and the head of the bow-shock are very similar in Pr and Po,
whereas the cocoon pressure and density in Pr are larger: by almost
an order of magnitude in the density ($\sim 1.5\times10^{-26}$
compared with $\sim 2\times10^{-27}$~g~cm$^{-3}$) and a factor of
two in the pressure ($\sim 4\times10^{-9}$ compared with $\sim
2\times10^{-9}$~erg~cm$^{-3}$) by the end of the simulations.  The
pressure is basically determined by the injected energy divided by
the volume of the shocked region, implying that the pressure in the
shocked ambient and jet material (i.e.\ the whole volume within the
bow-shock, excluding the jet) will decrease with time as it expands. The 
pressure in the cocoon follows the same power-law $P_{\rm
  c}\propto t^{-0.9}$ in both cases, but with a larger constant of
proportionality for Pr.  The radius of the bow-shock also increases
faster with time for Pr: by the end of the simulation the mean
radius of the shocked region is $\simeq 500$~pc compared with
$\simeq 400$~pc in Po. Another consequence of the higher pressure in
Pr is that the opening angle of the jet is smaller ($\simeq
0\fdg27$ versus $\simeq 1\fdg1$ in Po). The recollimation
shocks are therefore stronger in Po.

Po also shows a broader and faster backflow component than Pr, since
backflowing material can escape through the open boundary. In both
simulations, the backflow decelerates with the distance travelled from the
terminal shock. Typical velocities close to the terminal shock
($z\simeq2\,{\rm kpc}$) are similar in both cases: $\simeq 0.6\,c$ in
Pr and $\simeq 0.7\,c$ in Po. In Pr, the velocity drops to values
$\simeq 0.4\,c$ at $z\simeq 1.5\,{\rm kpc}$ and $\simeq 0.2\,c$ at
$z\simeq 1\,{\rm kpc}$, finally reaching zero at $z=80\,{\rm pc}$,
close to the reflecting boundary. In Po, the backflow velocity is
typically $\simeq 0.5\,c$ at $z\simeq1\,{\rm kpc}$ and drops to values
$\simeq 0.2\,c$ at $z=80\,{\rm pc}$.

As the jet in both simulations has the same injection conditions, we
expect that the jet and shock structures seen in Pr and Po should be very
similar if they are compared at the same cocoon pressure. In other
words, Pr at a given time will look quite similar to Po seen at some
earlier time. In particular, the pressure in the cocoon has a direct
influence on the position of a possible recollimation shock, and the
appearance of strong shocks in simulations with reflecting boundaries
will occur at later times than in simulations with open boundaries. We
thus conclude that the shock in the jet in PM07 (which used an open
boundary) would correspond to an older jet simulated with a reflecting
boundary condition, in which the bow shock would have propagated to
larger distances than the 15~kpc found in that work.

As the use of a reflecting boundary condition is likely to give a better
approximation to the true bipolar case, we use it for the remaining
simulations in this paper.  The differences introduced by the change
of boundary conditions should be borne in mind when comparing with the
simulations of PM07.

%%%%%%%%%%%%%%%%%%%%%%%%%%%%%%%%%%%%%%%%%%%%%%%%%%%%%%%%%%%%%%%%%%%%%%%%%%%%%%%%%%%%%%%%%%%%%%%%%%%%%%%%%%%%%%%%%%%%%%%%%%%%%%%
%
%
%%%%%%%%%%%%%%%%%%%%%%%%%%%%%%%%%%%%%%%%%%%%%%%%%%%%%%%%%%%%%%%%%%%%%%%%%%%%%%%%%%%%%%%%%%%%%%%%%%%%%%%%%%%%%%%%%%%%%%%%%%%%%%%

\subsection{Mass loading of powerful jets}
%                 -----------------------------
\label{ss:powerful}

\subsubsection{Absence of deceleration by mass loading}

  Neither of the simulations of powerful, mass-loaded jets show signs of strong deceleration over distances of 2\,kpc, but there is slow
deceleration due to a combination of mass loading and dissipation of kinetic energy
at internal shocks. The jet flow heats up at the shocks, increasing
its temperature by an order of magnitude (to $\sim3\times 10^{10}$\,K) between 
injection and the terminal shock. The role of mass loading is
revealed by a decrease in the jet mass fraction from 1 to around
0.9 on the axis (slightly larger for Pr and smaller for Po) and 
a decrease in the mean value over the jet cross section to about 0.6
before the jet terminal shock in both simulations. The mean jet flow velocity before the terminal 
shock is around 80\% of the injection value. 

The jet can only be
slowed efficiently by mass loading when the kinetic energy required to 
accelerate the loaded mass to the flow velocity of the jet 
($W_{\rm j} \dot{M}c^2$ per unit time, where $W_{\rm j}$ is the jet Lorentz factor, and
$\dot{M}$ is the mass-loading rate) is of the order or larger than the 
kinetic luminosity, $L_{\rm j}$ \citep{HB06}. If we assume a constant mass-loading
rate per unit volume $q$ and a jet of constant radius $R_{\rm j}$, the characteristic length for jet
deceleration is
\begin{equation}
l_{\rm d} \simeq \frac{1}{W_{\rm j}} \! \! \left(\frac{L_{\rm j}}{10^{44} {\rm erg\,s}^{-1}}\right)
\!\! \left(\frac{q}{10^{22} {\rm g\,yr}^{-1}{\rm pc}^{-3}}\right)^{-1} \!\!
\! \left(\frac{R_{\rm j}}{10 {\rm pc}}\right)^{-2} \!\! {\rm Mpc}.
\label{eq:ld}
\end{equation}
Even if we use a constant mass-loading rate equal to the maximum at
the centre of the galaxy, $q = q_0 = 4.95\times 10^{22} {\rm
g\,yr^{-1}\,pc^{-3}}$ (Table~\ref{tab1}), we still obtain deceleration
lengths of the order of hundreds of kpc. Such values
are only indicative, since: (i) part of the jet kinetic energy will be used to heat 
the loaded material; (ii) the mass loaded per unit time rises as the jet
gets longer and (iii) the true value of the mass-loading rate per
unit volume decreases very fast with increasing distance from the galaxy centre
(for $r \ga r_{\rm b}$ in our models), so the approximation $q \approx q_0$ is only
valid over distances $\la$1\,kpc. Whereas (i) and (ii) tend to
decrease the deceleration length, (iii) tends to increase it.
Nevertheless, this simple computation clearly explains the lack of
deceleration in models Po and Pr within the first 2\,kpc.

\subsubsection{Comparison between models with and without mass loading}
  It is interesting to compare  
Po and Pr with the simulation presented in PM07 (which has the same boundary condition as Po).
Note that the flow parameters at injection  
are quite different, even though the jet power is the same in all three cases. In PM07, the jet was injected at
500~pc, with a radius of 60~pc, a velocity $v_{\rm j}=0.87\,{\rm c}$ and an
overpressure factor of 7.8 with respect to the ambient value. The jets in Po
and Pr were injected at 80~pc, with a radius of 10~pc, a velocity
$v_{\rm j}=0.99\,{\rm c}$ and an overpressure factor of 17.6.
 
   In PM07, the first recollimation shock appeared at $z\simeq 1\,{\rm
kpc}$ after $t\simeq 8\times10^5{\rm yr}$. The jet was injected close to
the core radius of the galactic gas, and thus expanded into an
ambient medium in which the density and pressure fall rapidly with distance. Within this
environment the pressure of the cocoon evolves as $P_{\rm c}\propto
t^{-1.3}$. In the case of Po and Pr, the simulations covered much smaller durations ($t\simeq
2.5\times10^5~{\rm yr}$) and distances from the  
nucleus than that in PM07. The cocoon pressure evolves
following a flatter slope $P_{\rm c}\propto t^{-0.9}$, 
primarily because the jets in Po and Pr are entirely within the galactic core, where the ambient
density falls very slowly with distance. This results in a slower
expansion: the jets  in Po and Pr would need to be followed to larger 
distances and times to be directly
compared with the simulation in PM07.

\subsection{Mass loading of weak jets}

\subsubsection{Reference model without mass loading}
%                 ---------------------------------------------

\begin{figure*}
  \includegraphics[width=0.19\textwidth]{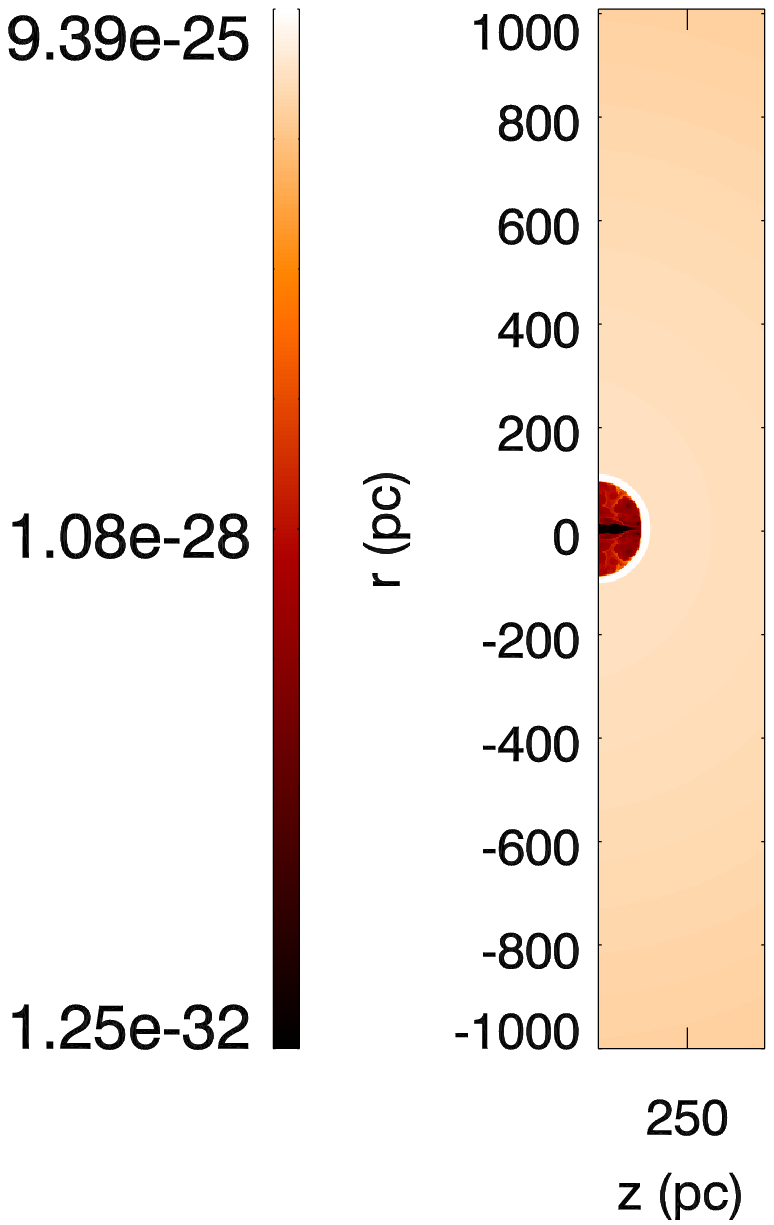}
  \includegraphics[width=0.19\textwidth]{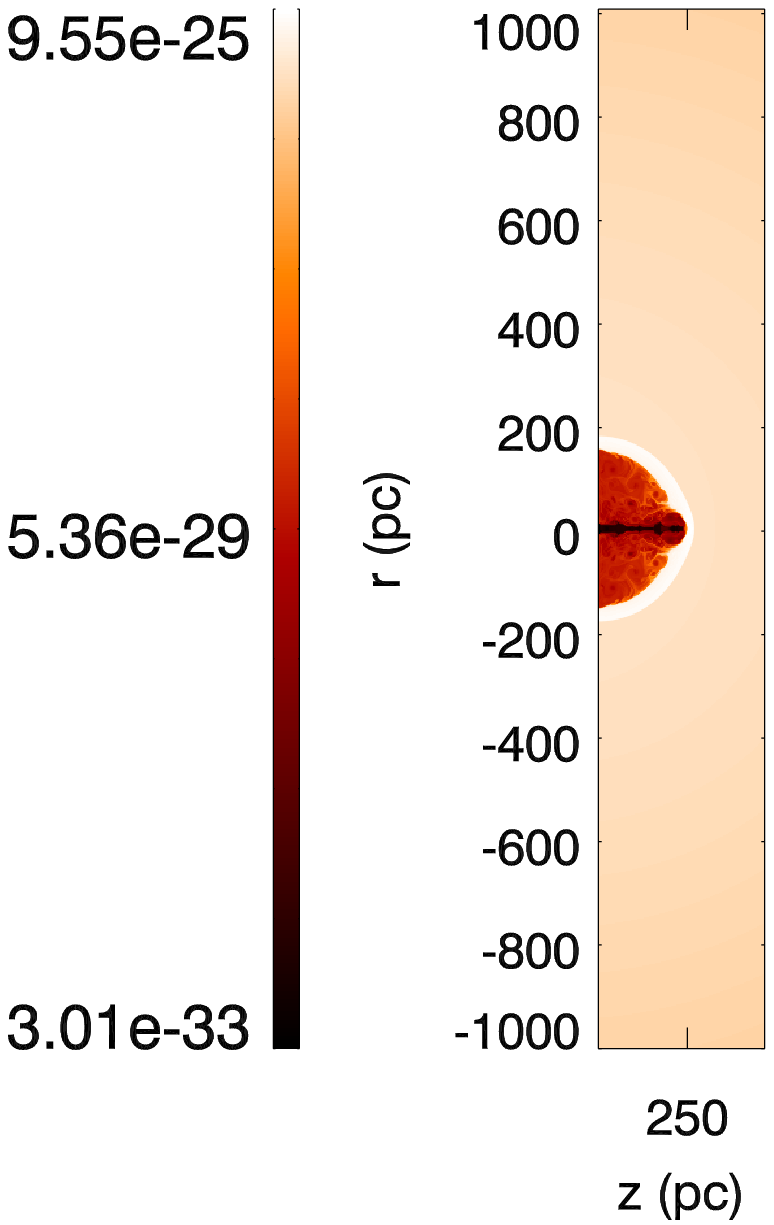}
  \includegraphics[width=0.19\textwidth]{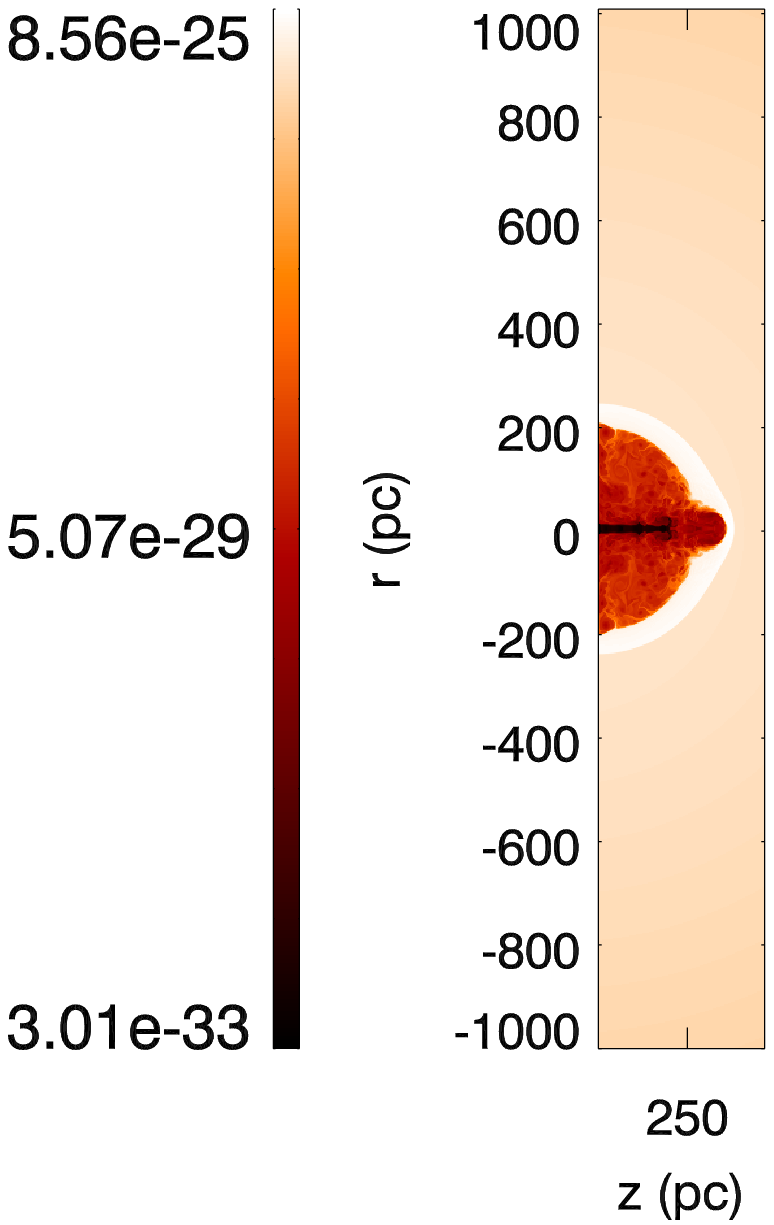}
  \includegraphics[width=0.19\textwidth]{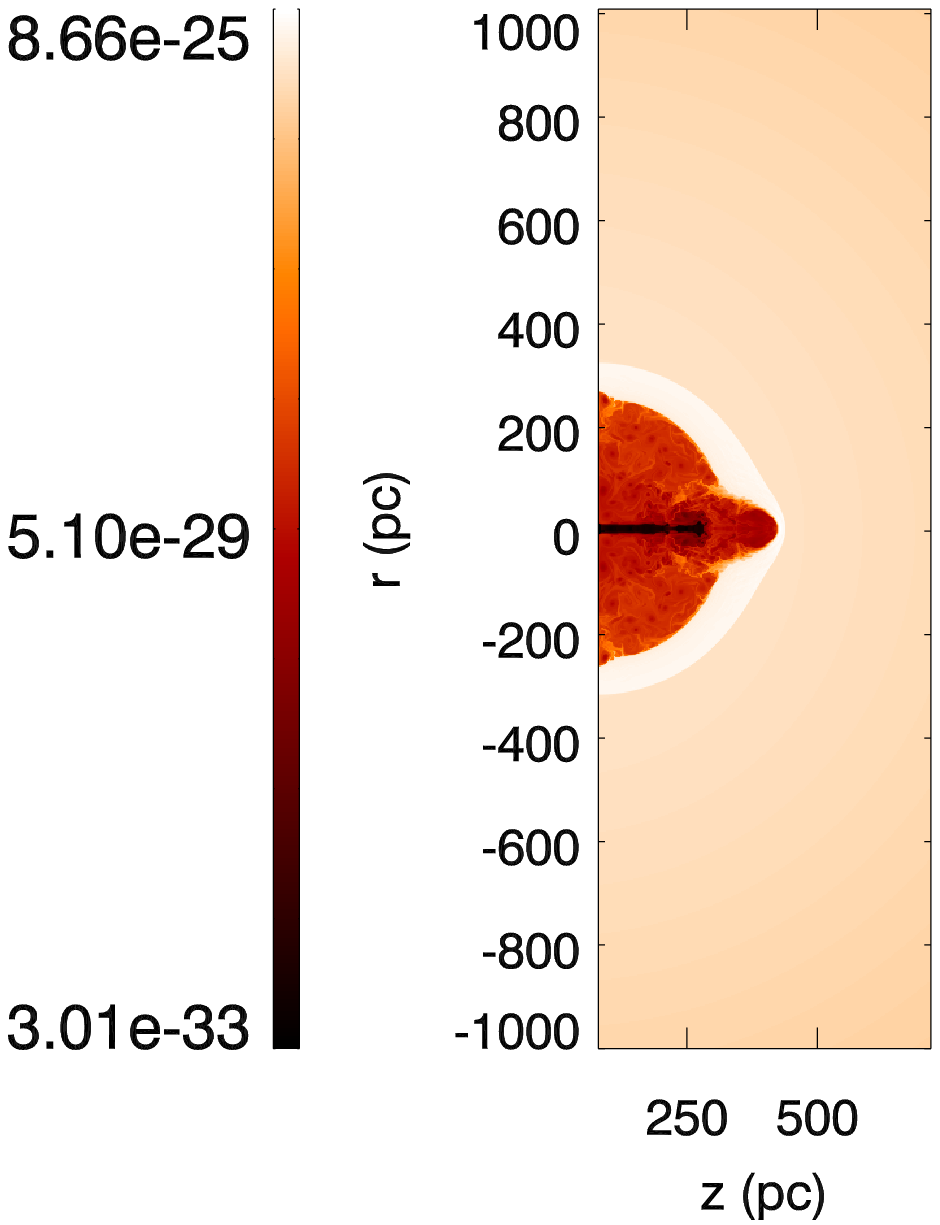}
  \includegraphics[width=0.19\textwidth]{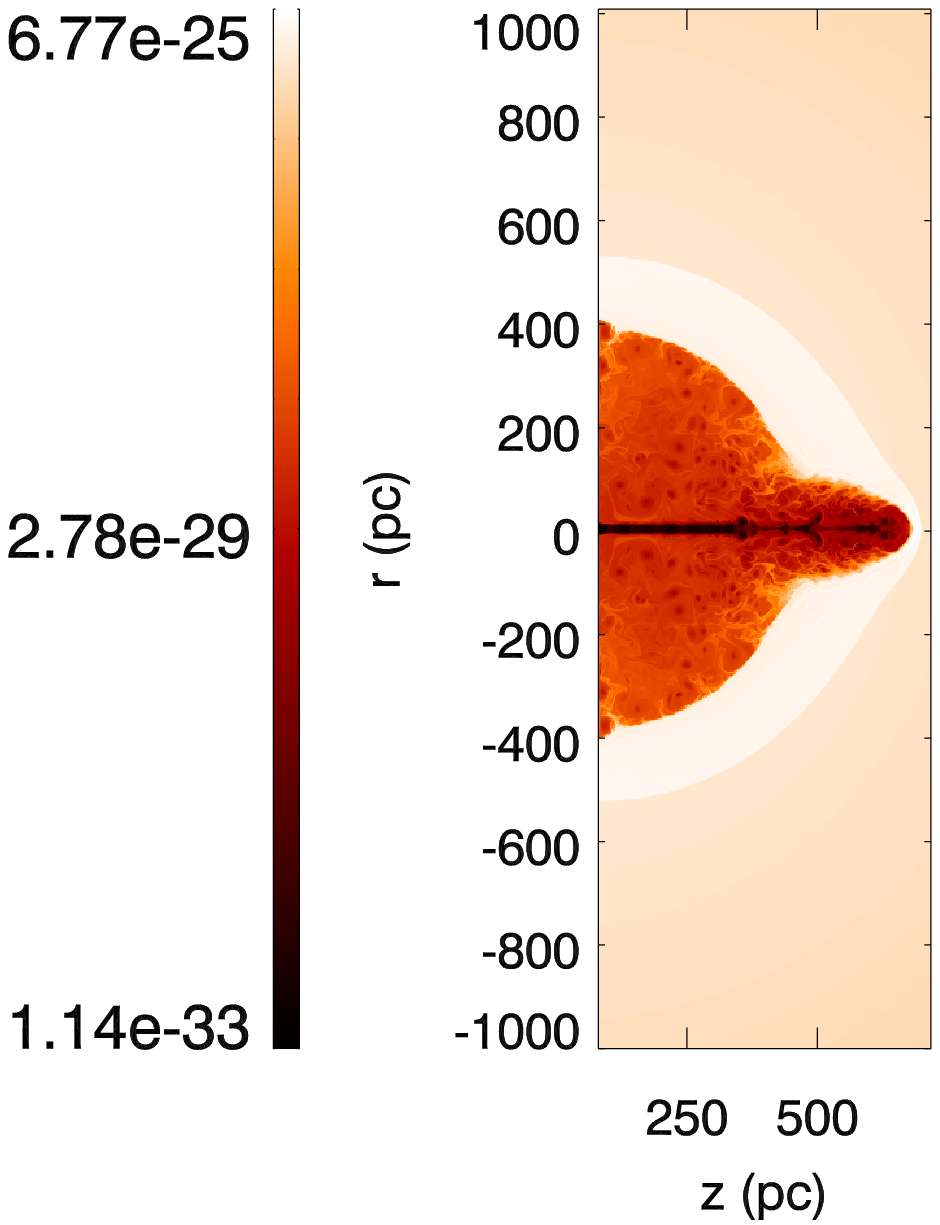}
 
  \caption{Evolution of the rest-mass density distribution in model A0.
From left to right, the times of the frames are $6.5 \times 10^4, 1.3
\times 10^5, 1.9 \times 10^5, 2.9 \times 10^5, 5.8 \times 10^5$ yr.}
  \label{fig:a0sequence}
 \end{figure*}

  Fig.~\ref{fig:a0sequence} displays a series of snapshots of the rest
mass density of model A0, in which there is no mass loading. The sequence shows the
propagation of the shock generated by the
jet through the ambient. The weakness of the jet is reflected in the almost 
spherical shape of this shock 
%(contoured by the outer transition between pale yellowish to greenish regions in the panels of Fig.~\ref{fig:a0sequence}) 
at very early times (Fig.~\ref{fig:a0sequence}, first panel), the low Mach number of the jet ($\approx$3 at the
beginning of the simulation) and the clear detachment of the shock from
the cocoon. 
%(outlined by the blue-green to pale yellowish transition in the panels). 
The contact discontinuity between the
cocoon and the shocked ambient soon develops a prominent `nose-cone' and the bow-shock becomes more extended in the 
 axial
direction. The beam\footnote{In the following, we will refer to the whole outflow structure (observed or simulated) as the \emph{jet}, and to the fast and collimated central spine of the (simulated) jet as the \emph{beam} in order to clarify the presentation of the results.} shows a series of internal conical shocks produced by the pressure mismatch between it and the cocoon.

\begin{figure*}
 \includegraphics[width=0.48\textwidth]{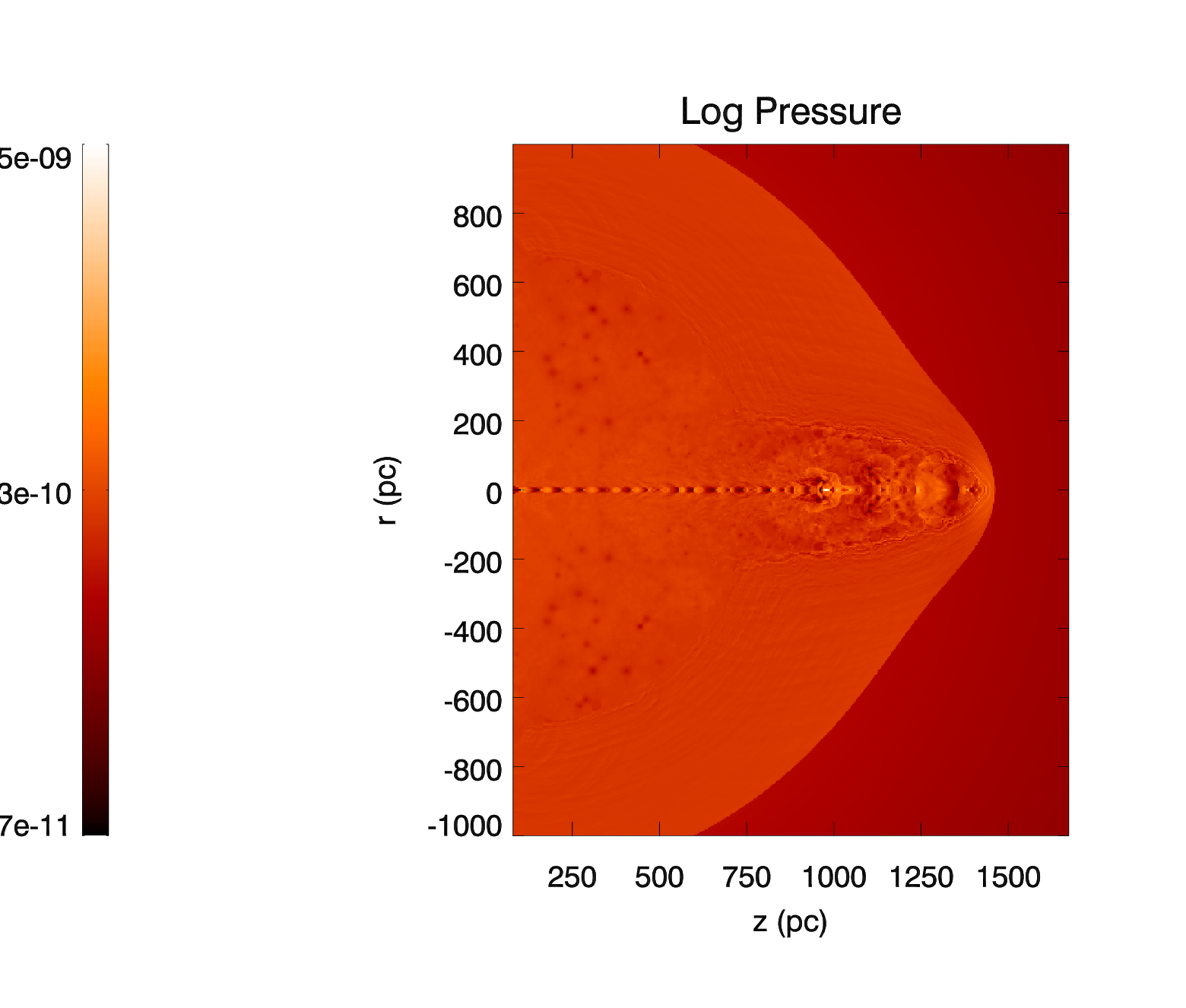}\hspace{0.5cm}
 \includegraphics[width=0.48\textwidth]{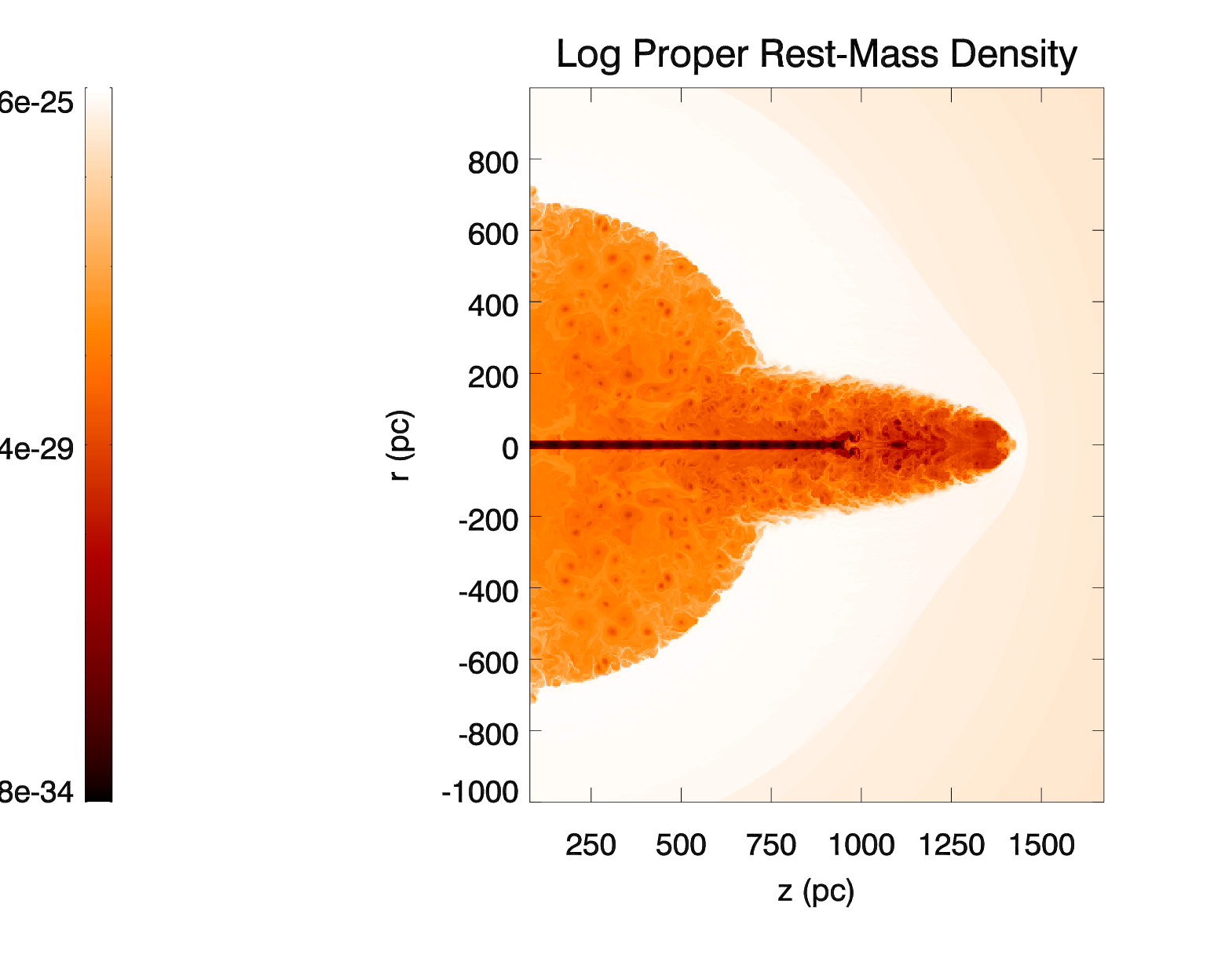} \\
 \includegraphics[width=0.48\textwidth]{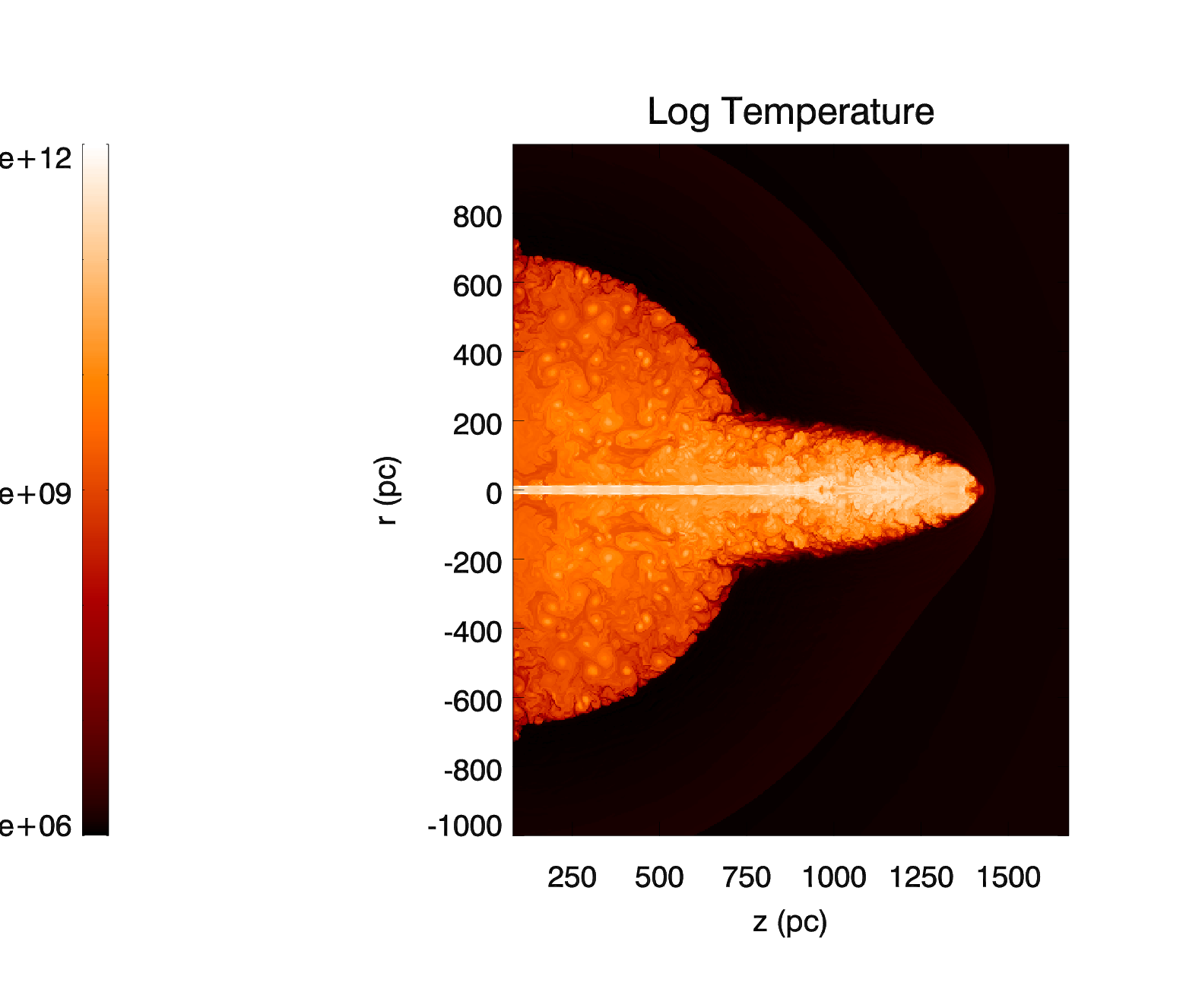}\hspace{0.5cm}
 \includegraphics[width=0.48\textwidth]{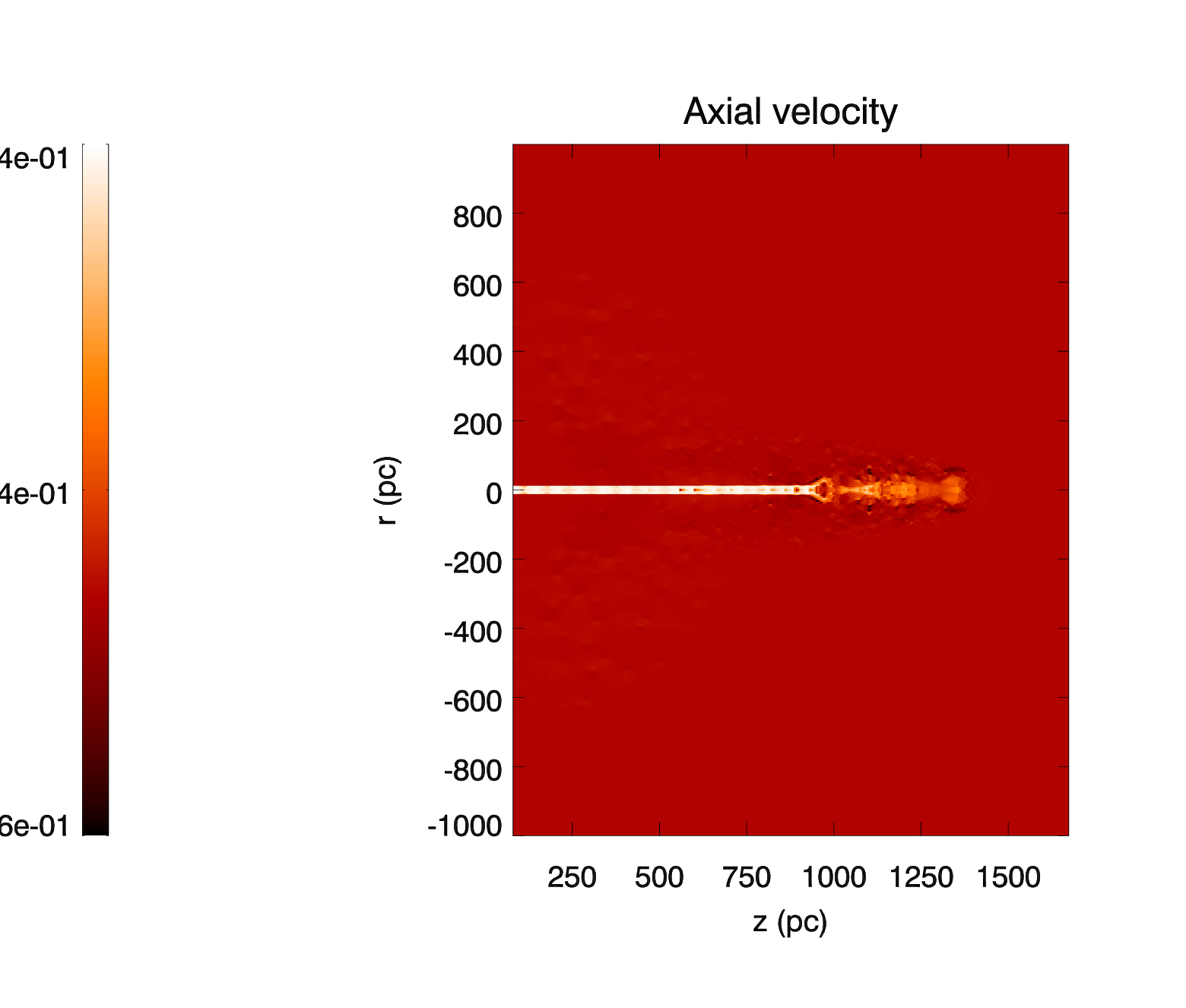}
 \caption{Pressure, rest-mass density, temperature and axial flow
velocity in model A0 at the end of the simulation ($t = 1.6 \times 10^6$
yr).}
 \label{fig:a0final}
 \end{figure*}

  Fig.~\ref{fig:a0final} shows the distributions of pressure,
rest-mass density, temperature and axial flow velocity at the end of the
simulation. The protrusion of the bow shock caused by the impact of the beam on
the originally spherical shock is clearly seen in the first three
panels. Conical shocks within the beam are apparent in the pressure, rest-mass density
and axial velocity panels. These eventually evolve into planar disrupting
shocks. The supersonic beam ends at a terminal shock
where the beam flow decelerates, compresses and heats. In 
Fig.~\ref{fig:a0final}, this shock is located at about $1$ kpc from the
injection point. 
%It is seen as the red to blue transition on the beam in the rest mass density panel, or the yellow to blue one in the axial velocity panel. 
The region between this shock and the bow shock forms
the so-called head of the jet, a very dynamic structure that governs the
propagation of the jet through the ambient. In  model A0,  the
region is quite broad with no substantial enhancement of the internal
energy density (Fig.~\ref{fig:a0final}; pressure panel); the terminal shock, and hence the jet, are still propagating at the end of the simulation, and have not stalled. The dynamics of the jet head is discussed in more detail in Section~\ref{ss:head_dyn}.
The cocoon is essentially isobaric, since the time-scale for the pressure to become constant is much shorter than the propagation time-scale.  The density and
temperature distributions display a clear transition between the almost
homogeneous values of the old jet material in the cocoon (made colder and denser due
to mixing with the shocked ambient gas through the contact
discontinuity) and the hot and dilute material newly injected at the nose cone. 
Finally, a hot shear layer develops between the beam
and cocoon material (this is not seen in the temperature panel of Fig.~\ref{fig:a0final} due to
its limited resolution). 

% Consider moving this into the Discussion section and making a forward reference at this point.
We emphasize that the end-points of all of the simulations presented
here correspond to very early stages of the evolution of low-power
radio galaxies. This is the main reason why the nearly spherical
region of gas surrounding the base of the jet in model A0
($\approx$700\,pc in radius in Fig.~\ref{fig:a0final}) appears much
hotter and less dense than the cool cores revealed by X-ray
observations in the host galaxies of FR\,I sources \citep{ha02,cr08},
and indeed included in the initial conditions for the simulations.
The hot cocoons will dilute and mix with the cooler ambient material
as the jet becomes transonic and the bow shock disappears. The
spherical shape that we find for the cocoon close to the injection is
due to the low advance velocity of the head of the jet and the action
of gravity: the dynamical time-scale of the gravitational field that
maintains the atmosphere in hydrostatic equilibrium is in the range
$t_{\rm D} \simeq 10^5\,{\rm yr}$ to $\simeq 3 \times 10^5\,{\rm yr}$ in
the simulated region; this is smaller than the simulation time\footnote{These numbers have been obtained
  using the dark-matter density profile that are needed to keep the
  atmosphere of gas in hydrostatic equilibrium and estimating $t_{\rm D}\sim
  (G\,\rho_{\rm DM})^{-1/2}$, where $G$ is the gravitational constant and
  $\rho_{\rm DM}$ is the density of dark matter within the simulated
  region.}.

%%%%%%%%%%%%%%%%%%%%%%%%%%%%%%%%%%%%%%%%%%%%%%%%%%%%%%%%%%%%%%%%%%%%%%%%%%%%%%%%%%%%%%%%%%%%%%%%%%%%%%%%%%%%%%%%%%%%%%%%%%%%%%%
%
%
%%%%%%%%%%%%%%%%%%%%%%%%%%%%%%%%%%%%%%%%%%%%%%%%%%%%%%%%%%%%%%%%%%%%%%%%%%%%%%%%%%%%%%%%%%%%%%%%%%%%%%%%%%%%%%%%%%%%%%%%%%%%%%%

%%%%%%%%%%%%%%%%%%%%%%%%%%%%%%%%%%%%%%%%%%%%%%%%%%%%%%%%%%%%%%%%%%%%%%%%%%%%%%%%%%%%%%%%%%%%%%%%%%%%%%%%%%%%%%%%%%%%%%%%%%%%%%%
%
%
%%%%%%%%%%%%%%%%%%%%%%%%%%%%%%%%%%%%%%%%%%%%%%%%%%%%%%%%%%%%%%%%%%%%%%%%%%%%%%%%%%%%%%%%%%%%%%%%%%%%%%%%%%%%%%%%%%%%%%%%%%%%%%%

\subsubsection{Weak jet models with mass loading: Model A}
%                        --------------------------------------

%\paragraph{Model A}
%                       ..............
\label{ss:modelA}

\begin{figure*}
  \includegraphics[width=0.19\textwidth]{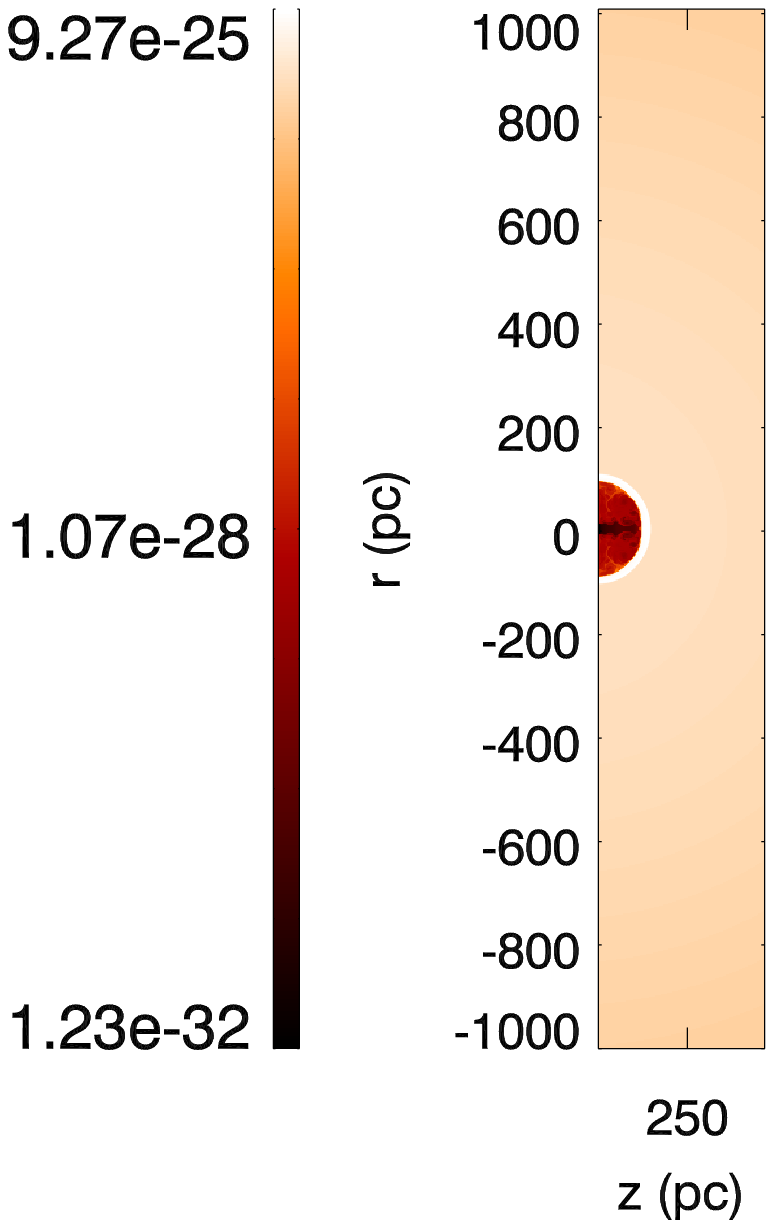}
  \includegraphics[width=0.19\textwidth]{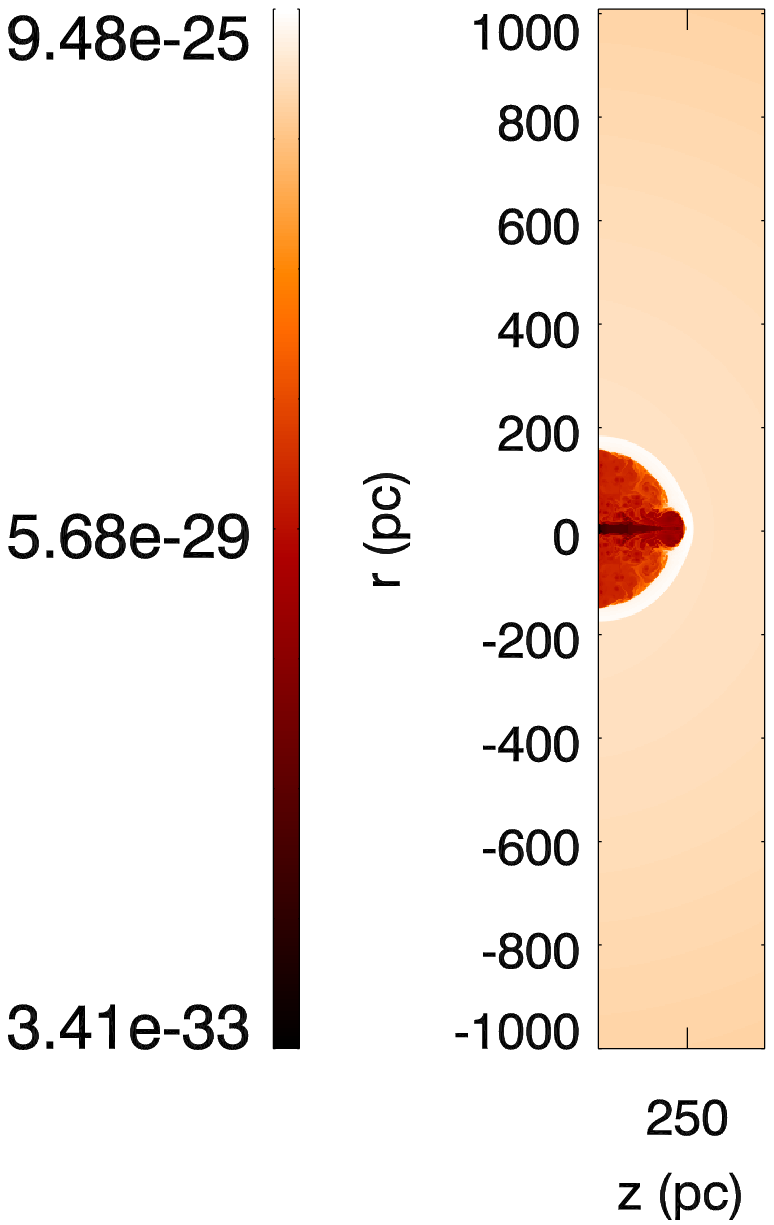}
  \includegraphics[width=0.19\textwidth]{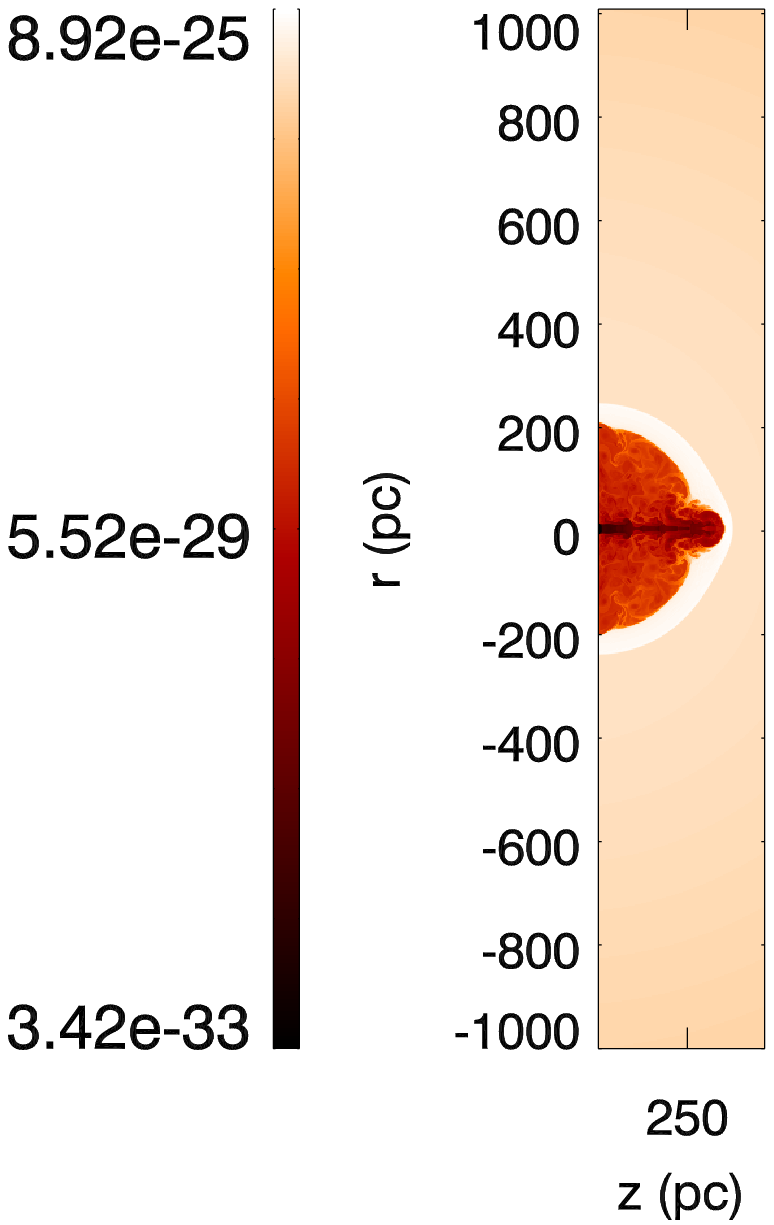}
  \includegraphics[width=0.19\textwidth]{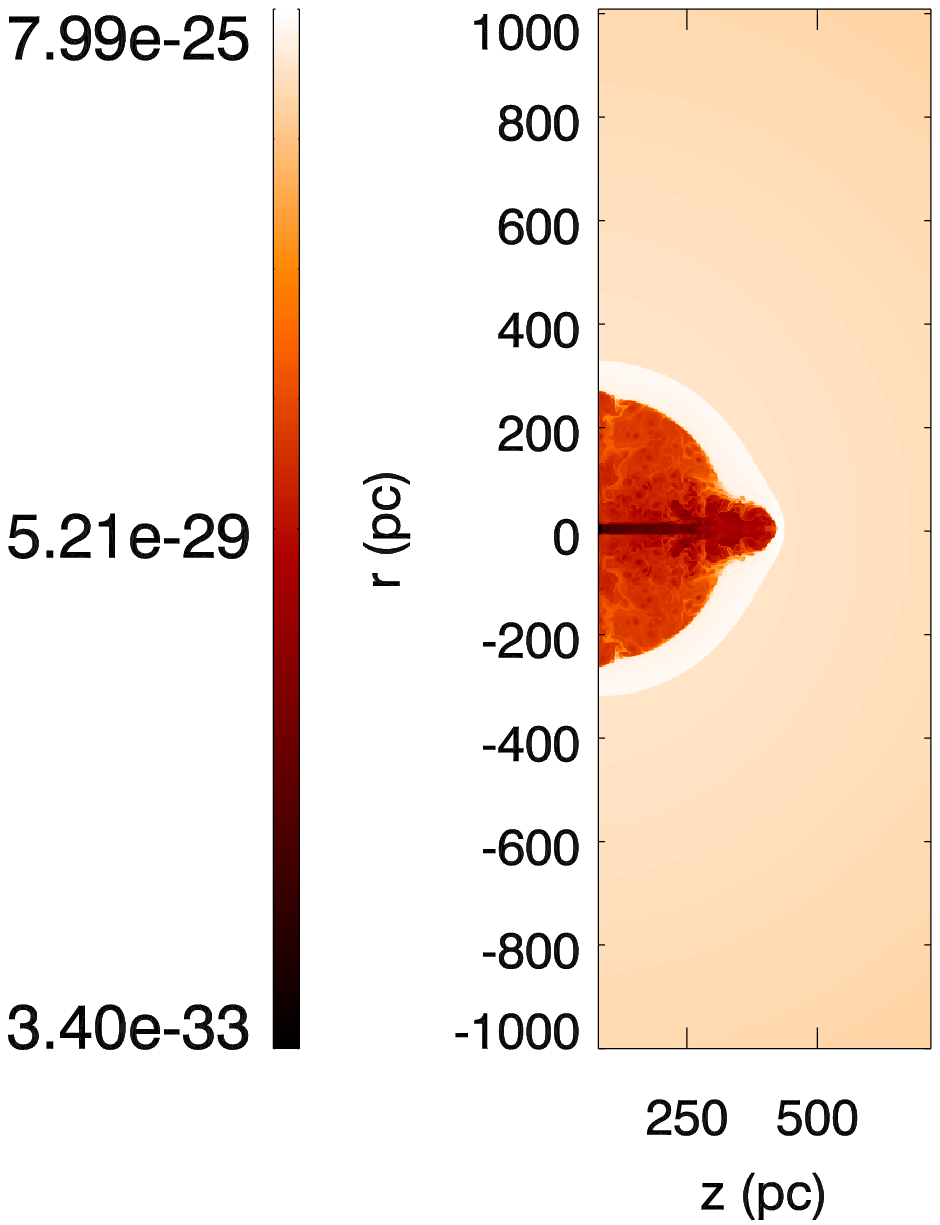}
  \includegraphics[width=0.19\textwidth]{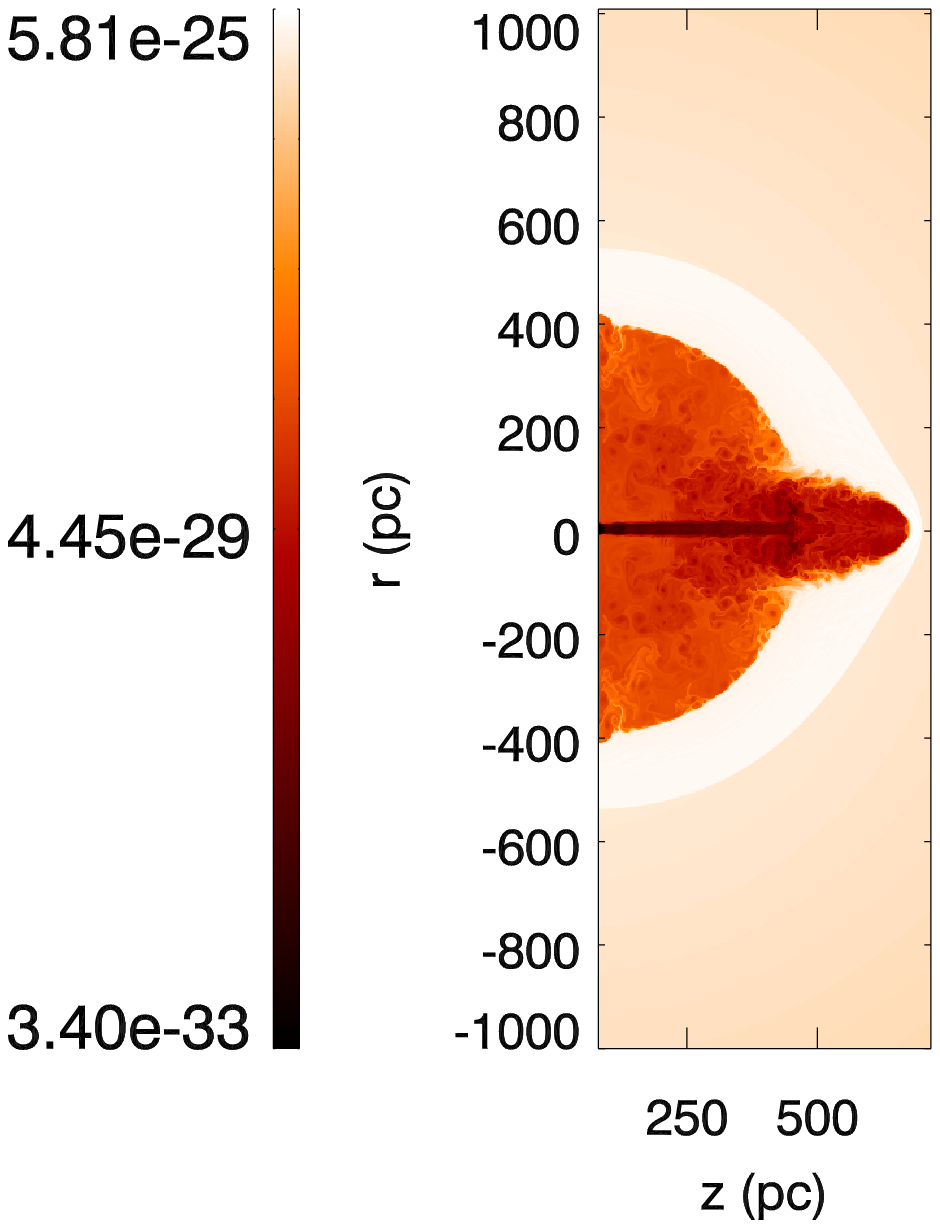}
 
  \caption{Evolution of the rest-mass density distribution in model A.
From left to right, the times of the frames are $6.5 \times 10^4, 1.3
\times 10^5, 1.9 \times 10^5, 2.9 \times 10^5, 6.0 \times 10^5$ yr.}
  \label{fig:asequence}
 \end{figure*}

\begin{figure*}
 \includegraphics[width=0.48\textwidth]{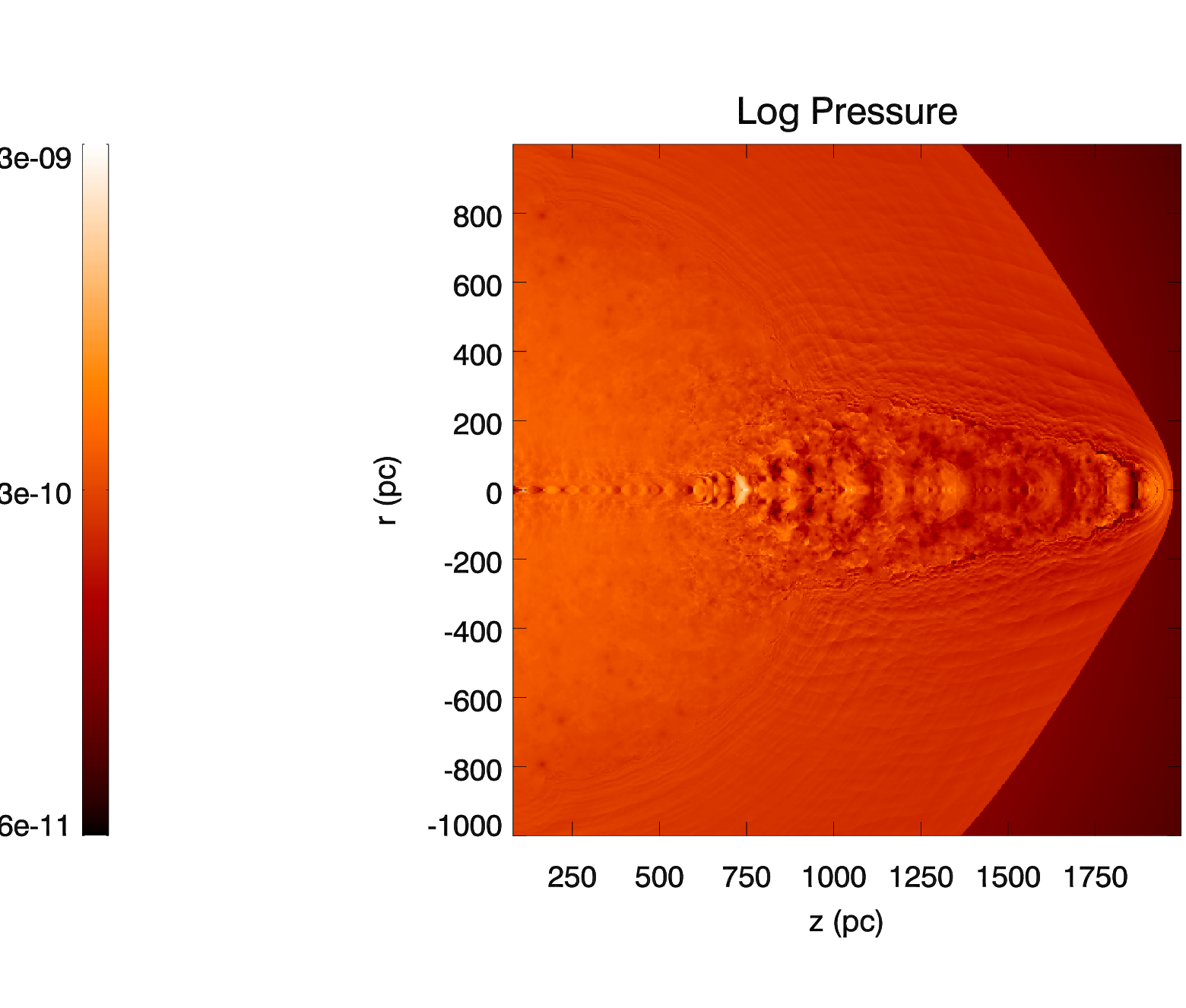}\hspace{0.5cm}
 \includegraphics[width=0.48\textwidth]{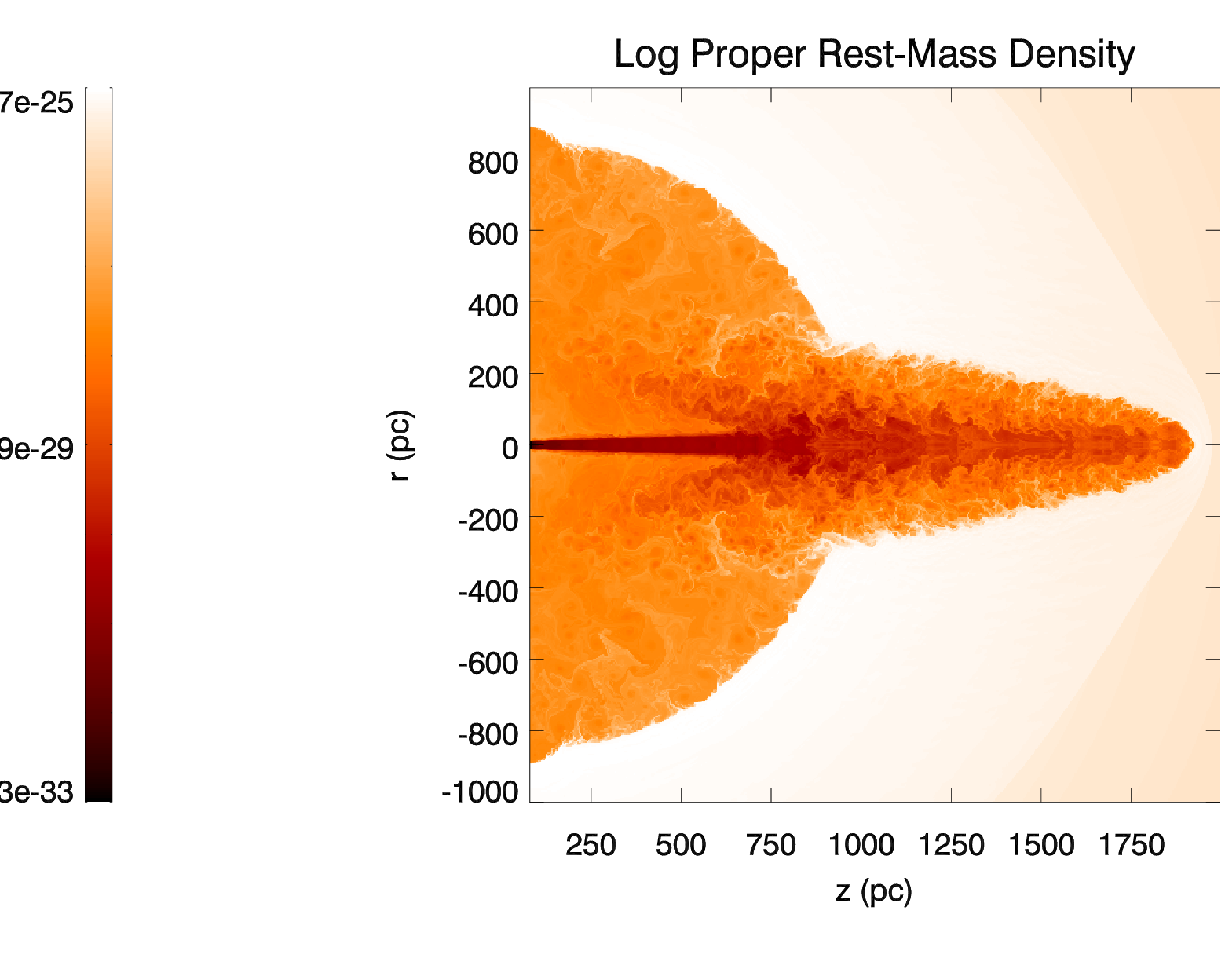} \\
 \includegraphics[width=0.48\textwidth]{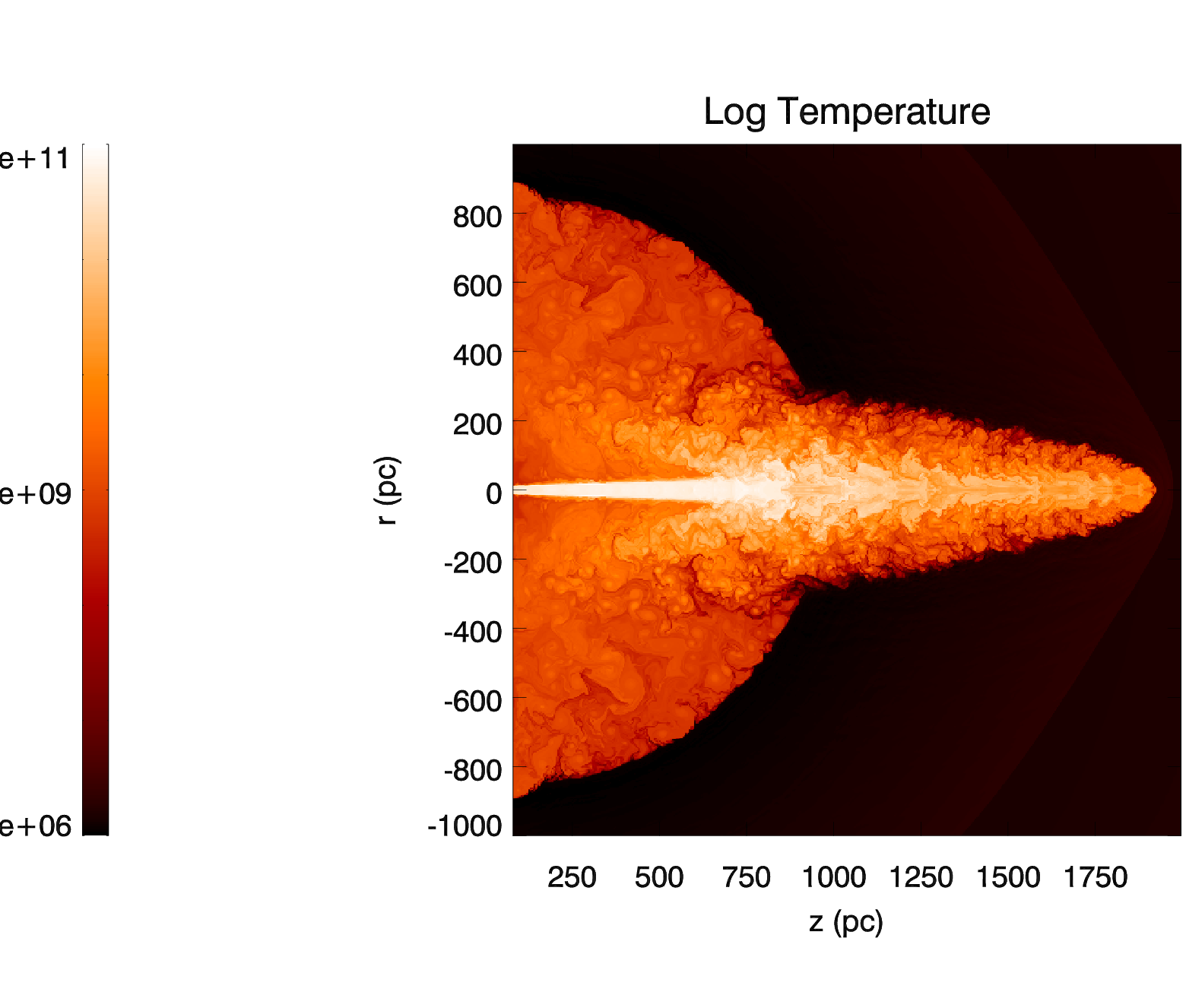}\hspace{0.5cm}
 \includegraphics[width=0.48\textwidth]{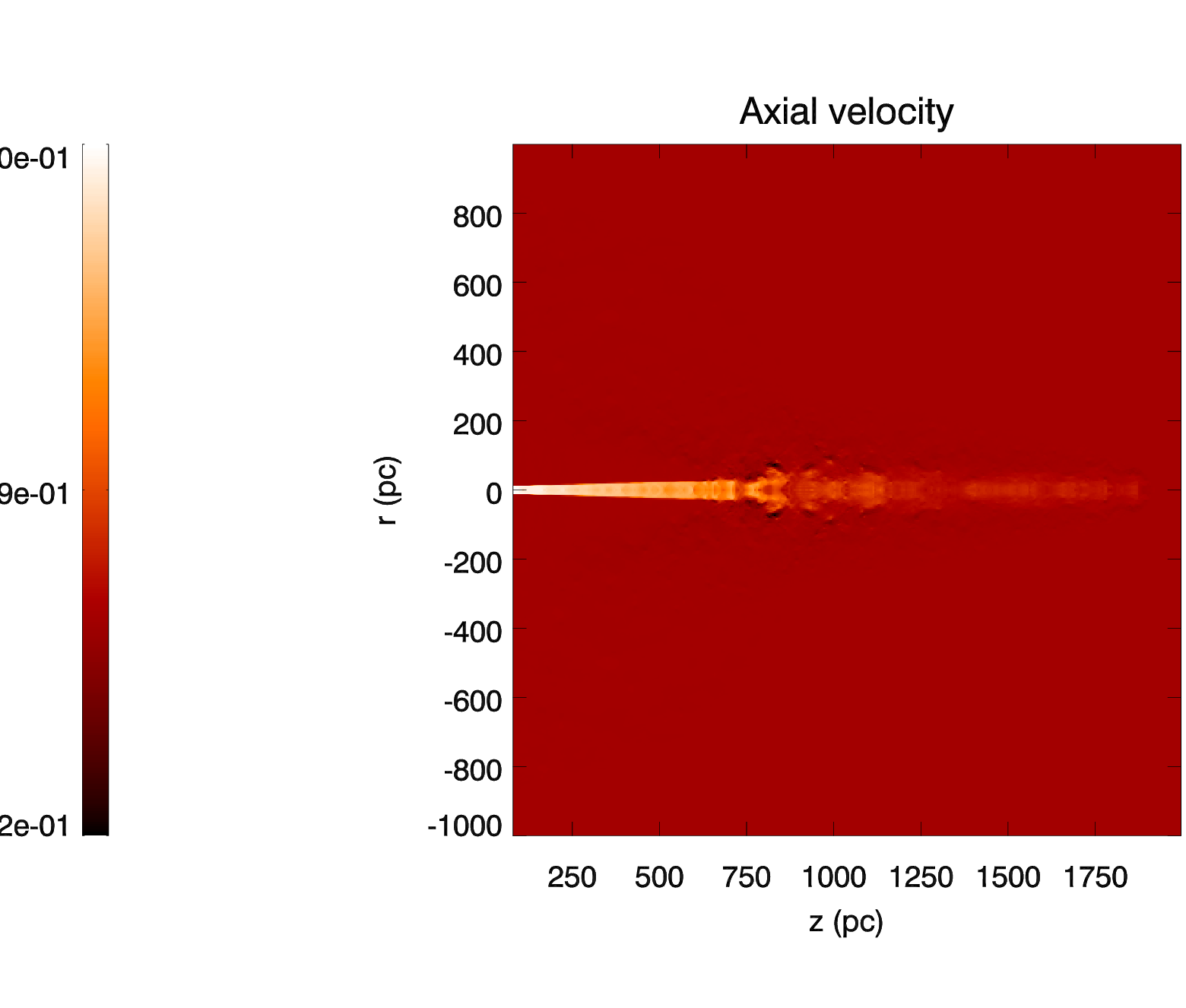}
 \caption{Pressure, rest-mass density, temperature and axial flow
velocity in model A at the end of the simulation ($t = 2.4 \times 10^6$ yr).}
 \label{fig:afinal}
 \end{figure*}

  Figs~\ref{fig:asequence} and \ref{fig:afinal} are the equivalents for model A of 
Figs~\ref{fig:a0sequence} and \ref{fig:a0final} for model A0. The times of
the snapshots in Figs~\ref{fig:a0sequence} and \ref{fig:asequence} are
almost identical, allowing a direct comparison of the density panels
for the two models. The main structural and dynamical features of
the cocoon/shocked ambient system found in model A0 (an almost spherical
cocoon, detachment of the bow shock from the cocoon and formation of
a nose cone at the head of the jet) also apply to model A. Comparison between 
the two sequences also shows very similar speeds for the propagation of the jet head and the expansion of the cocoon.  These similarities 
persist until the ends of the simulations (Figs~\ref{fig:a0final} and \ref{fig:afinal}).
The reasons for them are discussed in Section~\ref{ss:cocoon}. 

However, qualitative differences in the evolution of A0 and A appear
as a result of the accumulated effect of mass loading. In particular,
the mass loading causes an efficient deceleration of the plasma within
the beam in model A, which leads to expansion. The increase of the
beam cross section reduces the momentum transfer per unit area to the
ambient and causes the jet to decelerate. As a result, the beam
in model A is denser, cooler and shorter than the one in model A0 and
expands with an almost constant half-opening angle of $\approx
1\fdg5$ until its disruption at about 900\,pc from the nucleus
(Fig.~\ref{fig:afinal}).  Whereas the beam in model A0 propagates at
almost constant speed, the terminal shock in model A decelerates with
time and nearly stalls by the end of the simulation (see
Section~\ref{ss:head_dyn}).  Mass loading and the accompanying jet
expansion appear to reduce pinching along the jet in model A relative
to model A0, so the conical shocks are weaker in the former case 
(compare the pressure panels of Figs~\ref{fig:a0sequence} and
\ref{fig:asequence}).  However, pinching is more disruptive at the jet
head in model A and is the main cause of the flaring.

%%%%%%%%%%%%%%%%%%%%%%%%%%%%%%%%%%%%%%%%%%%%%%%%%%%%%%%%%%%%%%%%%%%%%%%%%%%%%%%%%%%%%%%%%%%%%%%%%%%%%%%%%%%%%%%%%%%%%%%%%%%%%%%
%
%
%%%%%%%%%%%%%%%%%%%%%%%%%%%%%%%%%%%%%%%%%%%%%%%%%%%%%%%%%%%%%%%%%%%%%%%%%%%%%%%%%%%%%%%%%%%%%%%%%%%%%%%%%%%%%%%%%%%%%%%%%%%%%%%

%%%%%%%%%%%%%%%%%%%%%%%%%%%%%%%%%%%%%%%%%%%%%%%%%%%%%%%%%%%%%%%%%%%%%%%%%%%%%%%%%%%%%%%%%%%%%%%%%%%%%%%%%%%%%%%%%%%%%%%%%%%%%%%
%
%
%%%%%%%%%%%%%%%%%%%%%%%%%%%%%%%%%%%%%%%%%%%%%%%%%%%%%%%%%%%%%%%%%%%%%%%%%%%%%%%%%%%%%%%%%%%%%%%%%%%%%%%%%%%%%%%%%%%%%%%%%%%%%%%

\subsubsection{Weak jet models with mass loading: Models B and C}
\label{ss:modelbc}
%                        --------------------------------------------

%\paragraph{Models B and C}
%                       ...........................

\begin{figure*}
 \includegraphics[width=0.48\textwidth]{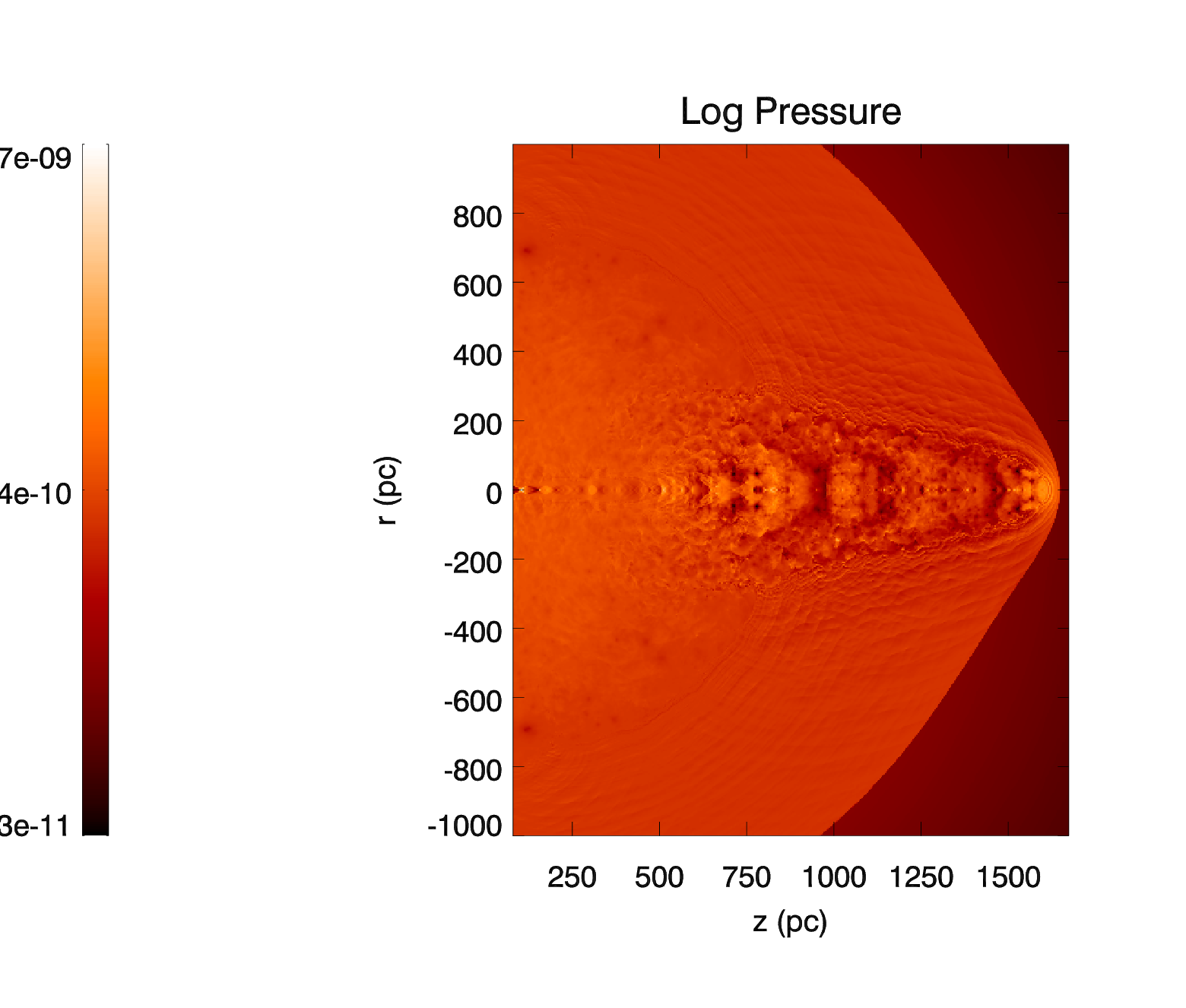}\hspace{0.5cm}
 \includegraphics[width=0.48\textwidth]{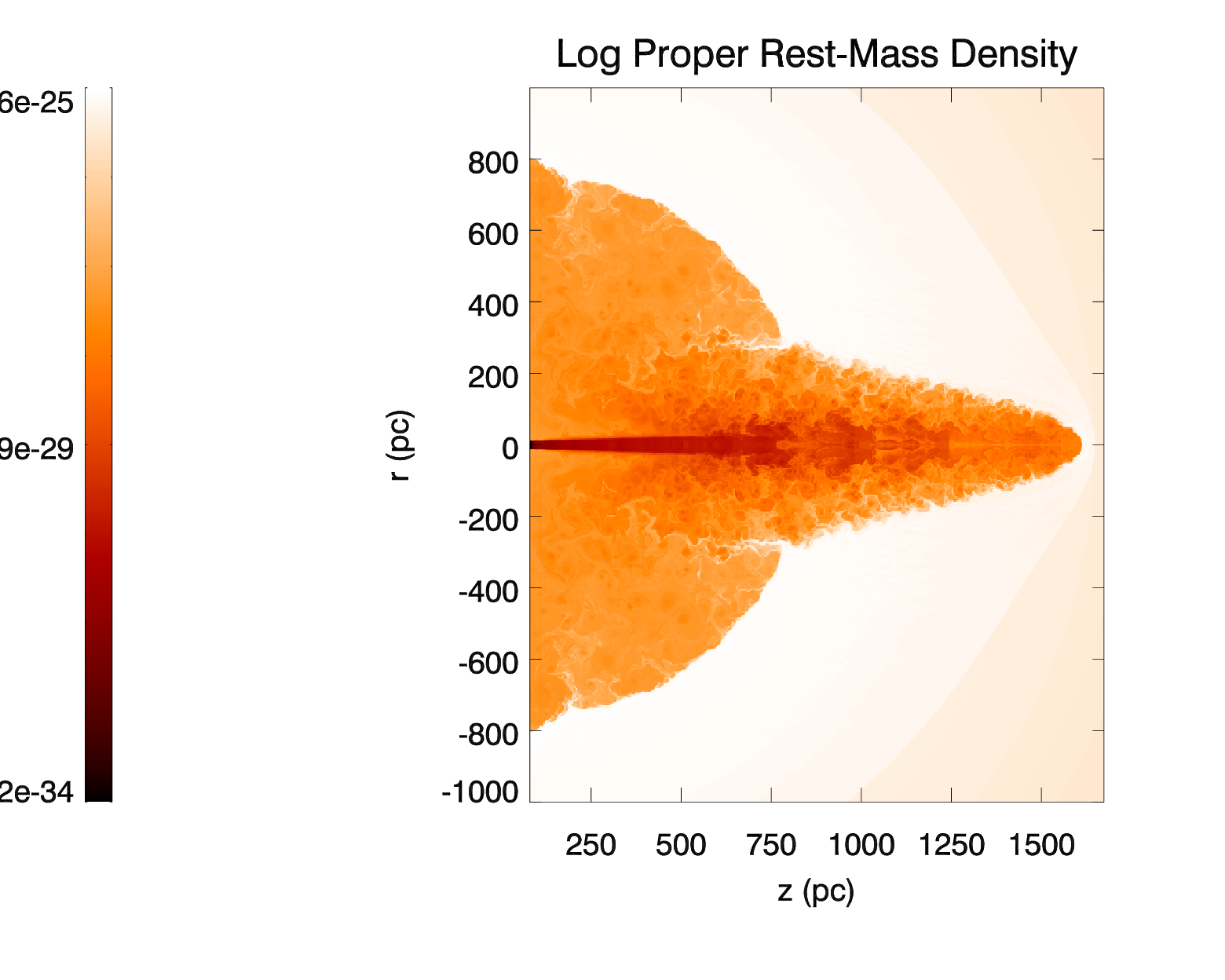} \\
 \includegraphics[width=0.48\textwidth]{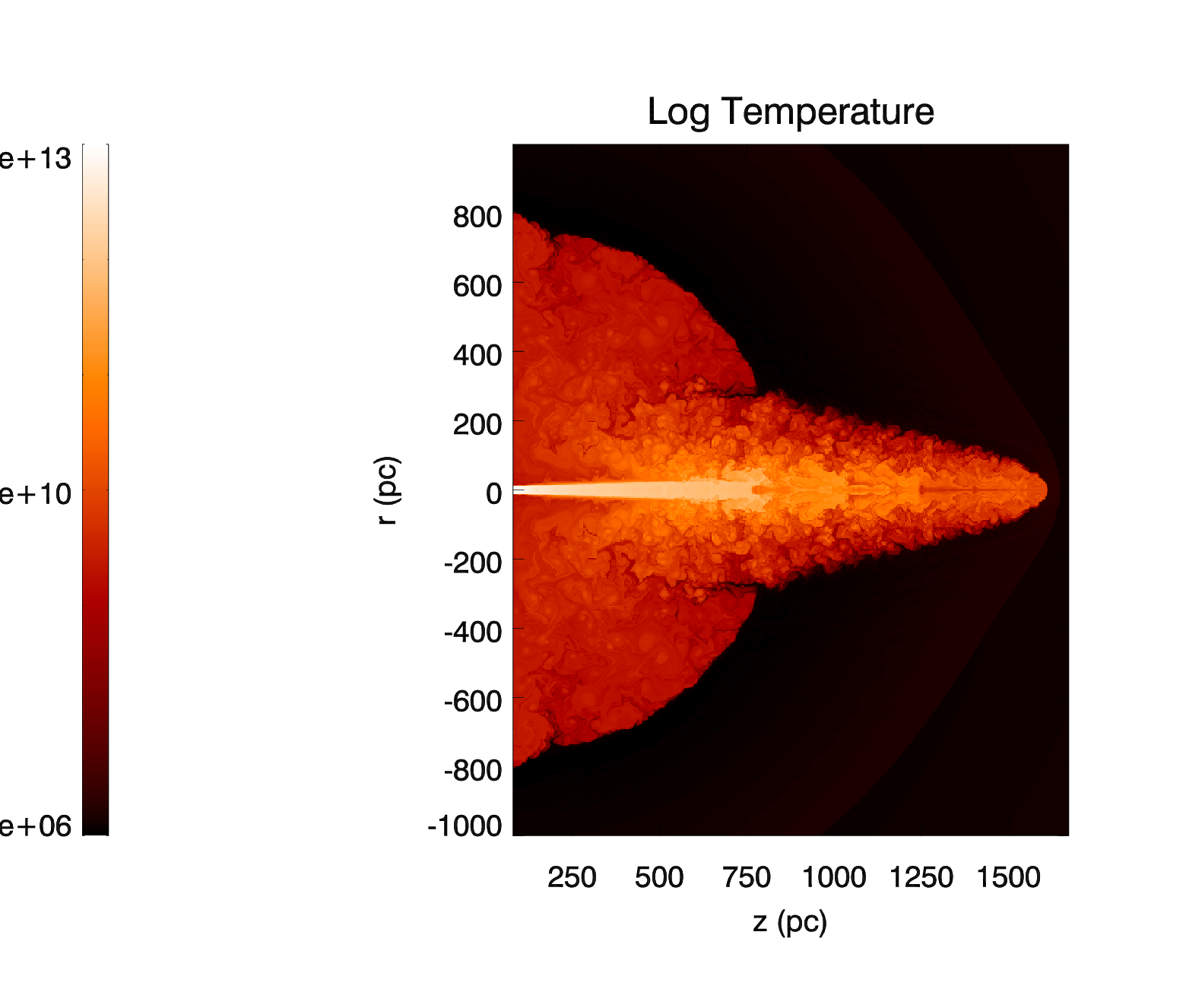}\hspace{0.5cm}
 \includegraphics[width=0.48\textwidth]{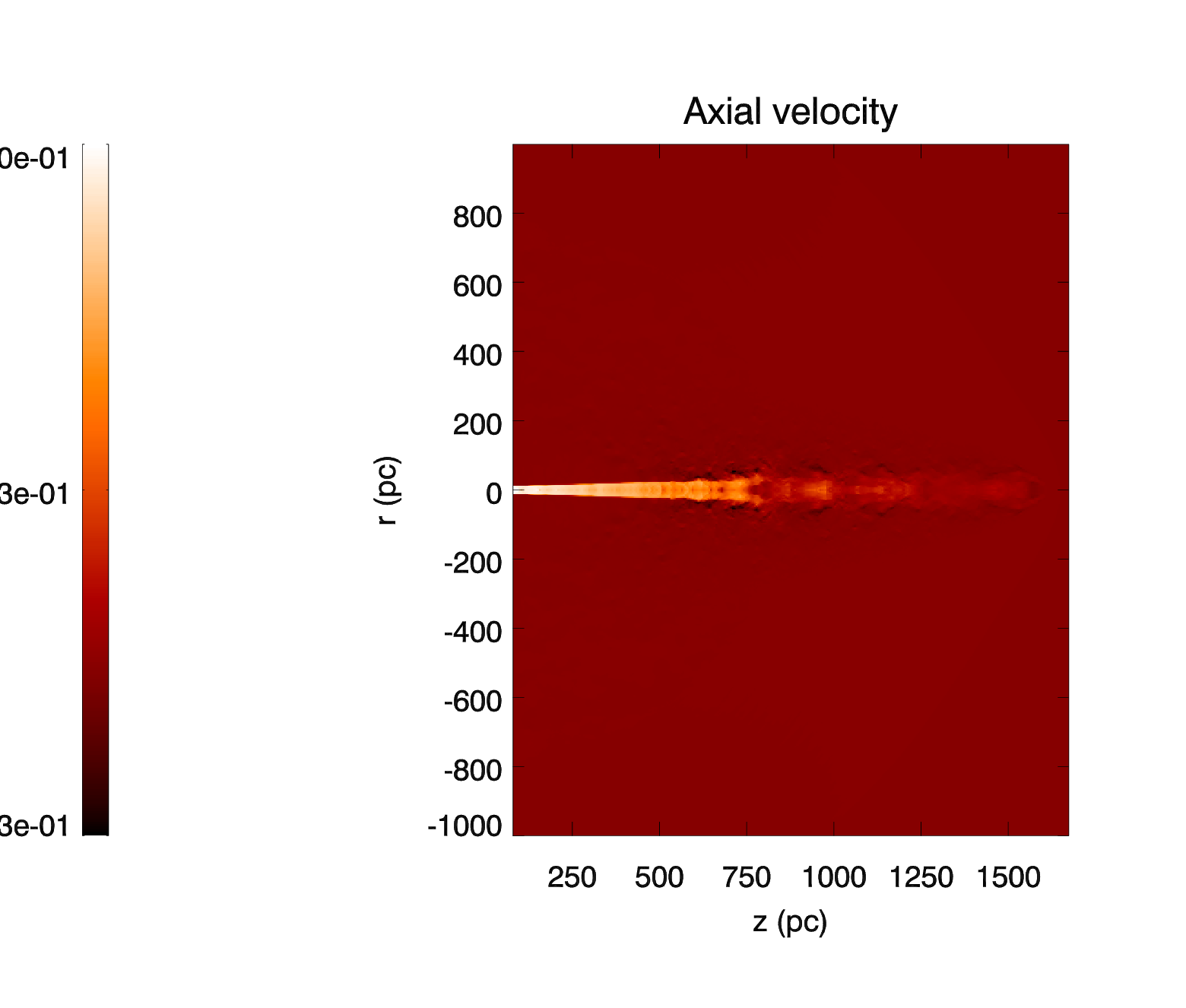}
 \caption{Pressure, rest-mass density, temperature and axial flow
velocity in model C at the end of the simulation ($t = 1.9 \times 10^6$ yr).}
 \label{fig:cfinal}
 \end{figure*}

  Models A, B and C (see Table~\ref{tab1}) form a sequence along which
  the density at injection decreases by a factor of one hundred and
  the specific internal energy increases by the same amount, keeping
  the same power (and thrust) for the jets. We do not show the early
  evolution of models B and C (very similar to that of model A), and
  concentrate on the distributions of pressure, rest-mass density,
  temperature and axial flow velocity at the end of simulation C
  (Fig.~\ref{fig:cfinal}), which show the largest differences compared
  with A.  The differences in the overall structure of the
  jet/cocoon/shocked ambient system between models A
  (Fig.~\ref{fig:afinal}) and C are very small
  (Section~\ref{ss:cocoon}). The similarity between these two models
  extends to the beam structure, with very similar distributions of
  density, pressure and flow velocity. The obvious conclusion
  (analysed in detail in Section~\ref{ss:beam}) is that the dynamics of the
  beams in models A, B and C are dominated by the process of
  mass loading with the thermodynamical properties playing a secondary
  role.

%
%%%%%%%%%%%%%%%%%%%%%%%%%%%%%%%%%%%%%%%%%%%%%%%%%%%%%%%%%%%%%%%%%%%%%%%%%%%%%%%%%%%%%%%%%%%%%%%%%%%%%%%%%%%%%%%%%%%%%%%%%%%%%%%

\subsubsection{Weak jet models with low mass loading: Model D}
\label{ss:modelD}
%                        -------------------------------------

%\paragraph{Model D}
%                       ..............

\begin{figure*}
 \includegraphics[width=0.48\textwidth]{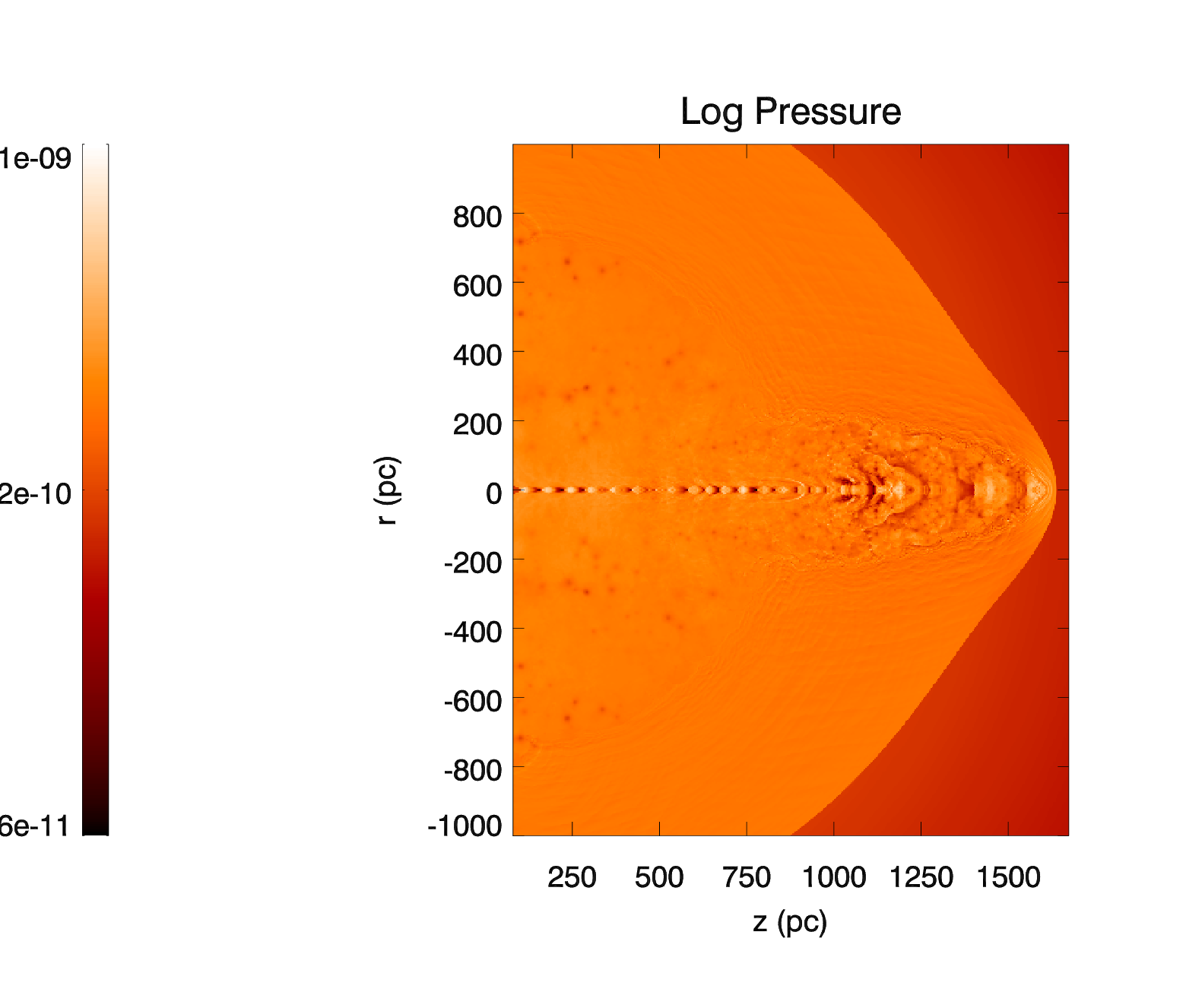}\hspace{0.5cm}
 \includegraphics[width=0.48\textwidth]{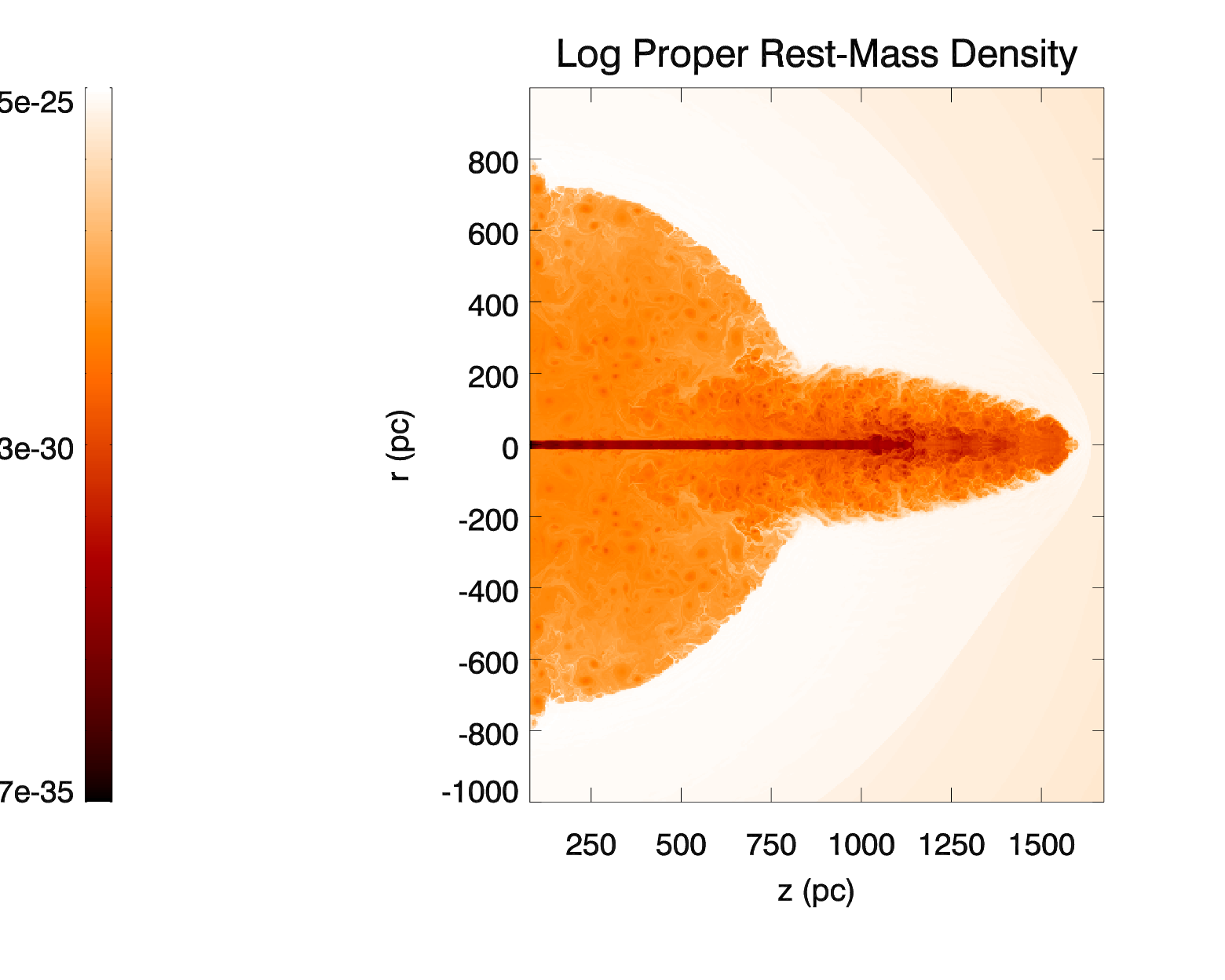} \\
 \includegraphics[width=0.48\textwidth]{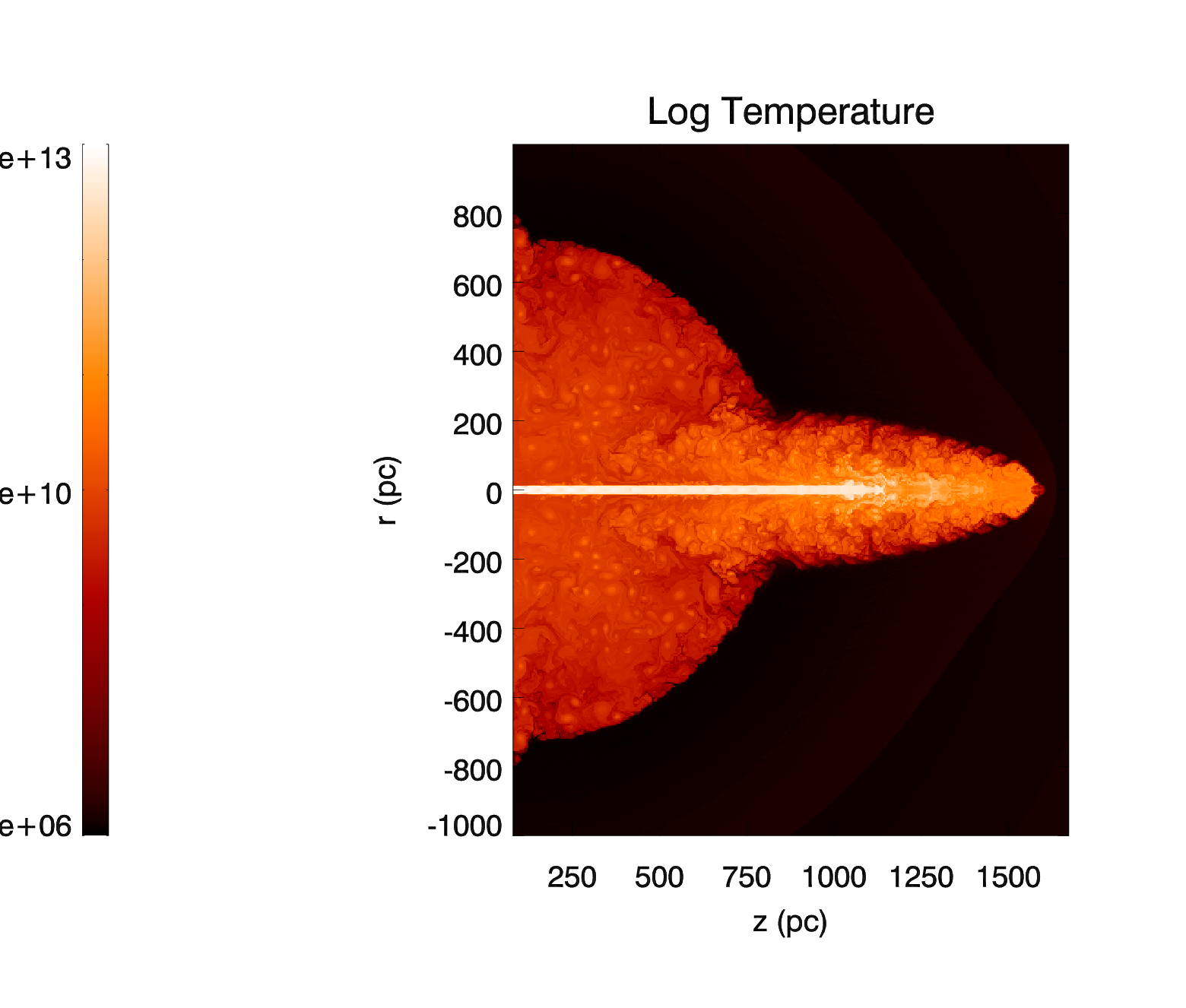}\hspace{0.5cm}
 \includegraphics[width=0.48\textwidth]{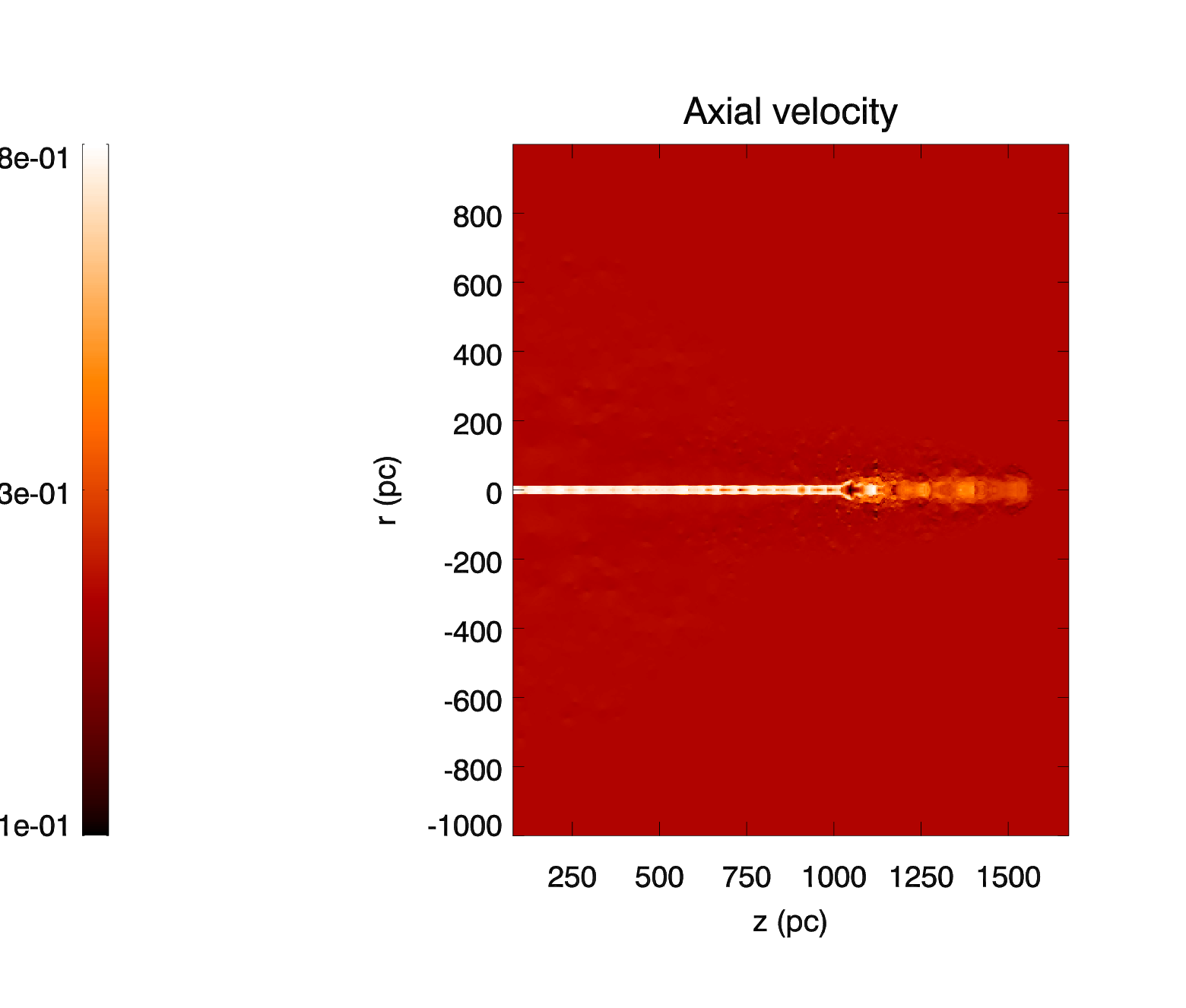}
 \caption{Pressure, rest-mass density, temperature and axial flow
velocity in model D at the end of the simulation ($t = 1.8 \times 10^6$ yr).}
 \label{fig:dfinal}
 \end{figure*}

  Model D has the same jet injection conditions as model C but a
  central mass-loading rate per unit volume, $q_0$, one order of magnitude
  smaller (Table~\ref{tab1}). Fig.~\ref{fig:dfinal} shows the
  distributions of pressure, rest-mass density, temperature and axial
  flow velocity of model D at the end of the simulation. The gross
  morphology of the jet/cocoon/shocked ambient ensemble is again
  almost identical to that of models A, B and C (and A0), as explained
  in Sect.~\ref{ss:modelA}. In spite of two orders of magnitude
  difference in the beam density (and temperature), model D resembles
  model A0 more than models A and C: the beam remains fast and
  well-collimated, with strong conical shocks. This result proves again
  the sensitivity of the beam structure and dynamics to the mass
  loading rate and the secondary role of the jet thermodynamics,
  within the range of parameters used in these simulations.

%%%%%%%%%%%%%%%%%%%%%%%%%%%%%%%%%%%%%%%%%%%%%%%%%%%%%%%%%%%%%%%%%%%%%%%%%%%%%%%%%%%%%%%%%%%%%%%%%%%%%%%%%%%%%%%%%%%%%%%%%%%%%%%
%
%
%%%%%%%%%%%%%%%%%%%%%%%%%%%%%%%%%%%%%%%%%%%%%%%%%%%%%%%%%%%%%%%%%%%%%%%%%%%%%%%%%%%%%%%%%%%%%%%%%%%%%%%%%%%%%%%%%%%%%%%%%%%%%%%

\section{Discussion}
%            %%%%%%%%%%
\label{disc}

\subsection{Comparison with previous work on stellar mass loading}

  The parameters defining the ambient medium, the mass-loss rates and 
the jets in models A, B, C and D are similar to those in BLK, 
allowing a comparison between the two sets of simulations. In
particular, the central mass-loss rate of the two reference models in
BLK ($2.36\times 10^{22} {\rm g\,yr^{-1}\,pc^{-3}}$) lies 
between those for models A, B and C ($4.95\times 10^{22} {\rm g\,yr^{-1}\,pc^{-3}}$), and 
model D ($4.95\times 10^{21} {\rm g\,yr^{-1}\,pc^{-3}}$). The kinetic luminosity per unit area at
injection of our models is about four times smaller than that of
the cold reference model in BLK, and just a bit larger
than that of their hot model. Finally, the Lorentz factor of the
flow at injection is 3 in models A, B, C and D, and 5 in BLK's reference models. The 
deceleration lengths, $l_{\rm d}$, predicted by equation~(\ref{eq:ld}) are therefore in the ratios: $l_{\rm{d,A,B,C}} : l_{\rm{d,hot}} : l_{\rm{d,cold}} : l_{\rm{d,D}}  \approx 1 : 1 : 4.5 : 10$. 
This prediction is in qualitative agreement with our results and those of BLK. Our models A, B and C all 
show deceleration, as do the hot and cold reference models of BLK, with the former having a shorter stopping distance (their Figs 3 and 4). Model D, by comparison, shows little deceleration.

The bulk parameters governing the propagation of the jets in our
simulations are similar to those of BLK, but differences are
expected from the disparity in the composition and thermodynamic
properties between the models and from the natures of the two kinds
of simulation. Our jets are composed of electron-positron pairs;
those of BLK contain electron-proton plasma. Thus, although the
temperatures of the two sets of models cover the same range
($\approx 3 \times 10^{11} - 3 \times 10^{13}$ K), all of our models
are thermodynamically hot. 
The simulations in BLK assume steady flows. They are therefore best
suited to describe a long, isolated jet propagating through a
prescribed (and undisturbed) ambient medium. Our simulations are
dynamical and allow us to study the complex head of the jet, the
formation of a cocoon and the process of flaring and disruption. They
are, however, restricted to the early evolutionary phases of FR\,I radio sources.

\subsection{Evolution of the cocoon and shocked ambient gas}
\label{ss:cocoon}
%                  ---------------------------------------------

\begin{figure}
 \includegraphics[width=0.48\textwidth]{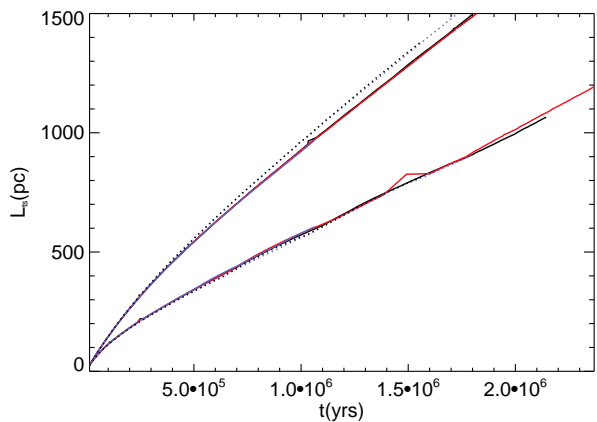}
 \caption{Length (top set of lines) and width (bottom set) of the
   shocked region as functions of time for models A, B, C, D and
   A0. Filled black:  A; filled red:  B; filled blue: 
   C. Dotted black:  A0; dotted blue:  D.}
 \label{fig:cavity_evol}
 \end{figure}

The gross morphological and dynamical properties of the shocked
ambient and cocoon in models A, B, C, D and A0 are all very
similar. The explanation is that all of the models
have the same kinetic power and thrust ($\approx L_{\rm j}/c = 
1.7 \times 10^{31}$ dyn) and propagate through the same
ambient medium, so the differences between their gross dynamical
properties are small.
 Fig.~\ref{fig:cavity_evol} shows the evolution of the length and
width of the shocked region with time for all of the models. The
differences among the models are tiny, particularly for the sideways
expansion. The axial expansions fall in two groups (although
with a difference in expansion speed of only a few percent), with
models A0 and D (not affected by the mass loading) undergoing the
faster expansion. In models A, B and C, second-order effects related
to the mass loading of the jet slow  the
expansion slightly (e.g.\ the increase in the jet cross
section reduces the jet thrust per unit area).

%%%%%%%%%%%%%%%%%%%%%%%%%%%%%%%%%%%%%%%%%%%%%%%%%%%%%%%%%%%%%%%%%%%%%%%%%%%%%%%%%%%%%%%%%%%%%%%%%%%%%%%%%%%%%%%%%%%%%%%%%%%%%%%
%
%
%%%%%%%%%%%%%%%%%%%%%%%%%%%%%%%%%%%%%%%%%%%%%%%%%%%%%%%%%%%%%%%%%%%%%%%%%%%%%%%%%%%%%%%%%%%%%%%%%%%%%%%%%%%%%%%%%%%%%%%%%%%%%%%

\subsection{Beam structure} \label{ss:beam}
%                  ---------------

  Although limited in duration, our simulations allow us to study the principal effects of mass loading on the beam evolution because the jet head is followed out to a distance of $1-2$~kpc, well beyond the stellar break radius $r_{\rm b} = 260$~pc. 

\begin{figure*}
 \includegraphics[trim=0cm 12.5cm 0cm 11.5cm,width=0.9\textwidth]{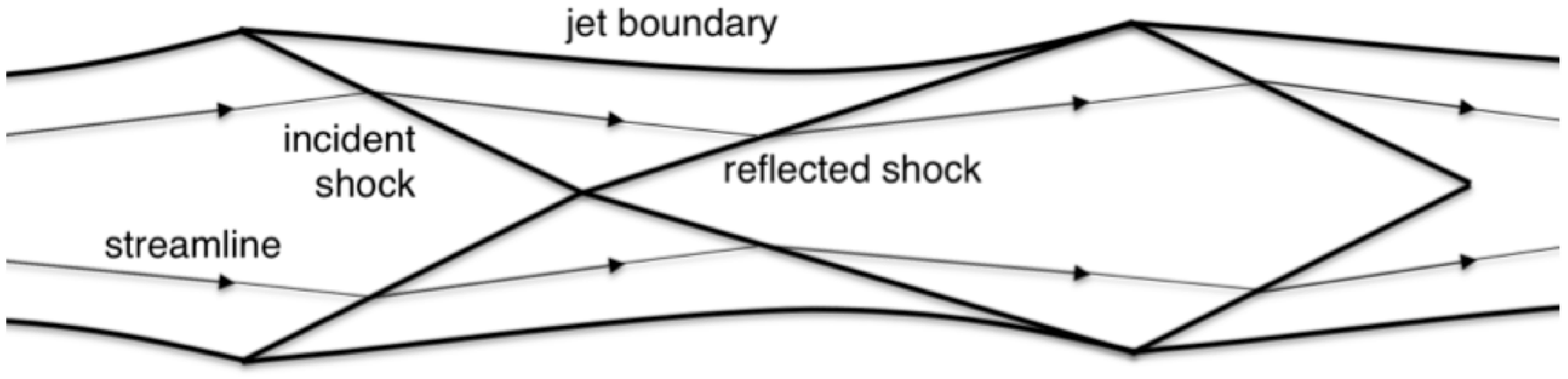}
 \caption{Sketch of the generation of a conical shock in a supersonic jet. For small pressure differences, the shocks are conical. If the initial jet overpressure is large or the fall in the ambient pressure is steep enough, the shock can produce a planar Mach disk, as observed in PM07.}
 \label{fig:sketch}
 \end{figure*}

 The internal structure of the jets is dominated by oblique shocks (conical shocks in axisymmetric models). Their origin is the pressure mismatch between the beam and its surroundings (i.e., the cocoon) and this produces periodic variations of the beam structure around some state of equilibrium. The wavelength of these variations is given approximately by ${\cal M}R$, where $\cal{M}$ is the relativistic Mach number and $R$ is the beam radius \citep{Wi87}. At the incident shock, the beam flow recollimates and decelerates whereas at the reflected shock, the flow is first compressed and then expands and accelerates, helped by the pressure gradient in the post-shock gas. The expansion continues until the pressure drops below the ambient value, when a new incident shock is initiated. Fig.~\ref{fig:sketch} shows a sketch of a conical shock. If the initial jet overpressure is large or the fall in the ambient pressure is steep enough, the shock can produce a planar Mach disk, as observed in PM07. The series of oblique shocks left behind by the jet head is present in all of the models but is especially prominent in models A0  (Fig.~\ref{fig:a0final}) and D (Fig.~\ref{fig:dfinal}). It is apparently unaffected by the small amount of mass loading in the latter case. In the more heavily mass loaded models A, B and C, the internal shocks are weaker (and the amplitudes of the pressure jumps are therefore smaller than in models A0 or D).

%%%%%%%%%%%%%%%%%%%%%%%%%%%%%%%%%%%%%%%%%%%%%%%%%%%%%%%%%%%%%%%%%%%%%%%%%%%%%%%%%%%%%%%%%%%%%%%%%%%%%%%%%%%%%%%%%%%%%%%%%%%%%%%
%
%
%%%%%%%%%%%%%%%%%%%%%%%%%%%%%%%%%%%%%%%%%%%%%%%%%%%%%%%%%%%%%%%%%%%%%%%%%%%%%%%%%%%%%%%%%%%%%%%%%%%%%%%%%%%%%%%%%%%%%%%%%%%%%%%

\begin{figure*}
  \includegraphics[width=0.45\textwidth]{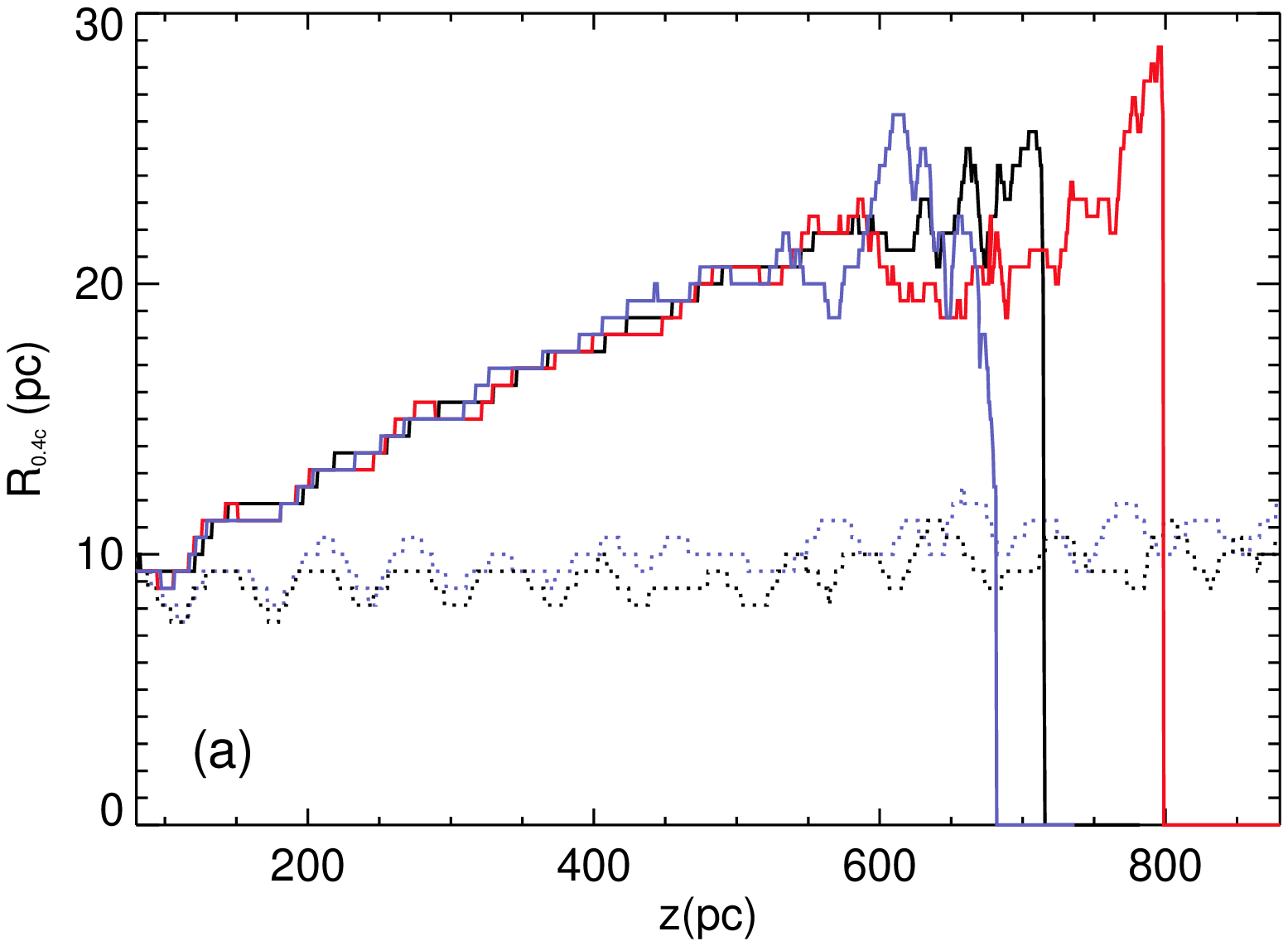}\hspace{1cm}
  \includegraphics[width=0.45\textwidth]{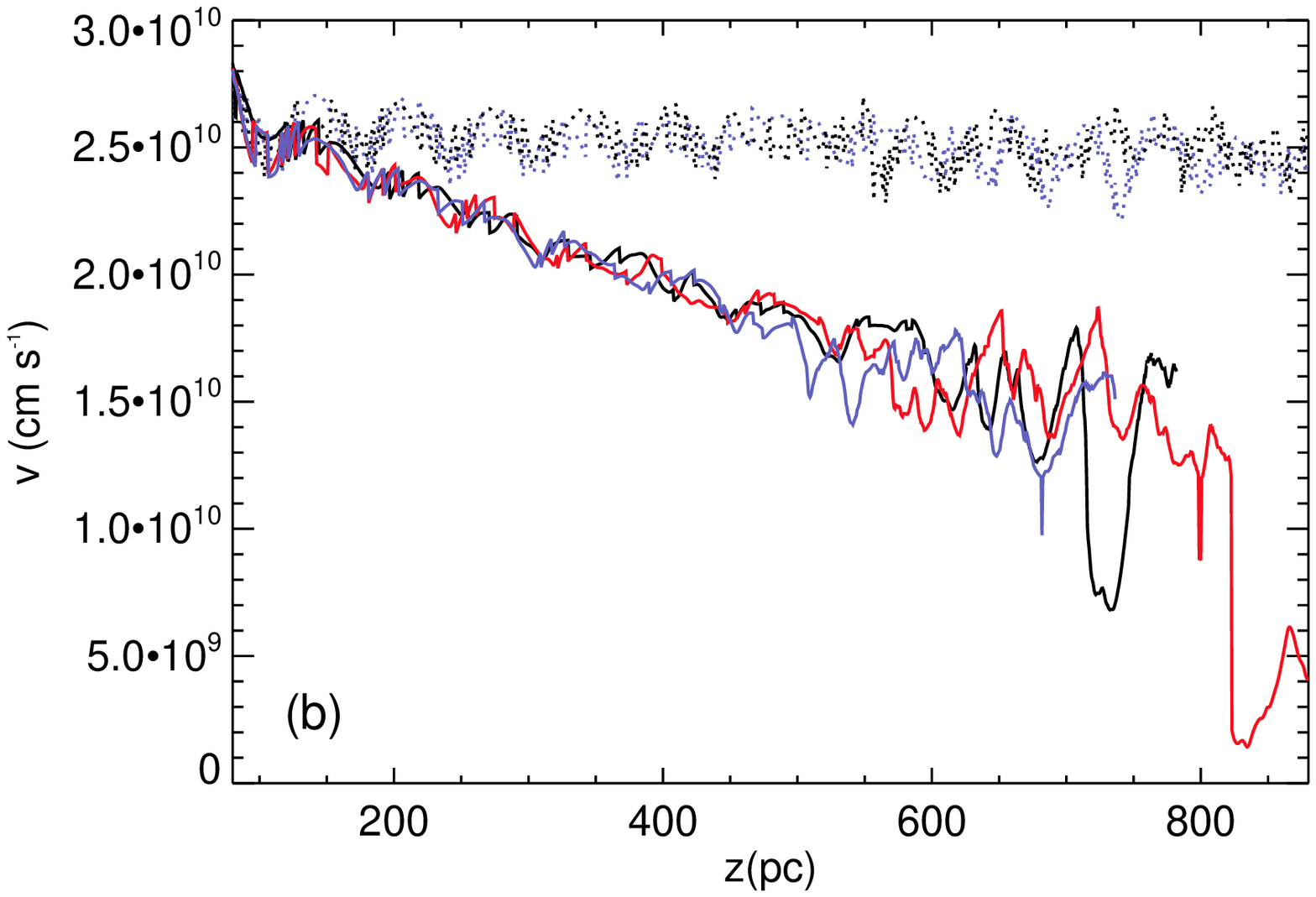}\\
  \includegraphics[width=0.45\textwidth]{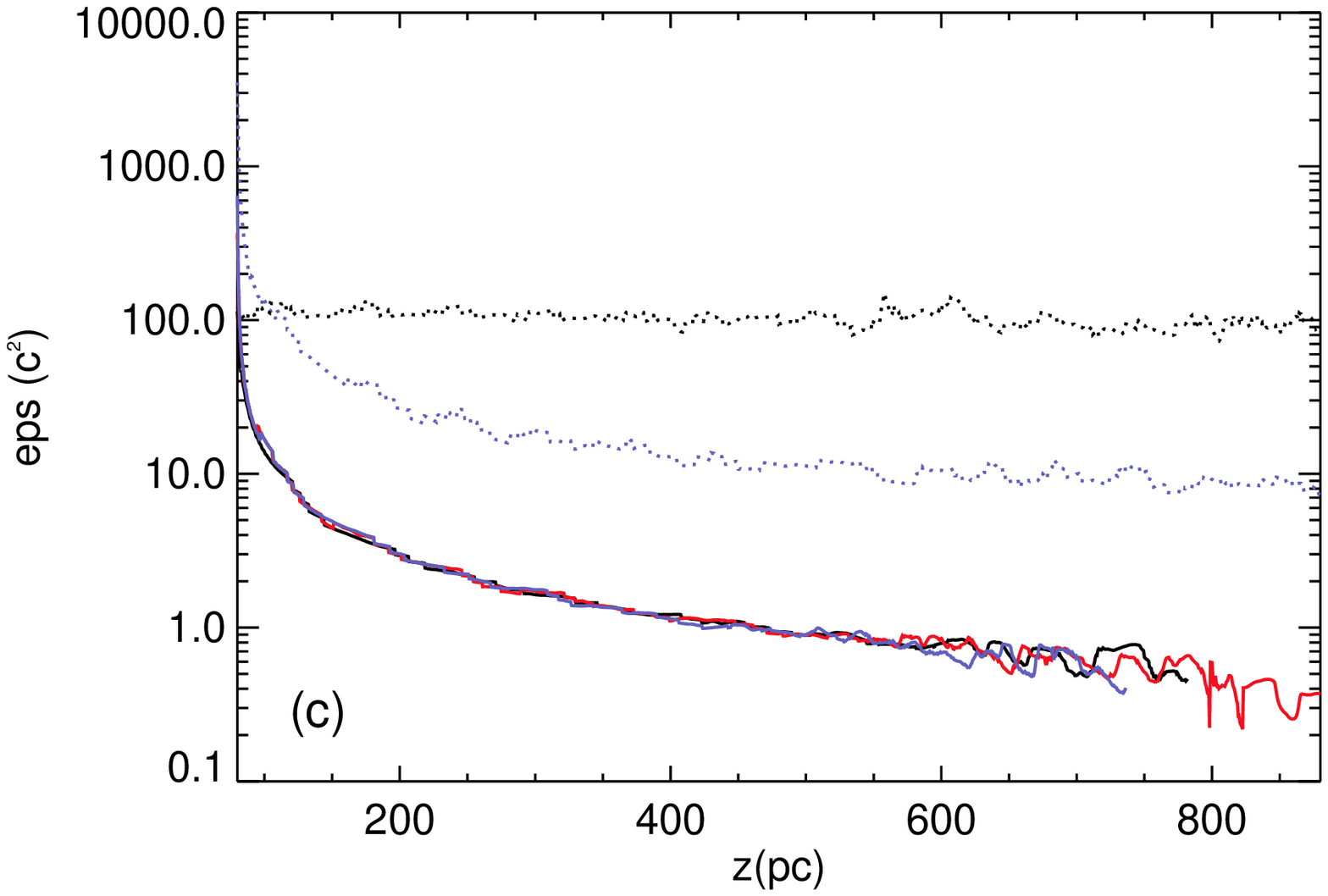}\hspace{1cm}
  \includegraphics[width=0.45\textwidth]{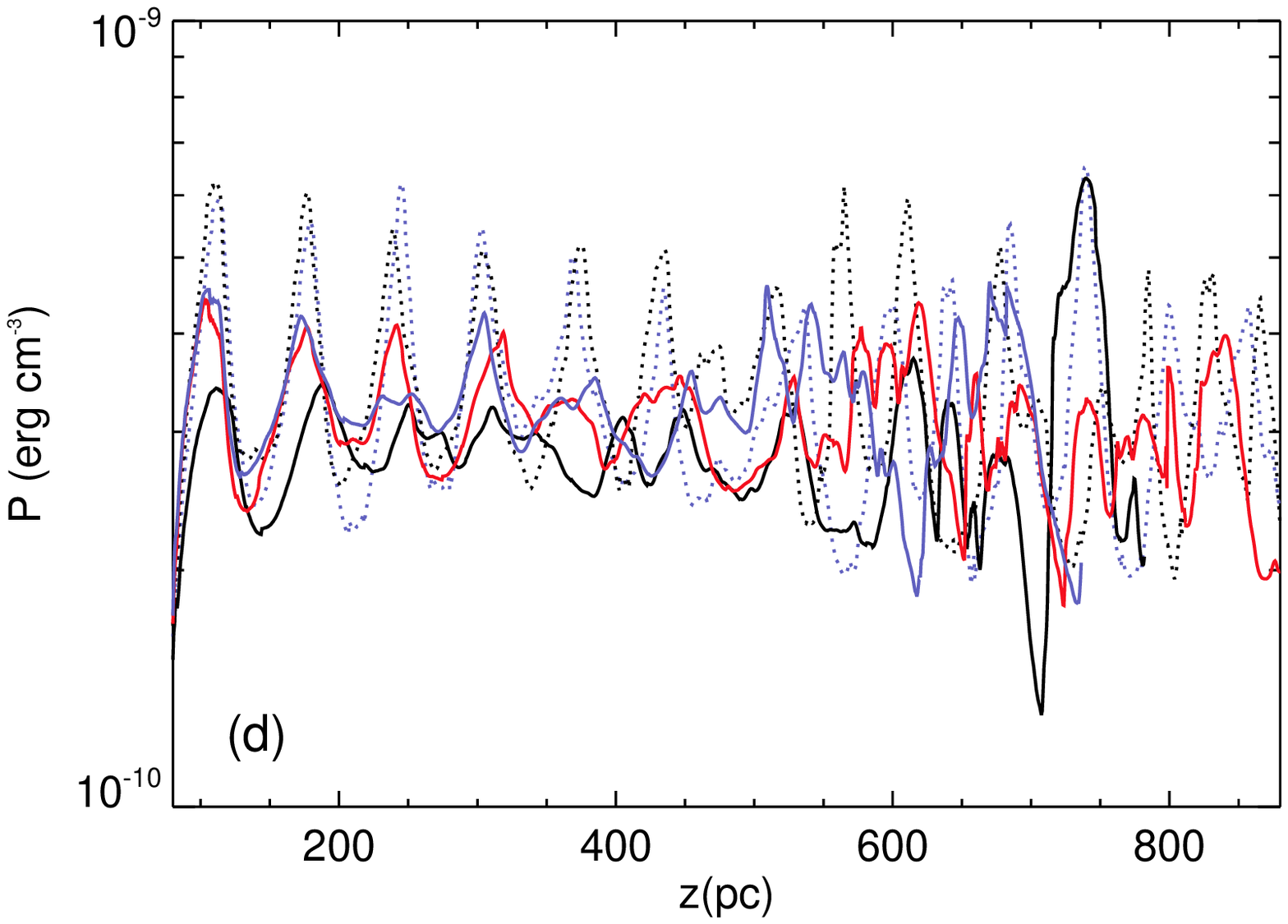}\\
  \includegraphics[width=0.45\textwidth]{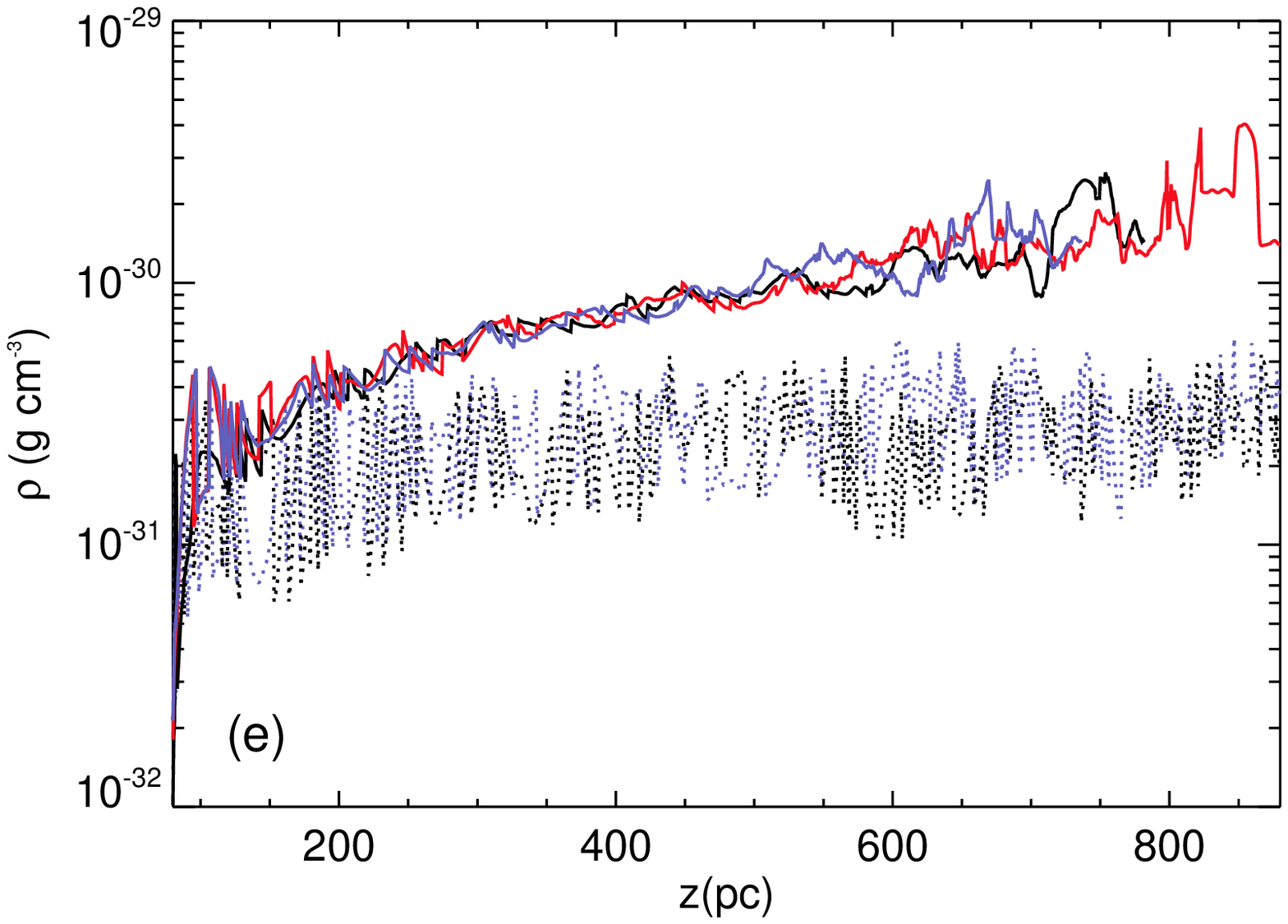}\hspace{1cm}
  \includegraphics[width=0.45\textwidth]{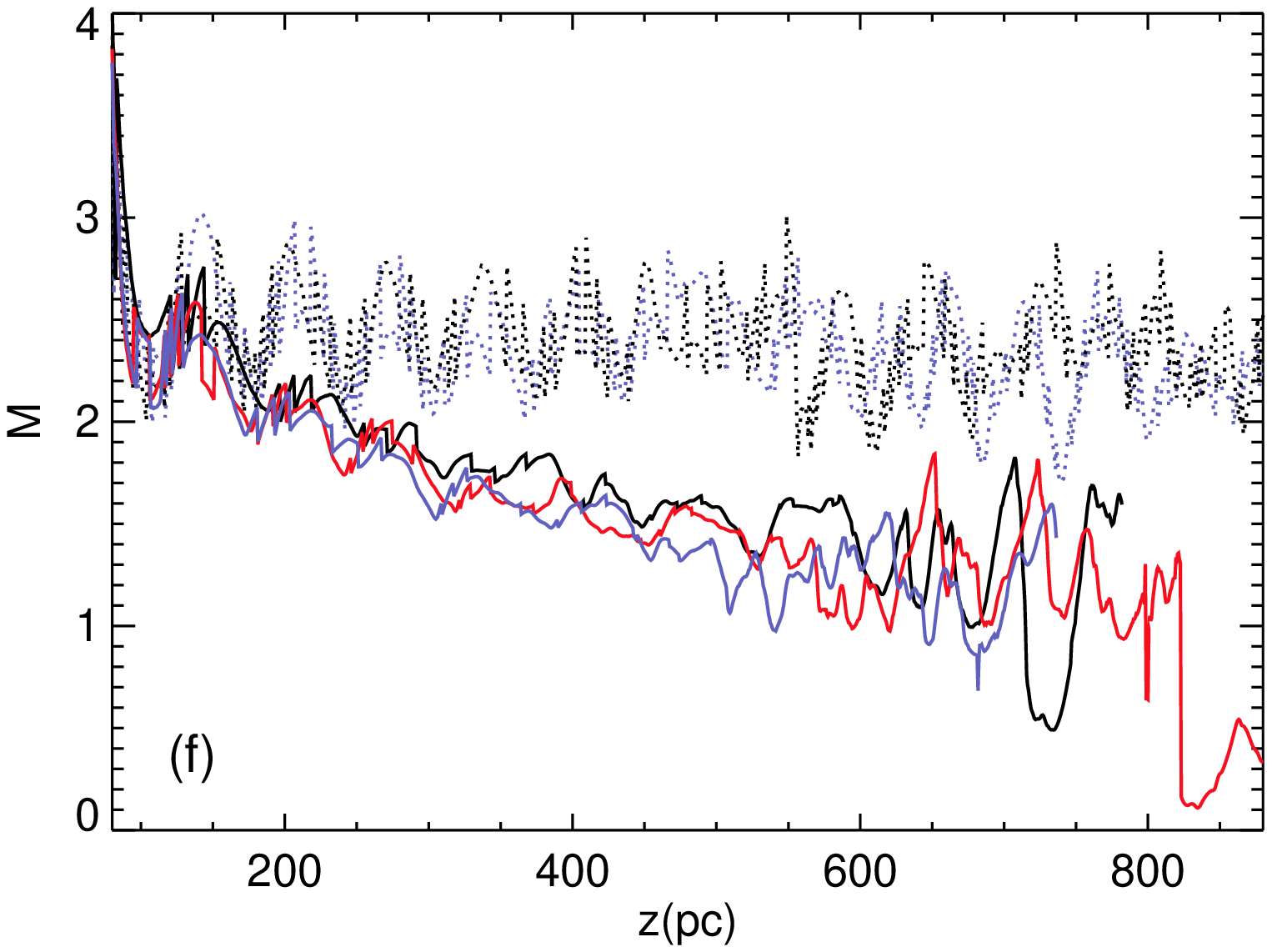}
  \caption{Plots of averaged beam parameters against axial distance, $z$, for models A, A0, B, C and D at the end of the simulation. (a) radius of the beam; (b) averaged value of axial velocity; (c) specific internal energy; (d) pressure; (e) density; (f) relativistic Mach number. Filled black:  A; filled red:  B; filled blue:  C; dotted black:  A0; dotted blue:  D.}
 \label{fig:beam_mean_values}
 \end{figure*}

  A simple analysis of the beam properties can be done by assuming stationarity from the injection point to the terminal shock. This is a fair assumption as long as the beam remains supersonic (which is certainly the case up to the terminal shock) and the injection boundary conditions and the mass loading are also stationary. The external pressure around the jet (the cocoon pressure) is quite homogeneous and varies slowly with time. The evolution of the averages across the beam of different relevant quantities along the axial direction can then be evaluated. To this end, the beam is defined as the domain encompassing all of the plasma with axial velocity larger than $0.4c$. The radius of this region is plotted against axial distance, $z$, in Fig.~\ref{fig:beam_mean_values}(a). The value of $0.4c$ represents a compromise that gives a well-defined beam profile up to the terminal shock at the cost of including a fraction of cocoon material accelerated by shear at the beam/cocoon surface in the analysis. As noted in Section~\ref{ss:modelbc}, the beam profiles  are remarkably similar for models A, B and C. This is primarily a consequence of mass loading dominating the beam structure, with the detailed thermodynamics of the different beam models playing a secondary role. The beam radius profile in model D remains close to that of model A0 and this confirms that the mass-load rate in model D is not strongly affecting the dynamics of the jet.

The primary effect of mass loading on the beam is the deceleration and
cooling of the flow as a result of the momentum transfer from the
original beam fluid to the newly incorporated material and the
reduction of the mean specific internal energy. Figs~\ref{fig:beam_mean_values}(b) and (c) show,
respectively, the averages of the flow velocity and the specific
internal energy across the beam as functions of axial distance for
models A, A0, B, C and D at the end of the simulations. The
deceleration of the beam flow in models A, B and C is clear. Despite
the differences in initial specific internal energy and density 
(two orders of magnitude between the extreme cases), the
deceleration rate in these models is roughly the same. This is because these initial values 
 have been
chosen so that all three models have the same initial internal energy per
unit volume and the same thrust (Table~\ref{tab1}).  The density
profiles are dominated by stellar mass input (except very close to the
injection point) and are also very similar (Fig.~\ref{fig:beam_mean_values}e).  The deceleration rate depends on the initial momentum of
the beam and the mass-loading rate, which are both the same in all
three models.  Similarly, the models all have the same amount of
internal energy to share with the cooler stellar wind plasma, so the
profiles of specific internal energy along the beam are also almost
identical. We conclude that the dissipation of kinetic energy into
internal energy along the beam is essentially the same in all three
mass-loaded models.  Finally, the similarity of the velocity and
specific internal energy profiles implies that the relativistic beam
Mach number, $\cal{M}$ must vary in the same way for models A, B and C.
It adjusts during the first $20$ pc after the injection to a value
around $2.4$, thereafter decreasing along the beam
(Fig.~\ref{fig:beam_mean_values}f).

In the contrasting case of low or zero mass loading (models A0 and D), the average
beam density profiles are dominated by cocoon material accelerated by
shear at the interface.  A comparison between the profiles for the
beam in model A0 (shown in Fig.~\ref{fig:beam_mean_values}) and the
values of the same quantities at injection (Table~\ref{tab1}) allows
us to quantify the effects of this boundary-layer entrainment. The
average flow velocity of the beam is $0.83c$, reduced by $12\%$ from its 
value of $0.95c$ at injection. The density however
increases by a factor of $70-100$ since the shear layer material included in the jet by our definition 
(cells with axial velocity larger than $0.4c$) dominates the average. The opposite happens in the case of the
specific internal energy, which is
several orders of magnitude larger in the beam than in the cocoon. The
average is dominated by the beam and shows a reduction of
about $30\%$ from the injection value. Increasing the velocity cut-off in the definition of the
beam reduces the beam radius and produces averages closer to the
 values at injection. A comparison between
the density profiles for  models A0 and A (Fig.~\ref{fig:beam_mean_values}e), confirms that mass loading by
stellar winds dominates over that from shear entrainment in
models A, B and C (but not in D).

We also note that the plasma in all of our models cools down along the
jets because our jets are initially thermodynamically hot. This
cooling is most pronounced in models A, B and C and least in A0, with
D being intermediate between them. Thus, the effect of thermal dilution
between the original beam plasma and the entrained stellar wind plasma
dominates over the heating produced by dissipation. This is not what
happens in some of the cases considered by BLK, notably their `cold'
reference model.

We can gain more insight into the behaviour of the jet radius and internal density 
from the constraints set by mass conservation and pressure balance.
  Time independence of the equation of continuity requires that 
\begin{equation}
\frac{d\,}{dz} (A \rho W v) = A q,
\label{eq:cont_steady}
\end{equation}
along the beam, where $A$ is the beam cross section and $\rho$, $q$, $v$ and $W$ are respectively the averages across the beam of the rest-mass density, the mass-entrainment rate, the flow velocity and the Lorentz factor. Hence, the product $A \rho$ must grow fast enough to balance the deceleration and still give a positive derivative in Eq.~(\ref{eq:cont_steady}). As a result, $\rho$, $A$ or both must increase along the beam.

  The pressure in the beam is governed by pressure balance with the cocoon. Since the  cocoon is almost isobaric, the beam pressure profile (averaged over fluctuations due to shocks) is almost constant (Fig.~\ref{fig:beam_mean_values}d). This panel also shows the smaller amplitude of the internal shocks in models A, B and C, mentioned at the beginning of this section. To counterbalance the effect of the decrease in the specific internal energy and keep the pressure constant, the density must increase along the beam (Fig.~\ref{fig:beam_mean_values}e). In practice, both $\rho$ and $A$ increase to ensure pressure equilibrium and to satisfy the equation of continuity. Note that beam expansion leads to further mass loading, which in turn leads to further deceleration and cooling. The final result is an effective deceleration of the jet head in models A, B and C, and associated decollimation and flaring of the jet beyond this point (see Section~\ref{ss:head_dyn}). On the other hand, the triggering of this runaway process occurs only for high enough mass-loading rates: model D (with a mass-loading rate only one order of magnitude lower than that of models A, B and C) evolves in essentially the same way as the unloaded model A0.

%%%%%%%%%%%%%%%%%%%%%%%%%%%%%%%%%%%%%%%%%%%%%%%%%%%%%%%%%%%%%%%%%%%%%%%%%%%%%%%%%%%%%%%%%%%%%%%%%%%%%%%%%%%%%%%%%%%%%%%%%%%%%%%
%
%
%%%%%%%%%%%%%%%%%%%%%%%%%%%%%%%%%%%%%%%%%%%%%%%%%%%%%%%%%%%%%%%%%%%%%%%%%%%%%%%%%%%%%%%%%%%%%%%%%%%%%%%%%%%%%%%%%%%%%%%%%%%%%%%

\subsection{Head dynamics and jet flaring}
%                  ----------------------------
\label{ss:head_dyn}

\begin{figure}
 \includegraphics[width=0.45\textwidth]{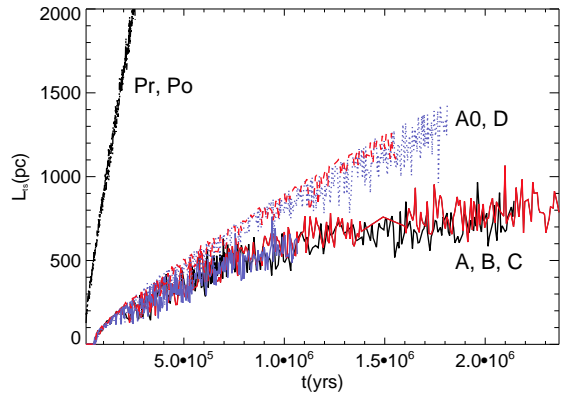}
  \caption{Position of the terminal shock as a function of time for all of the simulations discussed in this paper. Filled black:  A; filled red:  B; filled blue:  C; dotted black:  A0; dotted blue:  D; dash-dotted black:  Po; long-dashed black:  Pr}
 \label{fig:terminal_shock_with_time}
 \end{figure}

  The supersonic beam ends at a terminal shock where the beam flow decelerates and transfers part of its momentum to the ambient. In powerful jets, the head of the jet (the region between the terminal shock and the bow shock) forms a hot-spot. In the models presented here, due to the weakness of the jet, the region is quite broad and there is no substantial enhancement of the internal energy density.   Fig.~\ref{fig:terminal_shock_with_time} shows the position of the terminal shock as a function of time for all of the simulations in this paper. By the end of the simulations, the terminal shock in models A, B and C has nearly stalled at about $800$ pc, but those in models A0 and D are still propagating. 

In these axisymmetric simulations, the growth of KH pinching modes
also helps to trigger jet disruption via the process of external
mass loading. Although the pressure oscillations along the jet
(Fig.~\ref{fig:beam_mean_values}d) are larger in
models A0 and D, the growth of the pinch
mode surface amplitude is slower than in models A, B and C. These
modes are manifest for the latter group in the large-amplitude
oscillations of the averaged flow variables (beam axial flow velocity,
specific internal energy, pressure, rest-mass density and Mach number)
at distances between $\approx$500\,pc and disruption
(Fig.~\ref{fig:beam_mean_values}). The effects of pinching modes can also be seen for 
all of the simulations in the colour
pressure panels of Figs.~\ref{fig:a0final} and
\ref{fig:afinal}-\ref{fig:dfinal}. From linear relativistic KH theory, \citet{ha87} and
\citet{ha98} showed that the growth lengths (i.e.\ the $e$-folding
lengths of the growing modes) increase with increasing ${\cal M}R$. 
In the absence of deceleration due to mass loading, we would expect similar
growth lengths for the beams of models A, B, C, D and A0, since they 
have roughly the same values of ${\cal M}R$ at injection. However, the
continuous deceleration and dissipation in models A, B and C cause 
${\cal M}R$ to decrease with distance, even though the beams expand, so the 
growth lengths also decrease. Moreover, at low values of ${\cal M}R$, the
simple supersonic scaling breaks down and growth
lengths drop rapidly.
Thus, we conclude that the main cause of
the relative increase in the growth rates and the earlier disruption of
the jets in models A, B and C is the deceleration of the flow compared with that in A0 and D.
  
\begin{figure*}
 \includegraphics[width=0.45\textwidth]{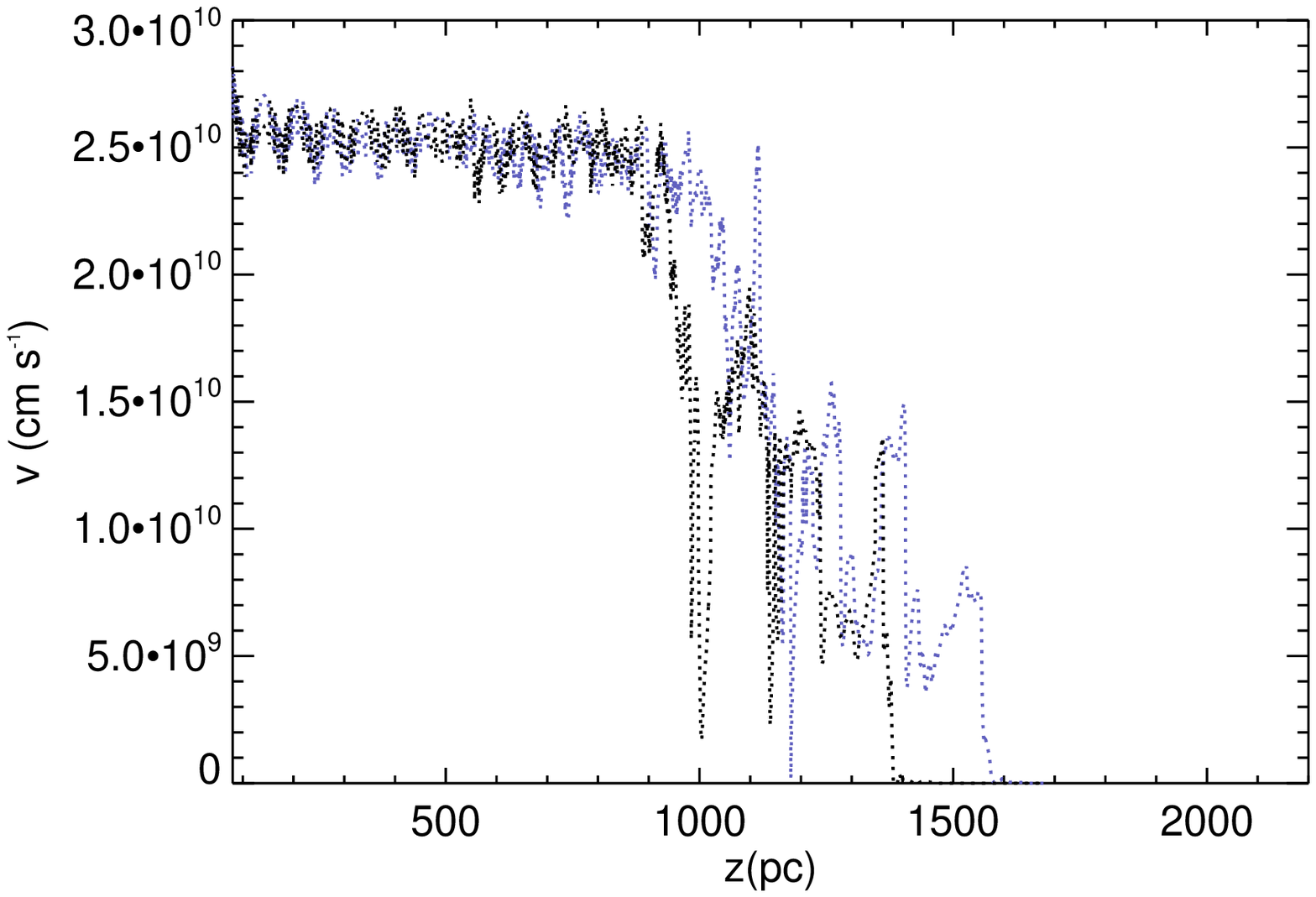}
 \includegraphics[width=0.45\textwidth]{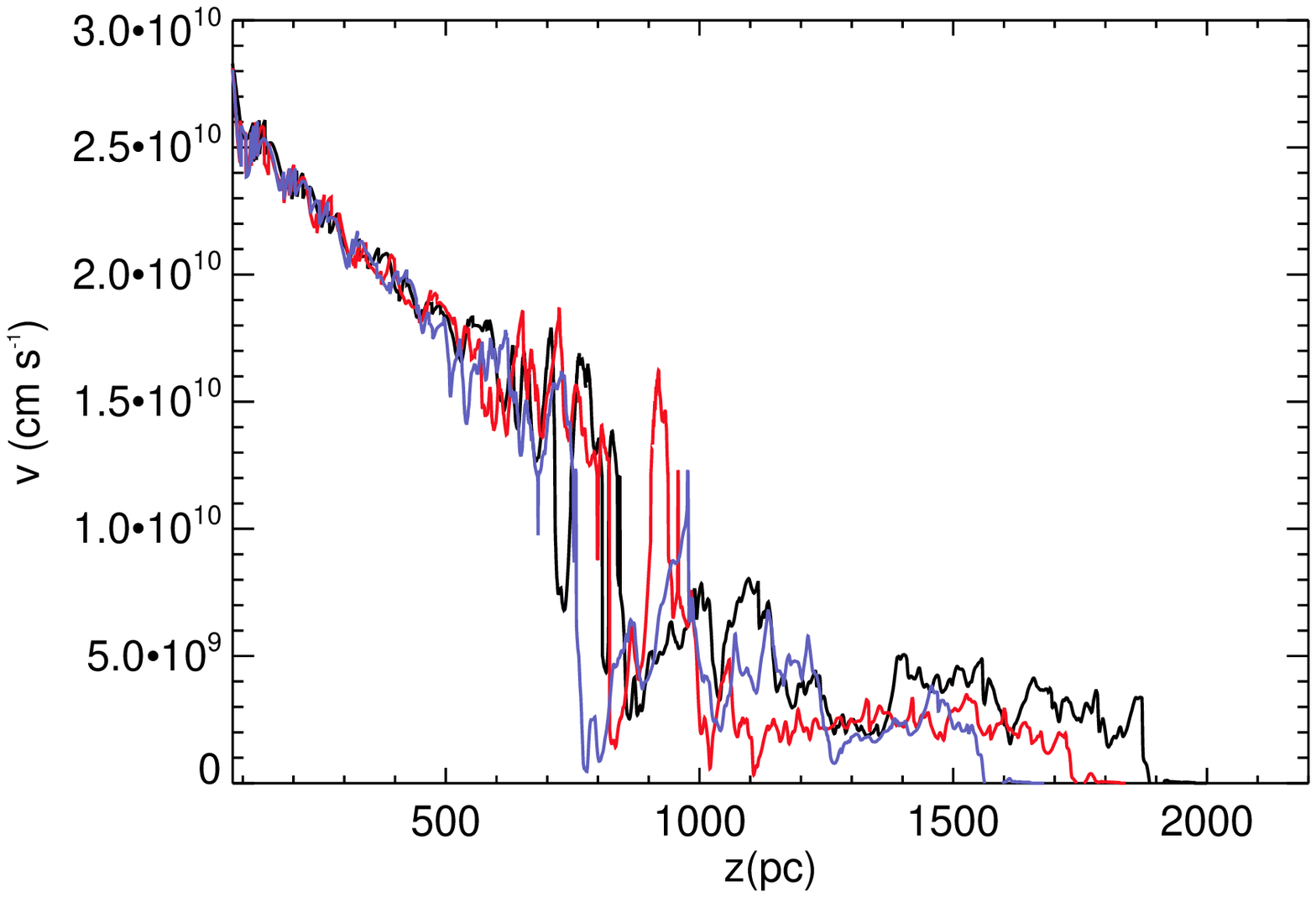}
  \caption{Profiles of average axial velocity covering the full
    simulation length. Upstream of the jet disruption point, these are
    exactly as plotted in Fig.~\ref{fig:beam_mean_values}. In the
    post-disruption region, the radius of the jet is held constant at
    its value immediately before disruption. Left panel: models A0
    (blue) and D (black).  Right panel: models A (black), B (red) and
    C (blue).}
 \label{fig:mean_velocity}
 \end{figure*}

In order to quantify the further deceleration after jet disruption,
Fig.~\ref{fig:mean_velocity} shows plots of average axial velocity
against distance continuing as far as the contact discontinuity with
the shocked ambient gas.  We cannot adopt the definition of the beam
used earlier (axial velocity $>0.4c$), as the velocities in the jet
head region are too low, and have held the radius of the averaging
region fixed at its value immediately before disruption.  The plots
therefore represent the on-axis velocity in the head region. The
models with zero or low mass loading (A0 and D;
Fig.~\ref{fig:mean_velocity} left panel) decelerate from $\approx
0.8c$ just before disruption to a mean speed of $\approx 0.3c$ over a
distance of $\approx 500$\,pc, with large fluctuations.  For the
models including mass loading (A, B and C:
Fig.~\ref{fig:mean_velocity} right panel), the beams decelerate from $\approx 0.8c$ to 
$\approx 0.5c$ before disruption and the final speed is
$\approx 0.1c$. 

%%%%%%%%%%%%%%%%%%%%%%%%%%%%%%%%%%%%%%%%%%%%%%%%%%%%%%%%%%%%%%%%%%%%%%%%%%%%%%%%%%%%%%%%%%%%%%%%%%%%%%%%%%%%%%%%%%%%%%%%%%%%%%%
%
%
%%%%%%%%%%%%%%%%%%%%%%%%%%%%%%%%%%%%%%%%%%%%%%%%%%%%%%%%%%%%%%%%%%%%%%%%%%%%%%%%%%%%%%%%%%%%%%%%%%%%%%%%%%%%%%%%%%%%%%%%%%%%%%%
\subsection{Transverse velocity profiles}
\label{transverse}

\begin{figure*}
  \includegraphics[width=0.45\textwidth]{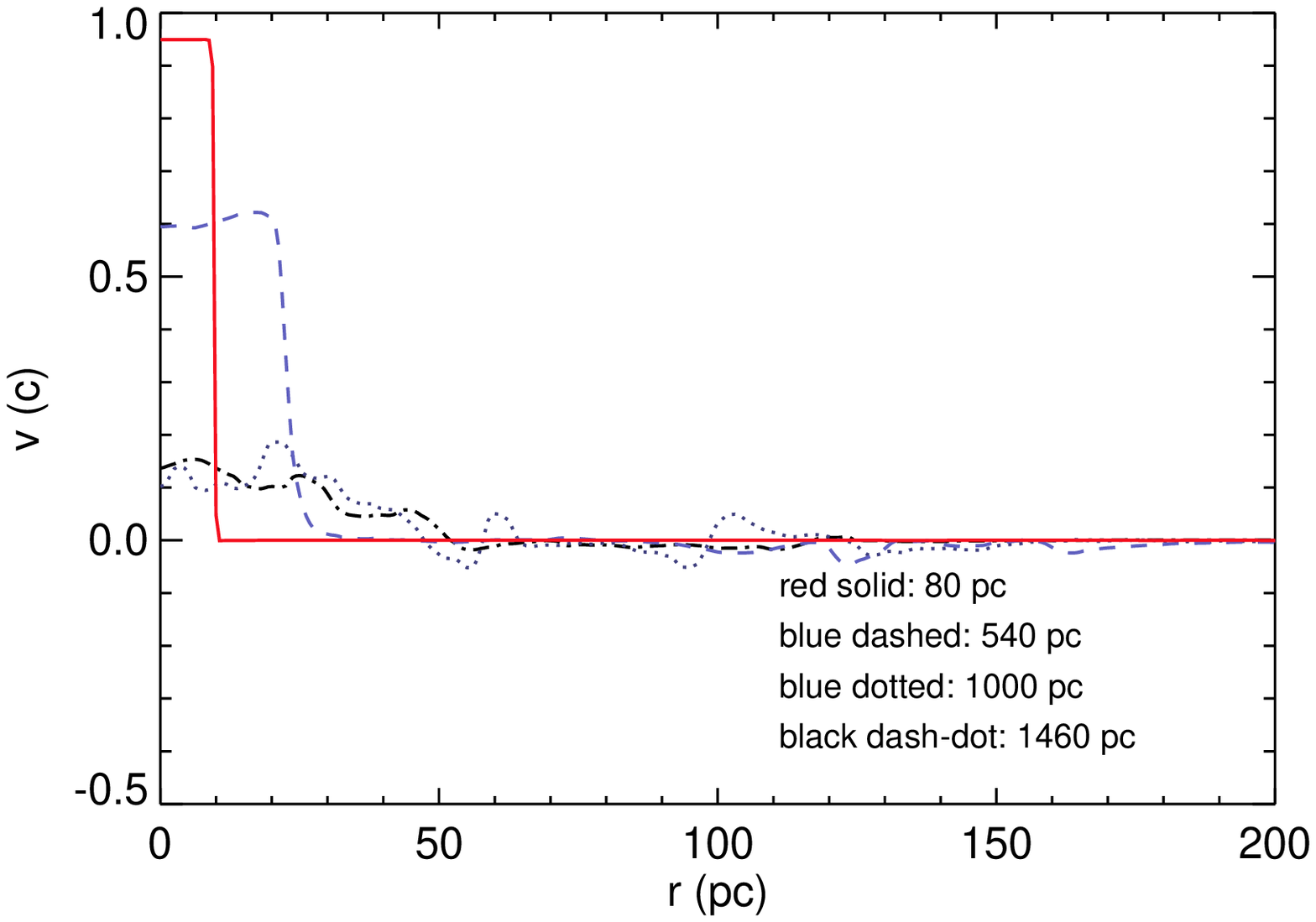}\hspace{1cm}
  \includegraphics[width=0.45\textwidth]{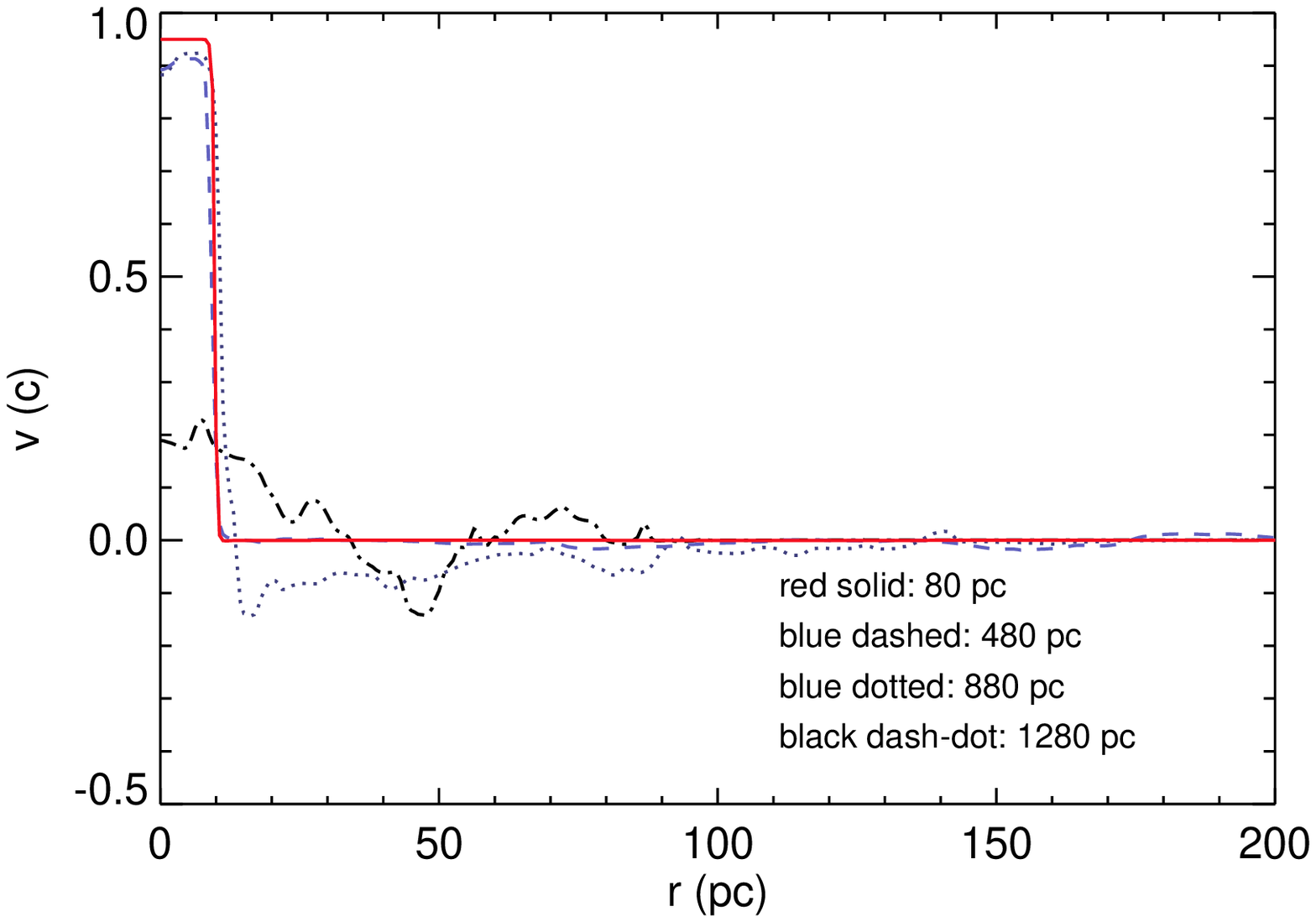}\\
  \includegraphics[width=0.45\textwidth]{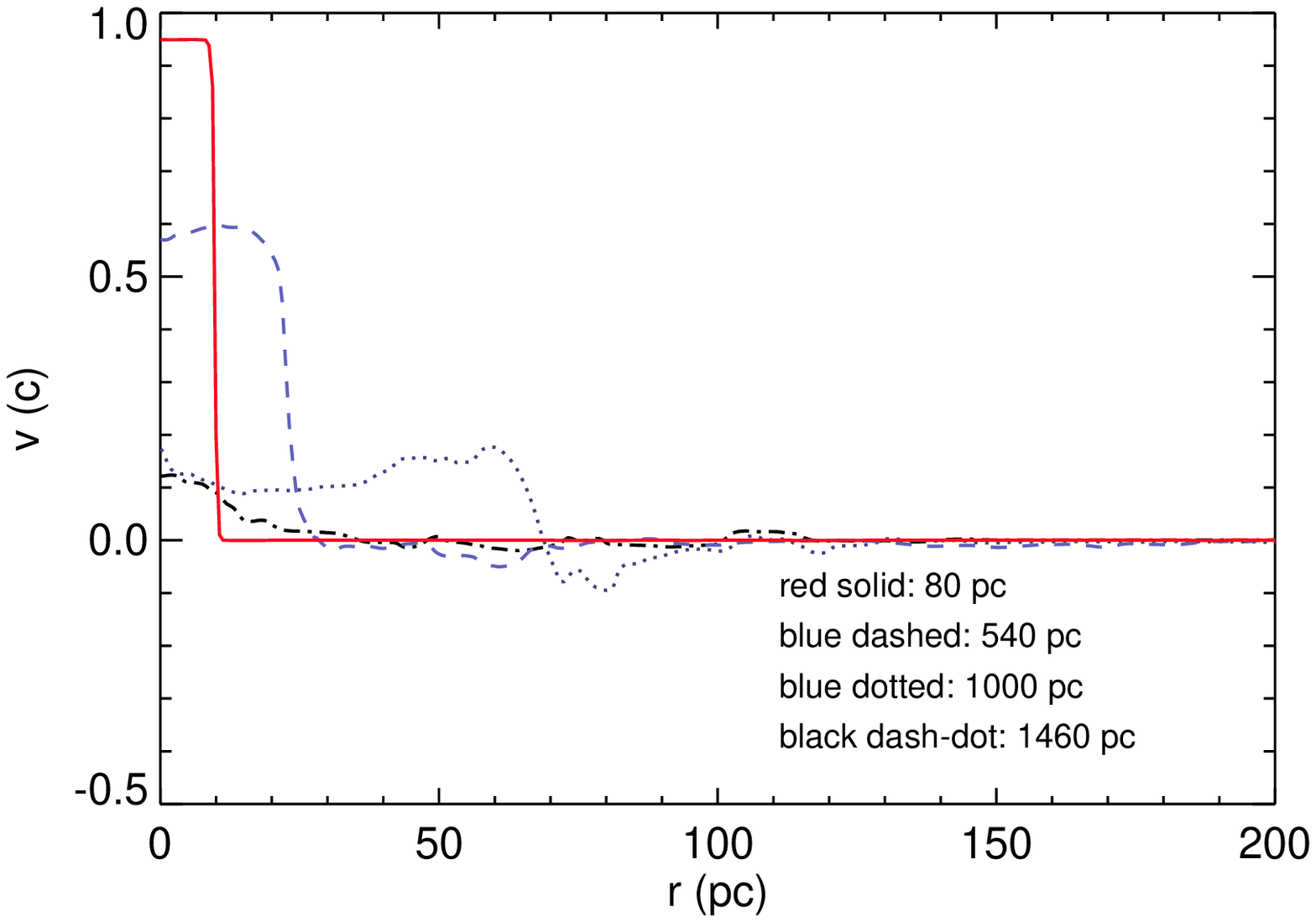}\hspace{1cm}
  \includegraphics[width=0.45\textwidth]{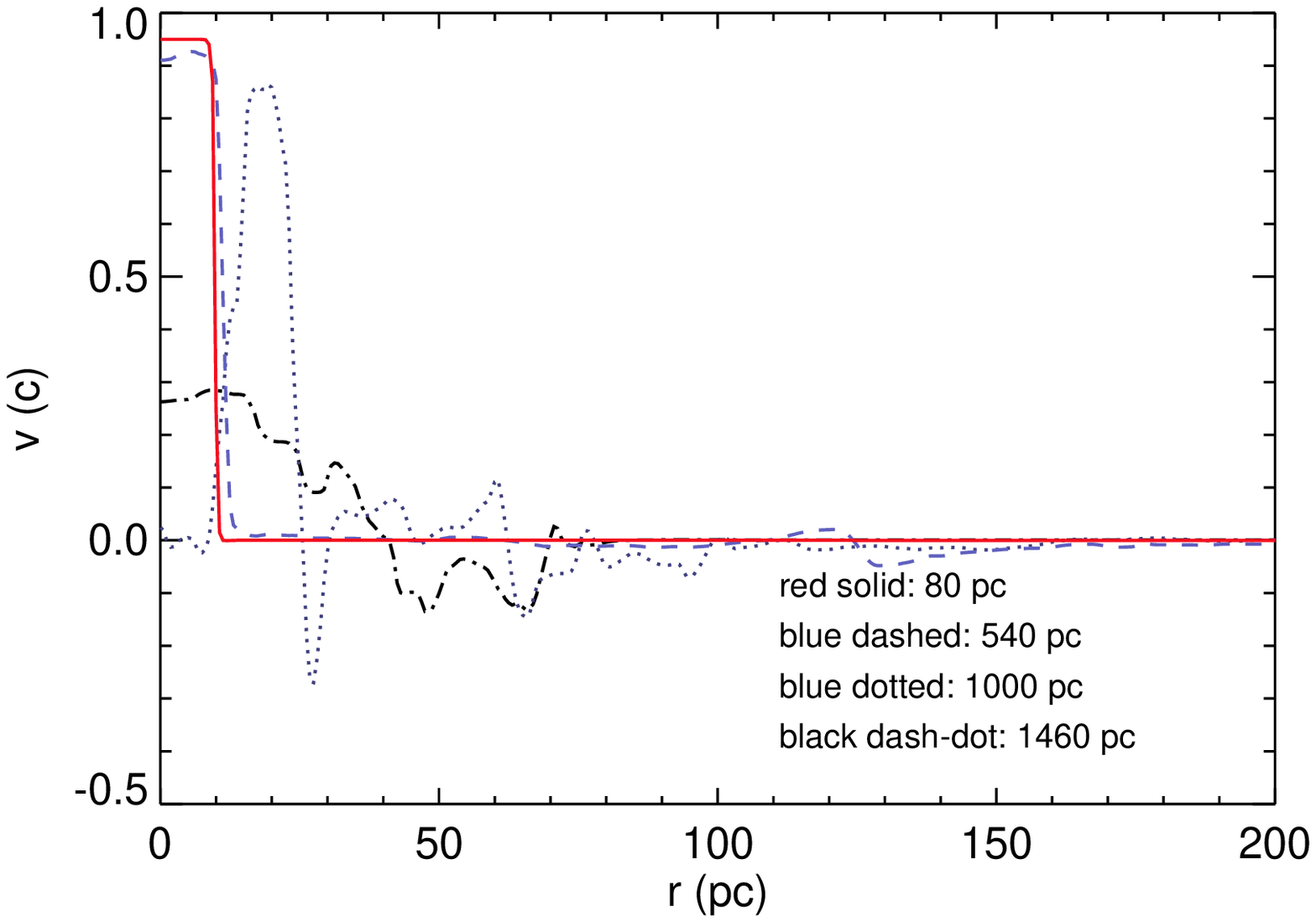}\\
   \caption{Radial cuts of axial velocity at different positions along the axial direction, typically at the injection position, one quarter, one half, and three quarters of the simulated grid in each case. The top left panel shows model A, the top right panel shows model A0, the bottom left panel shows model B, and the bottom right panel shows model D. The order of the lines, starting from the injection is: red solid, blue dashed, blue dotted, and black dash-dotted.}
 \label{fig:velocity_profiles}
 \end{figure*}

As noted in Section~\ref{intro}, \citet{lb14} asserted that the
development of a centrally-peaked transverse velocity profile was not
consistent with mass input distributed throughout the jet volume, as
would be expected for stellar mass loading. In agreement with this
statement, our simulations show that, although mass load by stellar
winds produces an expansion of the jet and a wide shear-layer, the
inner region of the jet preserves a flat transverse velocity profile
until the jet is disrupted. Profiles at 80 and 540\,pc for models A
and B show this effect clearly (Fig.~\ref{fig:velocity_profiles}).
After disruption, the transverse profiles become centrally peaked,
but here the deceleration is primarily due to interactions with the 
ambient medium rather than to stellar mass
loading, and the profiles are therefore similar (although with higher central
values) for models A0 and D, which have zero and low mass loading,
respectively (Fig.~\ref{fig:velocity_profiles}).

\subsection{Backflow}
\label{backflow}

Another interesting difference observed between the simulations with
and without mass loading relates to the backflow, with models A, B and
C showing small backflow velocities, $<0.1\,$c, whereas models
A0 and D show regions with velocities $\simeq 0.1\,$c. The cause of
this difference is related to the strength of the terminal shock,
which ultimately depends on the flow velocity immediately upstream of
the shock. The powerful jet simulation Pr is a more extreme case, with
backflow velocities $\simeq 0.6c$ close to the shock
(Section~\ref{ss:boundary}). In all cases, mass-loading at the contact discontinuity with the shocked ambient produced by KH instability decelerates the backflow. Recently, \citet{lb12} have claimed the
detection of backflows with velocities in the range of $0.05 -
0.35\,$c in two low-luminosity radio galaxies. If this result is
confirmed, it could be concluded from our results that the jet flow
velocity at the terminal shock is still large in those radio galaxies.
It is not clear, however, that the situations are analogous, since the
backflows modelled by \citet{lb12} are seen around the outer regions
of the jets, well downstream of the flaring region and initial
deceleration.

\subsection{Dynamical models of FR\,I jets }
\label{ss:dyn}
%                  ---------------------------

Over the years, a model for the jets in FR\,I radio sources has
emerged in which the main actors producing jet deceleration within the
first kiloparsec are (internal) mass loading and/or strong
recollimation shocks. In both cases, instabilities with post-linear or
non-linear amplitudes develop farther downstream and trigger external entrainment at the
edges of the jets \citep[][BLK,
  PM07]{bi84,la93,ko94,la96,lb02a,lb02b,ro08}. The important new
result from the simulations presented here is that a low-power
($L_{\rm j} \sim 10^{41}-10^{43}\,{\rm erg\,s^{-1}}$) FR\,I jet can be
decelerated efficiently as a result of mass loading from the winds of
the old stellar populations which dominate the low-excitation radio
galaxies typically associated with FR\,I radio sources, whereas a
high-power jet ($L_{\rm j} \sim 10^{44}\,{\rm erg\,s^{-1}}$) cannot.
There must be a range of variation between individual galaxies,
depending on the precise stellar distribution and population, but the
basic result should be robust.  

A corollary is that low-power FR\,I jets cannot be magnetically
dominated on kiloparsec scales: the stellar mass loading will
completely change the nature of the outflow within a few hundred pc.
There are also important implications for the lobe dynamics of weak
FR\,I sources and their ability to act as sources of ultra-high energy
cosmic rays: \citet{wy13} have recently suggested that the pressure in
the lobes of Centaurus A is dominated by heated plasma originating
from stellar mass input.

Other types of host galaxy may generate higher mass input rates,
however. For example, high-excitation radio galaxies typically have
lower stellar masses, but younger stellar populations with higher
average mass-loss rates (e.g.\ \citealt{bh12}). Individual young,
massive stars and starbursts may also be important \citep{HB06}.  The
high-excitation radio galaxies also have broad- and narrow-line
regions, which could provide mass loading sources for jets
if the filling factors are large enough to cause frequent encounters
between the jets and ionized gas clouds \citep*{AB10}.  Entrainment of
molecular clouds into the jets is another possibility \citep{h03}.
These sources of mass input may be able to decelerate more powerful
jets, and should be investigated in future simulations.

%%%%%%%%%%%%%%%%%%%%%%%%%%%%%%%%%%%%%%%%%%%%%%%%%%%%%%%%%%%%%%%%%%%%%%%%%%%%%%%%%%%%%%%%%%%%%%%%%%%%%%%%%%%%%%%%%%%%%%%%%%%%%%%
%
%
%%%%%%%%%%%%%%%%%%%%%%%%%%%%%%%%%%%%%%%%%%%%%%%%%%%%%%%%%%%%%%%%%%%%%%%%%%%%%%%%%%%%%%%%%%%%%%%%%%%%%%%%%%%%%%%%%%%%%%%%%%%%%%%

Recollimation shocks (Section~\ref{ss:beam}) result from the response of the
jet to a large pressure imbalance with its surroundings and have been
studied in the context of the collimation and structure of
extragalactic jets in both classical \citep{Sa83,Fa91} and
relativistic regimes \citep{Wi87,DM88,ko94}. If the jet is
instead under-pressured or only slightly over-pressured with respect to the
environment, small amplitude pinching will be induced, as
observed in models A0 and D. This pinching could then couple to a KH
unstable mode and grow with distance, triggering mixing and
deceleration. The spatial scale at which this process develops
depends on the jet and ambient properties but, in general, will be
longer than the kiloparsec scale. 
The most interesting case is,
however, when a recollimation shock reflects on the axis in a Mach
reflection rather than a regular reflection. In
this case, a planar Mach disk shock is formed, decelerating the flow very
efficiently and leading to a subsonic flow in the shock rest
frame. Based on the significant overpressure at the flaring point in
their models for 3C31, \cite{lb02b} proposed that the flaring point
is associated with a stationary shock system, even though
efficient Mach disks are only formed for jets with opening angles
much larger than the 8.5$^\circ$ found by \citet{lb02b} for the
inner region of 3C~31. In PM07, the authors tested such a possibility
based on axisymmetric numerical simulations of the 3C~31 jet and
environment. Three small Mach disks were produced within the 2\,kpc 
length of the flaring region in the model 
jet. Although the jet simulated in PM07 was injected with a large overpressure factor of 7.8 compared with the ambient,  jets generate their own environment
(the cocoon), which is also overpressured. The jet was in fact  
only marginally overpressured with respect to its immediate surroundings and had a relatively small
13.8$^\circ$ opening angle close to injection. Nevertheless, the
series of recollimation shocks was able to decelerate the flow to
subsonic speeds.

With the caveats noted in Section~\ref{intro}, kinematic models
of FR\,I jets \citep{lb14} imply that deceleration from $0.8c$ to
sub-relativistic speeds is a fairly gradual process, involving the
development of transverse velocity gradients and therefore associated
with instabilities and boundary-layer entrainment.  This
leaves open the question of whether recollimation shocks provide the
initial trigger for deceleration.  The high-resolution observations
summarized by \citet{lb14} present a challenge to the idea that
deceleration is initiated by recollimation shocks, at least in the two
best-resolved cases, 3C\,296 and NGC\,315 (their Fig.~17d and i).  Firstly,
neither of these jets show bright features crossing the jets at the start of their
flaring regions that might plausibly be identified with strong
reconfinement shocks. Instead, the brightness enhancements in the
flaring region are complex, non-axisymmetric and restricted to the
central parts of the jets.  Secondly, the expansion rates of the jets
increase monotonically with distance after the flaring point, with no
sign of recollimation until much larger distances.  On the other hand,
the bright knot at the base of the flaring region in 3C\,31 \citep[their Fig.~17e]{lb14} could
still plausibly be associated with a recollimation 
shock. Higher-resolution observations are needed to decide whether
such shocks are present in  FR\,I jets.

Another option considered to explain the deceleration of the flow is
the development of instabilities that grow to nonlinear amplitudes and
could give rise to the flaring of emission. \citet{lb14} suggested that the
growing modes have to be high-order because the observed jets flare
without disruption. All that is really necessary, however, is that they 
have short wavelengths. The process of slow deceleration could, instead, be related to the  
type 2 unstable flows (UST2) classified by  \citet{pe05}. These occupy a characteristic 
location in the relativistic Mach number - Lorentz factor plane and were confirmed using three-dimensional
simulations by \citet{PM10}. This type of deceleration corresponds to
a slow and progressive mixing, expansion and deceleration of jets that
is caused by short-wavelength, but low-order, body modes. \citet{pe05}
showed that this mechanism can be relevant in hot jets,
because the linear solutions give high growth-rates and short
wavelengths for the resonant values of KH body-modes, as opposed to
the solutions obtained for colder jets. Interestingly, \citet{lb14}
conclude that the jets in their sample could be formed by hot, low-Mach-number
flows. Further work is needed to establish whether this mechanism is effective  
for the parameter range applicable to FR\,I jets. The resolution used in our simulations is not high enough to 
allow for the development of short-wavelength unstable modes. However, the goal of this paper is to
determine the range of jet powers over which mass-loading by stars could be relevant. Our simulations allow us 
to achieve this goal and also to demonstrate the development of longer wavelength unstable modes. A detailed study to compare the 
effects of the growth of short-wavelength modes and mass-loading by stars would require specific high-resolution simulations.

It is important to note that the growth of instabilities from
the linear regime requires pressure equilibrium with the ambient: if
the jet is overpressured and expands, the expansion dominates the interaction between the
jet and the ambient, making it difficult for a growing instability to
become dynamically important. In addition, the wavelength of the
fastest-growing modes changes proportionally to the jet radius and the
growth rates are reduced \citep[see, e.g.,][]{ha87,ha00,ha11}.

%In our case, the disruption is produced by coupling to pinching modes (see Figs.~\ref{fig:mean_velocity} and% \ref{fig:velocity_profiles}). In this respect, \citet{lb14} point out that coupling to pinching modes and t%he consequent disruption could possibly be an artifact caused by axisymmetry. Nevertheless, pinching is a na%tural result of a symmetric pressure imbalance between the jet and the ambient\footnote{If the pressure imba%lance is asymmetric, then a helical mode can grow, and this cannot be reproduced by two-dimensional axisymme%tric simulations.}. Therefore, a supersonic jet in pressure equilibrium with the ambient will naturally gene%rate pinching if the ambient pressure drops, and the only way to avoid this is if the jet (or, at the least,% its edge) becomes transonic or subsonic upstream of the drop in ambient pressure. 

%%%%%%%%%%%%%%%%%%%%%%%%%%%%%%%%%%%%%%%%%%%%%%%%%%%%%%%%%%%%%%%%%%%%%%%%%%%%%%%%%%%%%%%%%%%%%%%%%%%%%%%%%%%%%%%%%%%%%%%%%%%%%%%
%
%
%%%%%%%%%%%%%%%%%%%%%%%%%%%%%%%%%%%%%%%%%%%%%%%%%%%%%%%%%%%%%%%%%%%%%%%%%%%%%%%%%%%%%%%%%%%%%%%%%%%%%%%%%%%%%%%%%%%%%%%%%%%%%%%

  We stress that our simulations have followed the evolution of the propagating jets for $\la 2\times10^6$~yr after they are initiated. Although we believe that the simulations capture several important aspects of jet propagation, the resulting morphologies  are very different from those observed in old FR\,I sources. Although the head of the jet in the decelerated models advances at a very low velocity by the end of the simulation, the shape and structure of the cocoon are likely to change radically when the bow-shock expands in the strong pressure gradient beyond the galactic core. The spherical structure observed around the injection region is expected to cool and become denser as the surrounding material falls back towards the galactic centre and is entrained across the contact discontinuity. This spherical cocoon  might be observable in young, low-power radio sources, but it is likely to disappear at later times due to mixing with the colder and denser material that surrounds it. Three-dimensional, long-term simulations should be performed to study the transition between the early stage studied here and the evolution of the whole structure when the bow-shock propagates out of the galactic core and becomes transonic.

\section{Summary and conclusions}
%            %%%%%%%%%%%%%%%%%%%%%%%
\label{conc}

  We have performed a series of axisymmetric simulations of the early
evolution (up to $2\times10^6$~yr) of FR\,I jets in a
realistic galactic environment to investigate the effects of mass loading by 
stellar winds. The simulations presented here allow us to
capture the effects of mass loading on beam evolution within the host
galaxy and are precise enough to discriminate between models 
differing by a factor of ten in the mass entrainment rate. Our results are
consistent with previous steady state simulations (BLK) and
theoretical estimates \citep{HB06}. 

 The overall structure and dynamics of the cocoon-shocked ambient
  system is very similar for all models with the same power and
  thrust. Slight differences in long-term evolution result from 
differences in the dynamics of the jet head (which are affected by the mass loading). The mass load 
carried by the beam affects its internal structure (internal shocks,
pinching, beam radius and jet opening angle) and properties (increase
in the rest mass density, cooling and deceleration).  

 We find that mass
entrainment rates consistent with present models of 
stellar mass loss in elliptical galaxies lead to deceleration of
the beam plasma and effective decollimation of weak ($L_{\rm j}\sim
10^{41}-10^{42}\,{\rm erg\, s^{-1}}$) FR\,I jets on scales of about 1
kpc. Deceleration  is accompanied by expansion and, in
axisymmetric simulations, the development of disruptive pinch modes
that lead to the decollimation and further deceleration of the jet
due to external entrainment. The composition of the jet is also completely changed 
by the process of mass loading, ruling out a dominant magnetic field on kiloparsec 
scales in low-power FR\,I jets. However, stellar mass loading  seems to be
unable to decelerate the most powerful ($L_{\rm j} \ga 10^{43}\,{\rm erg\, s^{-1}}$) FR\,I jets.
In these powerful jets, the
formation of strong recollimation shocks when the pressure in the
environment drops well below that of the jet or continuous mass entrainment 
produced by short-wavelength KH body modes, are more
plausible mechanisms for deceleration, entrainment and mixing \citep[see also][]{lb14}. 
Future work should include high-resolution three-dimensional simulations to allow for a detailed comparison 
between the different proposed mechanisms of deceleration. Our conclusions can also be tested observationally by modelling larger samples of weak and powerful FR\,I jets using the techniques developed by \citet{lb14}. We would expect transverse velocity gradients to develop in the powerful jets, but the weak jets should maintain their initial transverse velocity profiles as they decelerate.
  
%  The observed flaring of the FR\,I jets on kiloparsec scales should
%thus be triggered by stellar wind mass loading along weak FR\,I jets ($L_{\rm j} \leq
%10^{42}\,{\rm erg\, s^{-1}}$)
%and by continuous mass entrainment after the flaring point, caused probably 
%by shocks or instabilities, in powerful FR\,I jets ($L_{\rm j} \geq
%10^{43}\,{\rm erg\, s^{-1}}$).   

%  In the context of the FR\,I/FR\,II unification scheme, we conclude that,
%  for a given ambient medium, the long-term evolution of jets into FR\,I
%  or FR\,II morphology is basically controlled by the jet power and
%  thrust, the stellar mass-loss rate per unit volume within the
%  galaxy, and the jet overpressure with respect to its environment.  

\section*{Acknowledgments}
%             %%%%%%%%%%%%%%%%%

MP and JMM acknowledge financial support by the Spanish ``Ministerio de Ciencia e Innovaci\'on'' (MICINN) grants AYA2010-21322-C03-01, AYA2010-21097-C03-01 and CONSOLIDER2007-00050, and by the ``Generalitat Valenciana'' grant ``PROMETEO-2009-103''. 
PEH acknowledges support from NSF award AST-0908010 and 
NASA award NNX08AG83G to the University of Alabama. MP and JMM acknowledge Joan Ferrando for interesting discussions at the start of this project.

\label{lastpage}
\end{document}